\documentclass[a4paper,fleqn,usenatbib]{aa}

\usepackage[varg]{txfonts} 

\usepackage[T1]{fontenc}
\usepackage{ae,aecompl}

\usepackage{graphicx, rotating, booktabs} % Including figure files
\usepackage[verbose]{placeins}

\usepackage{amsmath}	% Advanced maths commands
\usepackage{amssymb}	% Extra maths symbols
\usepackage{float}
\usepackage{ragged2e}
\usepackage{breqn}

\usepackage{grffile} 
\usepackage{caption}
\usepackage{subcaption}
\usepackage{footnote}

\usepackage{geometry}
\usepackage{adjustbox}
\usepackage{mathrsfs}
\usepackage{siunitx}
\usepackage{mathtools}
\usepackage{pdflscape}
\usepackage{natbib}
\usepackage{stackengine}

\usepackage{array}
\newcolumntype{P}[1]{>{\centering\arraybackslash}p{#1}}
\newcolumntype{M}[1]{>{\centering\arraybackslash}m{#1}}
\usepackage{hyperref}
\usepackage[capitalise]{cleveref}
\usepackage{placeins}
\usepackage{appendix}

\usepackage[symbol]{footmisc}

\renewcommand{\thesubfigure}{\roman{subfigure}}

\begin{document}

%%%%%%%%%%%%%%%%%%% TITLE PAGE %%%%%%%%%%%%%%%%%%%

% Title of the paper, and the short title which is used in the headers.
% Keep the title short and informative.

\title{Multifrequency filter search for high redshift sources and lensing systems in \textit{Herschel}-ATLAS\thanks{Complete Tables \ref{tab:sample_370_a} and \ref{tab:sample_237_a} are available in electronic form at the CDS via \url{http://cdsweb.u-strasbg.fr/cgi-bin/qcat?J/A+A/}}}
\titlerunning{High-$z$ H-ATLAS sources}

\author{A. Manj\'on-Garc\'ia\inst{1}\thanks{E-mail: manjon@ifca.unican.es}
  \and D. Herranz\inst{1,2}
  \and J.~M. Diego\inst{1}
  \and L. Bonavera\inst{3}
  \and J. Gonz\'alez-Nuevo\inst{3}}

\institute{Instituto de F\'isica de Cantabria, CSIC-UC, Av. de Los Castros s/n, E-39005 Santander, Spain
  \and Departamento de F\'isica Moderna, Universidad de Cantabria, 39005-Santander, Spain
  \and Departamento de F\'isica, Universidad de Oviedo, C. Federico Garc\'ia Lorca 18, 33007 Oviedo, Spain} 

\authorrunning{Manj\'on-Garc\'ia et al.}

% These dates will be filled out by the publisher
\date{Received XXX / Accepted XXX}

% Don't change these lines
\hypersetup{draft} % Comentar al acabar

\label{firstpage}

\abstract{We present a new catalog of high-redshift candidate \textit{Herschel} sources. Our sample is obtained after applying a multifrequency filtering method (``matched multifilter''), which is designed to improve the signal-to-noise ratio (S/N) of faint extragalactic point sources. The method is tested against already-detected sources from the \textit{Herschel} Astrophysical Terahertz Large Area Survey (H-ATLAS) and used to search for new high-redshift candidates. The multifilter technique also produces an estimation of the photometric redshift of the sources. When compared with a sample of sources with known spectroscopic redshift, the photometric redshift returned from the multifilter is unbiased in the redshift range $0.8<z<4.3$. Using simulated data we reproduced the same unbiased result in roughly the same redshift range and determined the error (and bias above $z\approx4$) in the photometric redshifts. Based on the multifilter technique, and a selection based on color, flux, and agreement of fit between the observed photometry and assumed SED, we find 370 robust candidates to be relatively bright high-redshift sources. A second sample with 237 objects focuses on the faint end at high-redshift. These 237 sources were previously near the H-ATLAS detection limit but are now confirmed with our technique as high significance detections. Finally, we look for possible lensed \textit{Herschel} sources by cross-correlating the first sample of 370 objects with two different catalogs of known low-redshift objects, the redMaPPer Galaxy Cluster Catalog and a catalog of galaxies with spectroscopic redshift from the Sloan Digital Sky Survey Data Release 14. Our search renders a number of candidates to be lensed systems from the SDSS cross-correlation but none from the redMaPPeR confirming the more likely galactic nature of the lenses.}


\maketitle

%%%%%%%%%%%%%%%%% BODY OF PAPER %%%%%%%%%%%%%%%%%%%
\section{Introduction} \label{sec:intro}

%{\bf TO DO LIST:\\

%\noindent
%$\bullet$ Diego. Leer de arriba a abajo\\
%}

Over the last few decades, advances in the sensitivity of observations (specially in the IR part of the spectrum) and progress in data processing have allowed us to probe the high redshift universe in greater detail. The direct observation of galaxies in the redshift range $z\sim 1 - 10$ gives us the opportunity to study the cosmic history of star and galaxy formation at different cosmic epochs \citep[see for example][]{zotti10,Eales15}. However, despite the constant increase in diameter of the telescopes and increase of sensitivy of the detectors, observations of the distant universe are still flux-limited, rendering only those objects that are above the detection threshold. In a universe in which the inverse-square law prevails, a flux limit implies that the highest redshift galaxies accessible to any observatory will be among its faintest detectable objects. This situation is  alleviated for sources selected in the submillimetric range of the electromagnetic spectrum thanks to the strong,  
negative K correction, which leads to high-redshift galaxies being relatively easy to detect at submm wavelengths as compared with their low-redshift counterparts \citep{blain93}. In addition,
lucky alignments of background objects with foreground lenses can push the limits further by enhancing the flux of objects that could not be detected otherwise. But even with the aid of the negative K correction and  gravitational lensing, signal processing techniques are a fundamental tool to reach the faintest and most distant galaxies. This is particularly true in the microwave and far infrared (IR) parts of the electromagnetic spectrum, where the fluctuations from the cosmic infrared background (CIB) create a confusion noise whose level is comparable to the flux density of the typical high redshift galaxies. \\

The standard single-frequency detection methods for point sources in the CMB and far IR are based on wavelet techniques \citep{vielva03,barnard04,gnuevo06} or on the matched filter \citep[or MF hereafter,][see also \cite{herranz10} for a review.]{Tegmark1998,Herranz2002a,Barreiro2003,LopezCaniego2006}. Wavelets are well suited for the detection of compact sources due to their good position-scale determination properties, whereas the MF is the optimal linear detector-estimator because it provides the maximum signal-to-noise amplification for a source with a known shape (usually the point-spread function, or PSF hereafter, of the telescope) embedded in statistically homogeneous and spatially correlated noise. By default, these techniques are  applicable only to single-frequency sky images: even for multiwavelength observatories such as the \textit{Herschel} Space Observatory \citep{Pilbratt10} or \emph{Planck} \citep{Planck10}, the standard detection pipelines have produced individual source catalogs for each frequency band  \citep[see e.g.,][]{ERCSC,PCCS,PCCS2,Maddox2018}. The next logical step is to boost the signal of faint sources by combining the different bands into a single detection, that is, ``multifrequency detection''. Most of the blind component separation algorithms that are used for diffuse components in microwave and far IR astronomy can not deal with the high diversity of spectral behaviors associated to the different populations of extragalactic compact sources \citep[see for example][]{leach08}. However, over the last few years a number of multifrequency compact source detection techniques have been proposed in the literature \citep{MTXF1,MTXF2,Lanz2010,Lanz2013,PMNT}. A review on the topic can be found in \cite{Herranz12}. In particular, if the spatial profile and the spectral energy distribution (SED) of the sources are known, and if the cross-power spectrum is known,  or can be estimated from the data, the optimal linear detection method is the matched multifilter 
\citep[or MMF hereafter,][]{Herranz2002a}. \cite{Lanz2010} also showed that the MMF can be generalized for the case where the SED of the sources is not known. This generalization outperforms the single-frequency MF in terms of signal-to-noise ratio and can be used to infer the spectral index of synchrotron-dominated radio sources, as shown in \cite{Lanz2013}. However, in this paper we will incorporate a specific SED to the MMF in order to derive a photometric redshift estimation of dusty galaxies and high-redshift star forming galaxies detected in the IR part of the spectrum.\footnote{As measured at the observer's rest frame.}
We will do so by applying the multifrequency MMF filter to the first and second data releases of the \textit{Herschel} Astrophysical
Terahertz Large Area Survey \citep[the \textit{Herschel}-ATLAS or
H-ATLAS,][]{Eales10}, the largest single key project carried out in
open time with the \textit{Herschel} Space Observatory. We restrict our multifrequency analysis to the three wavelength bands covered by the SPIRE instrument aboard \textit{Herschel} \citep{SPIRE}, centered around 250, 350 and 500 $\mu$m. As discussed in \cite{Negrello2010}, \cite{Lapi11}, \cite{Gonzalez-Nuevo2012}, \cite{Pearson2013} and \cite{Donevski18}, the SPIRE bands are ideal for capturing the peak in the SED corresponding to dust emission of star-forming galaxies at $z \sim 2$, that is redshifted from its rest-frame wavelength around 70--100 $\mu$m to the SPIRE wavelengths: This is the redshift range where galaxies have formed most of their stars. At higher redshifts, dusty star-forming galaxies (DSFGs) occupy the most massive halos  
and are among the most luminous objects found at $z \gtrsim 4$
\citep{Michalowski14,Oteo16,Ikarashi17}. These high-redshift DSFGs have markedly red colors as seen by SPIRE, with rising flux densities from 250 to 500 $\mu$m (the so-called ``500 $\mu$m-risers"), and have received a great deal of attention in the recent years \citep[see for example][]{Ivison16,Negrello2017,strandet17}. The 
DSFGs, and particularly the 500 $\mu$m risers uncovered by \textit{Herschel}, are providing much insight into the early star forming history of the universe. However, sensitivity and limited angular resolution severely constrain the power of this type of objects as astrophysical probes. The sensitivity of SPIRE allows for the direct detection of only the brightest, and thus
rarest objects, at the bright end of the luminosity function. By means of our multifrequency MMF technique, we intend to enhance the detectability and statistical significance of very faint red objects in the H-ATLAS source catalog and so expand the list of reliable 500 $\mu$m-riser candidates.

\vspace{.5cm}

Although a non-negligible part of the faint H-ATLAS sources at $z$ > 1 could be detected thanks to having been amplified by weak lensing \citep{gnuevo14, gnuevo17}, most of the faint high-$z$ candidates in the H-ATLAS catalog have not been strongly lensed (with magnification factors larger than a few) by foreground halos \citep{Negrello2017}. In the other end of the flux density distribution, gravitational lensing plays an important role by magnifying distant galaxies that could be otherwise below the detection threshold or, at the very least, be observed with a significantly smaller flux  \citep{Negrello2007, Negrello2010,Negrello2014,Negrello2017,Cox2011,Conley2011,Lapi11,Gonzalez-Nuevo2012,Bussmann2012,Bussmann2013,Vieira2013,Wardlow2013,Canalog2014,Messias2014,Dye2015,Nayyeri2016,Spilker2016}. Gravitational lensing is a powerful astrophysical and cosmological probe particularly rewarding at submillimeter wavelengths. As mentioned before, submillimeter 
telescopes such as \textit{Herschel} have limited spatial resolution and consequently high source confusion, which makes it difficult to probe the dusty star forming galaxies. However, due to the relatively low probability of lensing (with typical magnification factors of a few), the identification of gravitational lenses is difficult and usually results in a few candidates. At high fluxes, wide-area submillimeter surveys can simply, and easily detect strong gravitational lensing events, with close to 100\% efficiency, as was proved by \cite{Negrello2010}. These are often strongly lensed galaxies (SLGs) with magnification factors of order ten that can be more easily detected owing to their magnified flux.  The identification of these lenses is of great interest for multiple reasons. They offer the possibility to study in greater detail distant galaxies and resolve some of their features. Also, the background galaxies can be used to reveal the internal structure of the lenses. Having a large catalog of SLGs will be important in future studies. For example, caustic crossing events on these galaxies can be used to study, not only distant luminous stars, but also the constituents of the lens itself. If a sizable fraction of dark matter is made of compact objects, caustic crossing events  can be used to set limits in their fraction on a range of masses from subsolar mass to tens of solar masses (through microlensing). This mass range can be difficult to probe otherwise \citep[see for instance][]{Diego18}.

\newpage

In this work we aim at producing a catalog of distant and faint IR sources. Also, we select from those the ones that are more likely to be gravitationally lensed. Our identification of lensed candidates takes advantage of our newly inferred redshifts. As it will be described in section \ref{sec:method} of this paper, the use of a parametrized SED template in the MMF technique allows us to provide a photometric estimation of the redshift of sources, that facilitates the identification of possible lensing systems (lens plus background source). Some of these systems will be confirmed (either as lens systems, or random alignments) in the near future with ground observations. Also, we should notice that our search for lensed systems was restricted to galaxies with known spectroscopic redshift from the SDSS. Extending the number of galaxies with redshift by using other data sets, like for instance with the data from the GAMA fields, would extend the number of candidates to lens systems not only contained in our various samples of high-$z$ \textit{Herschel} candidates. \\

The structure of this paper is as follows. Section~\ref{sec:method} describes the theoretical framework of
the MMF technique and the frequency dependence model used to estimate the
redshifts and fluxes of the SPIRE sources. In Section~\ref{sec:data} we describe the images and data of
the H-ATLAS survey on which we have applied the method. In Section~\ref{sec:simulations} we expose
the results of testing the method with simulations.
In Section~\ref{sec:calibration} we compare our MMF-derived flux densities and photometric redshifts with the H-ATLAS fluxes and spectroscopic redshifts of previously studied objects. Selection of high-$z$ \textit{Herschel} sources is explained in Section~\ref{sec:results} and the search for possible lensed \textit{Herschel} sources is reported in Section~\ref{candidate_lensed_sample}. Finally, conclusions are detailed in Section~\ref{sec:conclusions}. 

\section{Method} \label{sec:method}

The MMF is the optimal linear detection method when the frequency dependence and the 
spatial profile of the sources are known, and the cross-power spectrum of the noise is known or can be estimated from the data. In the Fourier space the MMF can be written as follows:

\vspace{-0.2cm}

\begin{equation}
  \begin{split}
    \boldsymbol{\mathrm{\Psi}} (q) &= \alpha \hspace{0.05cm} \mathbf{P}^{-1}   \mathbf{F}, \\
    \alpha^{-1} & \equiv \int dq \hspace{0.05cm} \mathbf{F}^{t} \hspace{0.05cm}  \mathbf{P}^{-1} \hspace{0.05cm} \mathbf{F}, \\
    \sigma^{2} &= \int dq \hspace{0.05cm} \boldsymbol{\mathrm{\Psi}}^{t} \hspace{0.05cm} \mathbf{P} \hspace{0.05cm} \boldsymbol{\mathrm{\Psi}} = \alpha
    \label{eq:mmf}
  \end{split}
\end{equation}
\noindent
where 
$\boldsymbol{\mathrm{\Psi}} (q)$ is the column vector of the filters $\boldsymbol{\mathrm{\Psi}} (q) = [\psi_{\nu} (q)]$, $\mathbf{F}$
is the column vector $\mathbf{F} = [f_{\nu} \tau_{\nu}] $, being $f_{\nu}$ the frequency dependence and $\tau_{\nu}$ the source profile at each
frequency $\nu$, $\mathbf{P}^{-1}$ is the inverse matrix of the cross-power spectrum $\mathbf{P}$ and $\sigma^{2}$ is the variance of the output-filtered image.
In Eq. (\ref{eq:mmf}) and in the following discussion, $q \equiv |\vec{q}|$ is the modulus of the Fourier wave vector; since we are assuming circularly symmetric source profiles, and since the cross-power spectrum only depends on the modulus $q$, all the formulas can be expressed in terms of $q$ instead of the full vector. However, it would be easy to generalize our formulas for non symmetric profiles just by replacing $q$ by $\vec{q}$ in the equations. Finally, $\alpha$ in Eq. (\ref{eq:mmf}) can be interpreted as the  normalization that is requested in order to guarantee that the filters $\Psi$ are unbiased estimators of the flux density of the sources under study. Further details can be found in \cite{Herranz2002a,Lanz2010,Lanz2013}.\\

Rewriting the vector $ \mathbf{F} = [f_{\nu} \tau_{\nu}] $ in the matrix form $ \mathbf{F} = \mathbf{T} (q) \boldsymbol{f} (\nu)$,
with diagonal matrix $ \mathbf{T} (q)$ = diag $[\tau_{1}(q)$, ... , $\tau_{N}(q)]$ and $ \boldsymbol{f} = [f_{\nu}]$
the vector of frequency dependence, we are able to include all the dependence in $q$ in the matrix $ \mathbf{T}$ pulling it completely apart from 
the dependence in $\nu$. This way Eq.~(\ref{eq:mmf}) can be rewritten as:

\begin{equation}
  \boldsymbol{\mathrm{\Psi}} (q) = \alpha \hspace{0.05cm} \mathbf{P}^{-1}  \mathbf{T} \boldsymbol{f},
  \hspace{0.4cm}  
  \alpha^{-1} = \int dq \hspace{0.05cm} \boldsymbol{f}^{t} \hspace{0.05cm} \mathbf{T} 
  \hspace{0.05cm}
  \mathbf{P}^{-1} \hspace{0.05cm} \mathbf{T}  \hspace{0.05cm} \boldsymbol{f} = 
  \boldsymbol{f}^{t} \hspace{0.05cm} \mathbf{H} \hspace{0.05cm} \boldsymbol{f}
  \label{eq:mmf2}
\end{equation}
\noindent
where matrix $\mathbf{H} = \int dq \hspace{0.05cm} \mathbf{T} \hspace{0.05cm} \mathbf{P}^{-1} 
\hspace{0.05cm} \mathbf{T}$ and we used the facts that $ \mathbf{T}^{t} = \mathbf{T} $ and that vector
$\boldsymbol{f}$ does not depend on $q$. \\

This reformulation of the Eq. (\ref{eq:mmf}) is very convenient for implementation of the MMF. The most time-consuming part of the filtering is the calculation of the matrices $\mathbf{P}$ and $\mathbf{T}$ since they must be calculated for all values of $q$. In the case we are considering in this paper the only quantity that varies during the maximization process is the redshift of the source we want to estimate. This allows us to compute the integrals of matrix $\mathbf{H}$ only once for each set of images of the source considered. \\

The MMF takes as argument a set of $N$ images corresponding to the same area of the sky observed simultaneously at $N$ different frequencies and returns a single filtered image where the source is optimally enhanced with respect to the noise. For $N$ images, the frequency dependence $f_{\nu}$ has $N$ degrees of freedom. Choosing one of the frequencies under
consideration as fiducial frequency of reference allows to reduce to $N$-1 the number of independent degrees of freedom. \\

The total filtered map is the result of a two-phase process. The first phase is the slowest one but, having separated the dependence in $q$ 
from the dependence in $\nu$ in Eq. (\ref{eq:mmf2}), it only needs to be done once for each set of images of the source considered. It consists on the calculation of a prefiltered map without any frequency dependence information, and for what is necessary to have previously calculated the Fourier transforms of the $N$ images and the filters without frequency dependence. The second phase is faster and requires only the calculations of the normalization $\alpha$ and the linear combination of prefiltered maps using a given frequency dependence $f_{\nu}$. The two necessary requirements to guarantee that the filtered field is optimal for the detection of point sources are that the filtered map is an unbiased estimator, on average, of the amplitude of the source (unbiased filter) and that the variance of the filtered map around that value is as small as possible, that is, that it is an efficient estimator of the amplitude of the source (maximum efficiency filter).\\

Summing up, in the first step, each individual frequency image is filtered with a linear filter, and in the second step all the resulting filtered maps
are combined so that the signal is boosted and the noise tends to cancel out.\\

The frequency dependence $f_{\nu}$ of the sources is not known a priori just with the information of the images. A template model from \cite{Pearson2013} developed to estimate redshifts using only the SPIRE fluxes from \textit{Herschel} has been used as frequency dependence for all the sources considered:

\begin{equation}
   S_{\nu} = A_{n} \Big [ B_{\nu(1+z)} \big(T_{h}\big) \big[\nu \cdot \big(1 + z\big) \big]^{\hspace{0.05cm}\beta}+a B_{\nu(1+z)}\big(T_{c}\big)\big[\nu \cdot \big(1+z\big) \big]^{\hspace{0.05cm}\beta} \Big]
   \label{eq:temp_model}
\end{equation}

\justify
where $S_{\nu}$ is the flux at a redshift frequency $\nu(1+z)$, $z$ is the unknown redshift of the source, $A_{n}$ is a normalization factor, $B_{\nu(1+z)}$ is the 
Planck function, $\beta$ = 2 is the emissivity index, $T_{h}$ = 46.9 K and $T_{c}$ = 23.9 K are the temperatures
of the hot and cold dust components, and $a$ = 30.1 is the ratio of the mass of cold dust to the mass of hot dust. \\

This template has emerged from a subset of 40 bright \textit{Herschel}-ATLAS sources with very well known redshifts in the range 0.5 < $z$ < 4.3. The redshifts of 25 of them, with $z$ < 1, were obtained through optical spectroscopy. The redshifts of the other 15 objects, in the range 0.8 < $z$ < 4.3, were estimated from CO observations. This SED has also already been used and studied in several previous works \citep{Eales15,Ivison16,Bianchini2016,Bianchini2018,Negrello2017,Fudamoto2017,Bakx2017,Donevski18}.\\

Given that all the sources used to build this model are among the most luminous H-ATLAS sources at their respective redshifts, a bias may arise from the fact that the model may not be representative of the less luminous sources. For instance, low-$z$ H-ATLAS sources have cooler SEDs than the template derived in \cite{Pearson2013} from their high-$z$ spectroscopic sample. It is important to bear in mind that the many different types of sources distributed in the sky constitute a very heterogeneous set of objects that do not have a common spectral behavior. This is the reason why the detection and estimation of the flux of point sources is a difficult task. In this sense, it should be noted that this template model is not expected to be a physically real SED but simply a representative model that can be used as a statistical tool for estimating redshifts from SPIRE fluxes. \\

We use source positions given by the H-ATLAS catalog and follow the procedure described above. 
For each source we maximize the signal-to-noise ratio (S/N) of the filtered map, defined as
\begin{equation}
  \mathrm{S/N} = \frac{A}{\sigma},
  \label{eq:snr}
\end{equation}
\noindent
with respect to the frequency dependence $f_{\nu}$. 
In the previous equation $A$ is the amplitude and $\sigma$ the standard deviation of the point source in the image.
Since for the frequency dependence we use the SED template (Eq. \ref{eq:temp_model})  with fixed $a$, $\beta$, $T_k$ and $T_c$ parameters, the only free parameters in the optimization are the source amplitude $A$ and its redshift $z$.
In fact, the amplitude is not really a variable, because for any given set of images it is determined by $z$ for any iteration of the filter through Eq. (\ref{eq:mmf}). 
By construction, the resulting $A$ coincides with the source's flux density when the optimization is completed. Therefore, the only variable in the optimization is $z$ and the maximization of the filtered S/N of a given source is tantamount to finding its redshift, provided Eq. (\ref{eq:temp_model}) is a valid description of its SED.
In the end, we have a maximized filtered image of the source, with an amplitude $A$ that corresponds to the flux density of the source in the chosen fiducial frequency. The fluxes of the source at the other frequencies can be obtained by multiplying this amplitude by the frequency dependence vector $f_{\nu}$, which is normalized to the fiducial frequency. This method is robust only in the case of point sources, that is, those whose spatial profile in each frequency agrees with the beam profile in that frequency.

\section{Data} \label{sec:data}

\textit{Herschel}-ATLAS is the extragalactic survey covering the widest area undertaken with \textit{Herschel} Space Observatory \citep{Pilbratt10}, imaging 659.25 $\mathrm{deg^{2}}$ of the sky distributed in five fields: three (GAMA9 with 53.43 $\mathrm{deg^{2}}$, GAMA12 with 53.56 $\mathrm{deg^{2}}$ and GAMA15 with 54.56 $\mathrm{deg^{2}}$) on the celestial equator, a large field (180.1 $\mathrm{deg^{2}}$) centered on the north Galactic pole (NGP) and an even larger field (317.6 $\mathrm{deg^{2}}$) centered on the south Galactic pole (SGP). Images have been taken in five far-infrared (far-IR) to submm photometric bands, 100, 160, 250, 350 and 500 $\mu$m, using the PACS and SPIRE instruments in parallel mode. PACS measurements have not been used in the main analysis of this work. The main reason is that the SED model (Eq.~\ref{eq:temp_model}) from \cite{Pearson2013} exploited to estimate the redshifts has been developed to use only the SPIRE fluxes. This was owing to that not all H-ATLAS sources have flux measurements at PACS wavelengths and only a few per cent of them were detected at greater that 5$\sigma$ in these bands. However, SPIRE bands themselves are ideal for capturing the emission peak belonging to the high-redshift sources aimed in this work. \\

Both Data Release 1 (DR1) and the recently released Data Release 2 (DR2) have been used in this analysis. \textit{Herschel}-ATLAS DR1 includes the three equatorial fields covered by the Galaxy And Mass Assembly (GAMA) \citep{Driver2009,Driver2016} spectroscopic survey. The three fields are $\sim$162 $\mathrm{deg^{2}}$ combined, and are approximatively located around 9h, 12h and 15h in $\alpha$. The associated catalog, described in \cite{Valiante2016} and \cite{Bourne2016}, covers all three regions and includes 120230 SPIRE sources, which have at least a S/N $=$ 4$\sigma$ (including confusion noise) in any of the 250, 350 or 500 $\mu$m maps. \textit{Herschel}-ATLAS DR2 covers the two fields centered in the NGP and SGP, which are about 450 $\mathrm{deg^{2}}$ combined. The maps are described in \cite{Smith2017} while the submillimeter catalog is described in \cite{Maddox2018}\footnote{H-ATLAS catalogs are public and available at \url{http://www.h-atlas.org/public-data/download}.} and include 118980 (NGP) and 193527 (SGP) sources, respectively. These sources have also at least a S/N $=$ 4$\sigma$ detection in all of the SPIRE bands. The complete H-ATLAS catalog contains a total of 432737 sources, most of them being point sources. After removing extended sources and stars, the catalog contains 410997 sources. \\

As explained in greater detail in \cite{Valiante2016}, sources were detected using the MADX algorithm (Multiband Algorithm for source Detection and eXtraction) applied to the SPIRE maps. The first step of this method is to use \textit{Nebuliser} to remove the diffuse Galactic dust emission from all maps in the three bands, resulting in raw images with the local large-scale background subtracted (``backsub" maps). Then the images are convolved with a proper matched-filter for each band \citep{Chapin2011}. Maps of the variance in each of these convolved maps are also created. During the convolution, the contribution of each pixel of the input image is weighted by the inverse of the square of the instrumental noise in that pixel. The resulting maps are background subtracted maps and noise-weighted maps filtered with a customized matched filter (``fbacksub" maps). In the next step (in MADX), the maps at 350 and 500 $\mu$m are interpolated onto images with the same pixel scale as the  250 $\mu$m one, and the three images and their corresponding variance maps are then combined together to form a single signal-to-noise or `detection' image. In practice, images at 350 and 500 $\mu$m are given a zero weighting regarding source detection, that is, the detection image in MADX is simply the 250 $\mu$m image. The position of the source in this image will be used to estimate the fluxes of the source in the 350 and 500 $\mu$m maps. A list of potential sources is produced by finding all peaks in the detection image with S/N > 2.5$\sigma$. A Gaussian fit is carried out in each of these peaks to provide an estimate of the source position and their flux densities are measured at the positions of those peaks in all the SPIRE bands. Finally, only sources with S/N > 4$\sigma$ in at least one of the three SPIRE bands are kept in the final catalog.\\

We have worked with the backsub maps instead of the fbacksub ones in order to test our own multifrequency matched-filter's performance without any other alteration but the subtraction of the large-scale background emission. The method used to subtract this large-scale emission does not affect the flux density of point sources. The units of the maps are Jy/beam. We converted these fluxes to Jy/pixel by dividing the values in the maps by the ratio between the beam area and the pixel area in $\mathrm{arcsec^{2}}$ (469/36, 831/64 and 1804/144 at 250, 350 and 500 $\mu$m, respectively). The maps have pixel sizes of 6, 8 and 12 arcsec at 250, 350 and 500 $\mu$m, respectively. All maps must have the same pixel size so as to be able to combine the three-channel images of a source into one single filtered image. Thus we re-binned 350 and 500 $\mu$m maps to a pixel size of 6 arcsec, the same pixel size than the 250 $\mu$m map. This repixelization may cause small alignment errors between the pixel positions of the center of the source in the different channels. These pixel misalignments have already been considered and monitored in the method. We have achieved a perfect alignment for $\sim$ 90\% of the H-ATLAS sources leaving the rest with deviations not greater than 2 pixels in one or some of the channels.

Once all maps have the same pixel size and are in units of Jy/pixel we apply our new algorithm on the positions of the 4$\sigma$ detections produced by the MADX algorithm in order to test our method. Positions ($\alpha$ and $\delta$) of all point sources identified in the maps are known and taken from the H-ATLAS catalog, converted from degrees into pixels and used to select the objects in the backsub maps. We extract square patches of 256x256 pixels centered on the position of the source for the three bands (250, 350 and 500 $\mu$m) and align them to run our MMF algorithm. When a H-ATLAS source is close to the edge of the H-ATLAS footprint, the zeros in the map are replaced by white noise generated with the same mean and standard deviation than the rest of the map (within the footprint). \\

Prior to the filtering step, a prefiltered map without any frequency dependence is built using the Fourier transforms of the three image patches and the matrix $\mathbf{\mathrm{H}}$. Cross-power spectrum of the images and matrices T are needed to get matrix $\mathbf{\mathrm{H}}$. $\mathbf{\mathrm{T}}$ matrices are calculated using the point spread functions (PSF) at 250, 350 and 500 $\mu$m as source profiles at each frequency. And the inverse of the cross-power spectrum matrix is calculated for each position of a pixel in the images. In the final phase of the filtering process, we introduce the frequency dependence shown in Eq. (\ref{eq:temp_model}). The 250 $\mu$m channel is chosen as the frequency of reference. Since the redshift $z$ of the source is unknown, this last step is repeated for equally distributed redshift values in the range 0 $\leq$ $z$ $\leq$ 7, with a step of 0.007 until the redshift which maximizes the S/N, that is, the optimal redshift, for the considered source is found. The result is an estimation of the redshift of the source, its frequency dependence vector $f_{\nu}$ derived from Eq. (\ref{eq:temp_model}) and a maximized filtered image of the source. The flux at 250 $\mu$m is directly the amplitude of the filtered image and the fluxes at 350 and 500 $\mu$m can be estimated multiplying this amplitude by the corresponding components of $f_{\nu}$.

\section{Simulations} \label{sec:simulations}
%%%%%%%%%%%%%%%%%%%%%%%%%%%%%%%%%%%%%%%%%%%%%%
In order to test our method, we used simulated data with a well known SED. Simulations are useful for both,  identifying possible biases and also to estimate the errors in the reconstructed redshift. Simulations were done using only GAMA's backsub maps from H-ATLAS DR1. The recently released NGP and SGP fields from H-ATLAS DR2 were not used for the simulations but this should have no impact on our results. \\

\begin{figure}
	\includegraphics[width=\columnwidth]{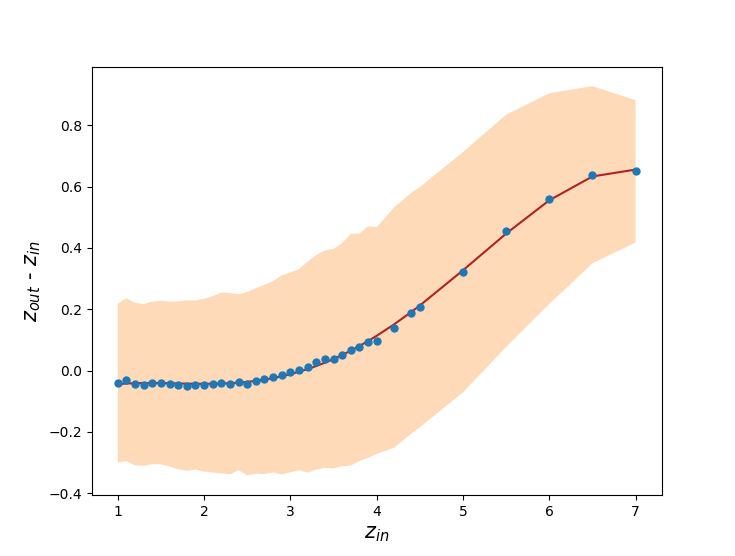}
    \caption{Difference between the redshift recovered with our MMF method ($z_{out}$) and the input redshift ($z_{in}$) as a function of $z_{in}$. 5000 simulations are run for each $z_{in}$ value in the range 1 $\leq$ $z_{in}$ $\leq$ 7. The mean value ($z_{out}$) is computed and shown as blue dots. The solid line shows a polynomial fit to these mean values. This fit is later used to correct for this bias. $z_{out}$-$z_{in}$ errors (1$\sigma$) are plotted as a shaded region.}
    \label{fig:log_normal}
\end{figure}

We started each simulation with a randomly chosen square patch of the desired size (256 x 256 pixels in our case) from any of the three equatorial fields surveyed (GAMA 9, GAMA 12 or GAMA 15). The same patch region was selected for the three submm photometric bands. Since the three SPIRE channels have different pixel sizes, and the MMF needs to work with a common pixel size, we re-bin the 350 and 500 $\mu$m maps to have the same pixel size than the 250 $\mu$m map. Alignment errors between the pixel positions of the source in the different channels (which may harm the MMF filtering result), can take place due to this repixelization but, as we already explained in Section~\ref{sec:data}, they have already been considered for the H-ATLAS sources, as they are for simulated sources. Thus, all maps used in simulations have a pixel size of 6 arcsec and are in units of Jy/pixel. Then a source with the corresponding beam profile (according to the PSF of the channel), an adequate amplitude (in order to obtain fluxes like the H-ATLAS ones), and a fixed redshift and SED (Eq.~\ref{eq:temp_model}), is placed in the middle of each one of the three patches. From this moment we followed the same procedure, described in Section~\ref{sec:data}, as with any H-ATLAS source. As it is done with the H-ATLAS sources, if the selected map patch contains zeros (i.e, it is near the edge of one of the GAMA fields), these are replaced by white noise with dispersion given by the map background. 
We performed 5000 simulations, as described before, for each one of the redshifts considered within the ranges 1 $\leq$ z $\leq$ 4.5, with a 0.1 step, and 4.5 $\leq$ z $\leq$ 7, with a 0.5 step. For each input redshift value considered, $z_{in}$, we compute the mean value of the 5000 output redshift values, $z_{out}$, and the standard deviation. The difference between the redshifts estimated after applying our MMF method to the simulated sources and the input redshift is shown in Fig.~\ref{fig:log_normal} as a function of $z_{in}$. \\

The bias observed above $z\approx4$ could be due (in part) to the fact that the Pearson model (Eq.~\ref{eq:temp_model}) is built based on some of the most luminous H-ATLAS sources and a restricted range in redshift (0.5 $\leq$ z $\leq$ 4.3). However, more importantly, photometric redshifts derived from a SED have problems when the peak of the IR emission is not bracketed by the three SPIRE bands. This peak falls in the SPIRE bands between $z$ $\sim$ 1 and $z$ $\sim$ 4, and it is precisely in this redshift range where our method seems more robust returning unbiased redshift estimates. Beyond $z\approx4$ a positive bias can be appreciated which can be as high as $\approx 0.6$ at $z\approx 7$. Using a polynomial fit, we find that our estimations of the redshift after applying the MMF can be corrected through:

\begin{dmath}
  z_{true} = 0.0013 \cdot z_{out}^{4} - 0.018 \cdot z_{out}^{3} \\ + 0.056 \cdot z_{out}^{2} + 0.943 \cdot z_{out} + 0.057
  \label{eq:bias_fit}
\end{dmath}
\noindent
where $z_{true}$ is the unbiased redshift estimation of the corresponding H-ATLAS source.

\section{Comparison with known-redshift H-ATLAS sources} \label{sec:calibration}
%%%%%%%%%%%%%%%%%%%%%%%%%%%%%%%%%%%%%%%%%%%%%%%%%%%%%%%%%%%%%%%%%%%%%%%%%%%%%%%%%%%%
We compare the redshifts obtained by the MMF method with a set of 32 \textit{Herschel}-ATLAS sources with known spectroscopic redshifts from \cite{Pearson2013}, \cite{Negrello2017} and \cite{Bakx2017}. 
Several of the sources selected are ubiquitous in all these references. Ten of these 32 sources are chosen from \cite{Negrello2017}, 17 are sources with $z_{spec}$ $>$ 0.8 used in \cite{Pearson2013} to build their template and five are taken from \cite{Bakx2017}. Redshifts and flux densities estimated with the MMF for these sources are shown in Table~\ref{tab:spectroscopic}. \\

\begin{figure}
	\includegraphics[width=\columnwidth]{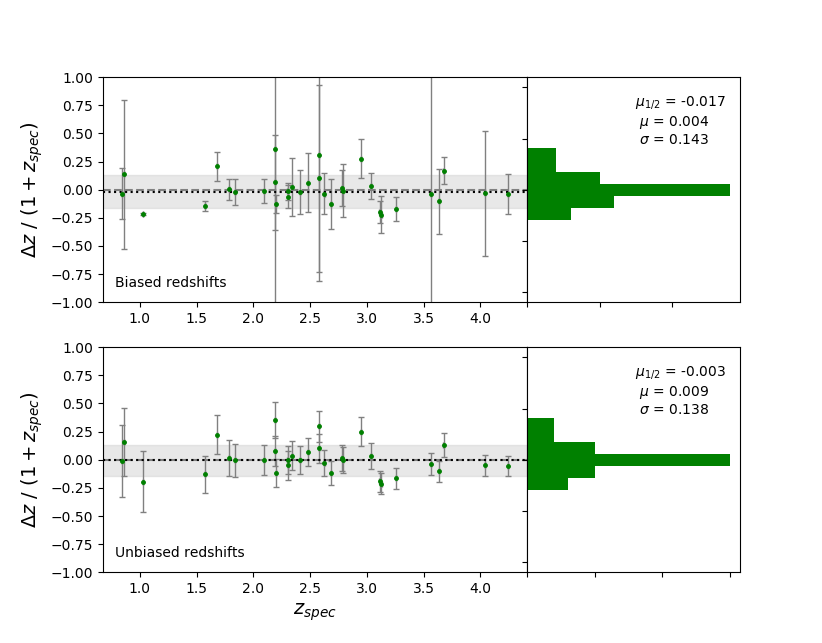}
    \caption{
    Difference, $\Delta z/(1+z_{spec})$, as a function of $z_{spec}$ between the biased (top) or unbiased (bottom) photometric redshifts estimated with our matched multifilter (MMF) and the spectroscopic redshifts from sources in Table~\ref{tab:spectroscopic}. The statistical parameters noted illustrate the systematic overestimates or underestimates, mean $\mu$ and median $\mu_{1/2}$, and the degree of scatter, $\sigma$, of the photometric redshifts ($z_{MMF}^{biased}$ and  $z_{MMF}^{unbiased}$) found using the MMF and the \protect\cite{Pearson2013} SED. Error bars in the top panel are calculated from using Eq. (\ref{eq:mmf}) and Eq. (\ref{eq:temp_model}) while error bars in the bottom panel are derived from simulations described in Section~\ref{sec:simulations}. The gray band depicts the 1$\sigma$ region centered in the median $\mu_{1/2}$ value for each distribution.}
    \label{fig:photo_vs_spec}
\end{figure}

The differences between photometric redshifts estimated with the MMF and the measured spectroscopic redshifts for these 32 objects are shown in Fig.~\ref{fig:photo_vs_spec}. The top plot shows $\Delta z/(1 + z_{spec}) = (z_{photo}-$z$_{spec})/(1+z_{spec})$ before the bias correction. The mean and median are $\mu$ = 0.004 and $\mu_{1/2}$ = -0.017 respectively, with an rms scatter of $\sigma$ = 0.143. The bottom plot shows the same quantity after the bias correction. Since at $z<4$ the bias correction is small, the improvement is small in this redshift range. Nevertheless, the mean and median ($\mu$ = 0.009, $\mu_{1/2}$ = -0.003), and scatter ($\sigma$ = 0.138) are slightly better than in the sample without bias correction. These statistical parameters are also included in the corresponding redshift column of Table~\ref{tab:spectroscopic}. If we take the definition for outliers (those with $\lvert\Delta$z / (1+$z_{spec}$)$\lvert$ > 0.3) used in \cite{Ivison16}, only one of the objects considered is identified as an outlier both for the analysis with biased and unbiased redshifts. Error bars in the top panel are calculated from using Eq. (\ref{eq:mmf}) and Eq. (\ref{eq:temp_model}) while error bars in the bottom panel are derived from simulations described in Section~\ref{sec:simulations}. We note how the error bars in the bottom panel are more representative of the dispersion around the zero value than the error bars in the top plot. This result indirectly confirms that the error bars derived from the simulation are the most meaningful ones for our estimated redshifts. For the meaning of the error bars in the top panel, see the following note:

{\textbf{A note on error bars:} from Eq. (\ref{eq:mmf}), and using a parametric SED such as Eq. (\ref{eq:temp_model}) it is possible, under some general (but not necessarily true) assumptions, such as the statistical independence of the background noise at the three different SPIRE channels, to estimate the degree of uncertainty of our photometric redshift $z_{MMF}$. The error bars of all redshift estimations from \cref{tab:spectroscopic,tab:spectroscopic1,tab:spectroscopic2}, except for $z_{MMF}^{unbiased}$, have been obtained this way. However the statistical \emph{uncertainty} of an estimator, and its actual \emph{error} with respect to the groundtruth are not necessarily the same thing. The uncertainty given to an estimator can be under, or overestimated depending on the validity of the statistical assumptions made. On the other hand, the estimator may be biased and this bias may not be accounted for in the calculation of the uncertainty. When possible, it is preferable to calculate the error of the estimator using real, already known values of $z_{spec}$ or, if few spectroscopic redshifts are available, by means of realistic simulations. This is the approach followed in this section to obtain the errors of the unbiased redshifts $z_{MMF}^{unbiased}$ from Table~\ref{tab:spectroscopic}, showing that for SPIRE, and in the redshift range $1 \leq z \leq 7$, the actual average error of the estimation of $z$ is under control and typically smaller than the uncertainty calculated from Eq. (\ref{eq:mmf}) and Eq. (\ref{eq:temp_model}).} \\

In order to test the robustness of our results, we have repeated the comparison with the spectroscopic sources, but changing the maps from which we extract these sources and the convolution functions used. We run our method, but using several combinations of the backsub and fbacksub maps as well as SPIRE's PSFs and MFs. The results obtained with the different configurations are shown in Table~\ref{tab:spectroscopic1}. We also obtain redshift estimates for these configurations by applying the Pearson's $\chi^2$ test statistic, but without using the MMF, and taking into account only the flux density measurements from different maps or the tabulated fluxes from the H-ATLAS catalog, and comparing them with the fluxes predicted by the \cite{Pearson2013} SED. These last results are shown in Table~\ref{tab:spectroscopic2}. \\

As can be seen by comparing the redshift estimations for the 32 known-spectroscopic sources from \cref{tab:spectroscopic,tab:spectroscopic1,tab:spectroscopic2}, both our unbiased and biased estimates obtained with the MMF method on the backsub maps and using the SPIRE PSFs outperform all the other redshift estimations derived with the alternative combinations of maps and convolution functions. Only redshifts estimated by applying a Pearson's $\chi^2$ test statistic with flux measurements from fbacksub maps ($z_{\chi^2}^{fb}$) and with tabulated fluxes from the H-ATLAS catalog ($z_{\chi^2}^{t}$) get comparable results. Nevertheless, our unbiased MMF redshifts are the ones which get lower offsets and scatter, and agree with spectroscopic redshifts with the greatest accuracy. \\

\begin{sidewaystable*}
  \centering      
  \caption{List of the 32 sources used to test the matched multifilter (MMF) method which includes a combination of sources from \protect\cite{Pearson2013}, \protect\cite{Negrello2017} and \protect\cite{Bakx2017}. For each source the following information is provided: measured spectroscopic redshift, \textit{Herschel}/SPIRE fluxes as tabulated in the H-ATLAS catalog ($S_{SPIRE}$), our redshift estimate before ($z_{MMF}$) and after ($z_{MMF}^{unbiased}$) bias correction, and the MMF flux densities. The statistical parameters $\mu$, $\mu_{1/2}$ and $\sigma$ at the bottom of the table are the mean, median and rms scatter of the quantity $\Delta z$/(1 + $z_{spec}$) for the corresponding photometric redshift.}
  \label{tab:spectroscopic}
        \tabcolsep=14pt
        \resizebox{\textwidth}{!}{\begin{tabular}{cccccccccccr}
        \hline
		H-ATLAS name & $\alpha$ ($deg$) & $\delta$ ($deg$) & $z_{spec}$ & & $S_{SPIRE}$ (mJy) & & $z_{MMF}$ & $z_{MMF}^{unbiased}$ & & $S_{MMF}$ (mJy) & \\
        \cline{5-7} \cline{10-12}
        & & & & 250 $\mu$m & 350 $\mu$m & 500 $\mu$m & & & 250 $\mu$m & 350 $\mu$m & 500 $\mu$m \\
		\hline
        \textbf{GAMA-9} &  & \\
        \textbf{H-ATLAS J083051.0+013225}  & 127.71268 & 1.54028 & $3.634^{\diamond}$ & 249 $\pm$ 7 & 305 $\pm$ 8 & 269 $\pm$ 9 & 3.15 $\pm$ 1.05 & 3.15 $\pm$ 0.34 & 230 $\pm$ 5 & 288 $\pm$ 7 & 265 $\pm$ 6 \\
        \vspace{.05cm}
        H-ATLAS J085358.9+015537  & 133.49574 & 1.92712 & $2.091^{\star}$ & 396 $\pm$ 8 & 368 $\pm$ 8 & 228 $\pm$ 9 & 2.05 $\pm$ 0.22 & 2.09 $\pm$ 0.28 & 365 $\pm$ 7 & 361 $\pm$ 7 & 243 $\pm$ 5 \\
        \vspace{.05cm}
        \textbf{H-ATLAS J091043.1-000321}  & 137.67949 & -0.0562 & $1.784^{\star}$ & 421 $\pm$ 7 & 371 $\pm$ 7 & 221 $\pm$ 8 & 1.79 $\pm$ 0.17 & 1.83 $\pm$ 0.28 & 388 $\pm$ 7 & 354 $\pm$ 6 & 218 $\pm$ 4 \\
        \vspace{.05cm}
        \textbf{H-ATLAS J090302.9-014127}  & 135.76235 & -1.69088 & $2.308^{\star}$ & 354 $\pm$ 7 & 339 $\pm$ 8 & 220 $\pm$ 9 & 2.10 $\pm$ 0.24 & 2.14 $\pm$ 0.29 & 313 $\pm$ 6 & 314 $\pm$ 7 & 215 $\pm$ 4 \\
        \vspace{.05cm}
        \textbf{H-ATLAS J090740.0-004200}  & 136.91691 & -0.70018 & $1.577^{\star}$ & 478 $\pm$ 7 & 328 $\pm$ 8 & 171 $\pm$ 9 & 1.20 $\pm$ 0.07 & 1.24 $\pm$ 0.26 & 450 $\pm$ 8  & 325 $\pm$ 6 & 160 $\pm$ 3 \\
        \vspace{.05cm}
        \textbf{H-ATLAS J090311.6+003906} & 135.79868 & 0.65206 & $3.037^{\star}$ & 133 $\pm$ 7 & 186 $\pm$ 8 & 165 $\pm$ 9 & 3.17 $\pm$ 0.34 & 3.17 $\pm$ 0.34 & 139 $\pm$ 5 & 174 $\pm$ 6 & 162 $\pm$ 6 \\
        \vspace{.05cm}
        \textbf{H-ATLAS J091305.0-005343} & 138.27077 & -0.89563 & $2.626^{\star}$ & 118 $\pm$ 6 & 137 $\pm$ 7 & 104 $\pm$ 8 & 2.50 $\pm$ 0.47 & 2.53 $\pm$ 0.30 & 123 $\pm$ 5 & 136 $\pm$ 6 & 105 $\pm$ 4 \\
        \vspace{.05cm}
        H-ATLAS J084933.4+021443  & 132.38934 & 2.24526 & $2.410^{\star}$ & 217 $\pm$ 8 & 249 $\pm$ 8 & 209 $\pm$ 9 & 2.35 $\pm$ 0.47 & 2.39 $\pm$ 0.30 & 201 $\pm$ 6 & 216 $\pm$ 7 & 160 $\pm$ 5 \\
        \vspace{.05cm}
        \textbf{H-ATLAS J091840.8+023047} & 139.67015 & 2.51348 & $2.581^{\star}$ & 126 $\pm$ 7 & 151 $\pm$ 8 & 128 $\pm$ 9 & 2.94 $\pm$ 2.14 & 2.95 $\pm$ 0.33 & 109 $\pm$ 6 & 132 $\pm$ 7 & 116 $\pm$ 6 \\ 
        
        \textbf{GAMA-12} &  & \\
        \textbf{H-ATLAS J114637.9-001132} & 176.65816 & -0.19226 & $3.259^{\star}$ & 316 $\pm$ 7 & 358 $\pm$ 7 & 292 $\pm$ 8 & 2.53 $\pm$ 0.36 & 2.56 $\pm$ 0.31 & 289 $\pm$ 6 & 322 $\pm$ 6 & 252 $\pm$ 5 \\
        \vspace{.05cm}
        H-ATLAS J114023.0-001043  & 175.09587 & -0.17810 & $0.844^{\star}$ & 68 $\pm$ 7 & 40 $\pm$ 8 & 20 $\pm$ 9 & 0.78 $\pm$ 0.19 & 0.82 $\pm$ 0.27 & 64 $\pm$ 8 & 38 $\pm$ 5 & 16 $\pm$ 2 \\
        \vspace{.05cm}
        H-ATLAS J115820.2-013753  & 179.58398 & -1.6313 & $2.191^{\star}$ & 120 $\pm$ 7 & 124 $\pm$ 8 & 101 $\pm$ 8 & 2.40 $\pm$ 0.93 & 2.44 $\pm$ 0.30 & 94 $\pm$ 7 & 102 $\pm$ 7 & 77 $\pm$ 5 \\
        \vspace{.05cm}
		H-ATLAS J113243.0-005108  & 173.17953 & -0.85248 & $2.578^{\star}$ & 68 $\pm$ 7 & 106 $\pm$ 8 & 100 $\pm$ 9 & 3.69 $\pm$ 2.90 & 3.64 $\pm$ 0.36 & 59 $\pm$ 4 & 81 $\pm$ 6 & 83 $\pm$ 6 \\
        \vspace{.05cm}
		\textbf{H-ATLAS J113526.3-014605} & 173.85958 & -1.76854 & $3.128^{\star}$ & 279 $\pm$ 7 & 283 $\pm$ 8 & 204 $\pm$ 9 & 2.21 $\pm$ 0.51 & 2.25 $\pm$ 0.29 & 260 $\pm$ 6 & 269 $\pm$ 6 & 191 $\pm$ 5 \\
        
        \textbf{GAMA-15} &  & \\
        H-ATLAS J142935.3-002836  & 217.39717 & -0.47682 & $1.026^{\star}$ & 801 $\pm$ 7 & 438 $\pm$ 7 & 200 $\pm$ 8 & 0.59 $\pm$ 0.01 & 0.63 $\pm$ 0.28 & 769 $\pm$ 7 & 413 $\pm$ 4 & 161 $\pm$ 2 \\
        \vspace{.05cm}
        \textbf{H-ATLAS J142413.9+022303} & 216.0582 & 2.38424 & $4.243^{\star}$ & 112 $\pm$ 7 & 182 $\pm$ 8 & 193 $\pm$ 8 & 4.04 $\pm$ 0.76 & 3.94 $\pm$ 0.38 & 117 $\pm$ 4 & 169 $\pm$ 6 & 186 $\pm$ 7 \\
        \vspace{.05cm}
        \textbf{H-ATLAS J141351.9-000026}  & 213.46663 & -0.00745 & $2.478^{\star}$ & 189 $\pm$ 7 & 217 $\pm$ 8 & 176 $\pm$ 9 & 2.70 $\pm$ 0.65 & 2.73 $\pm$ 0.31 & 166 $\pm$ 5 & 192 $\pm$ 6 & 158 $\pm$ 5 \\
        \vspace{.05cm}
        H-ATLAS J141148.9-011439  & 212.95392 & -1.24429 & $0.857^{\star}$ & 61 $\pm$ 7 & 44 $\pm$ 8 & 17 $\pm$ 9 & 1.11 $\pm$ 0.57 & 1.15 $\pm$ 0.26 & 53 $\pm$ 12 & 37 $\pm$ 9 & 17 $\pm$ 4 \\
        
        \textbf{NGP} &  & \\
        \textbf{H-ATLAS J134429.4+303034}  & 206.12285 & 30.50948 & $2.301^{\diamond}$ & 462 $\pm$ 8 & 466 $\pm$ 9 & 343 $\pm$ 9 & 2.25 $\pm$ 0.18 & 2.29 $\pm$ 0.29 & 443 $\pm$ 6 & 463 $\pm$ 7 & 332 $\pm$ 5 \\
        \vspace{.05cm}
        \textbf{H-ATLAS J132630.1+334407}  & 201.6255  & 33.73548 & $2.951^{\diamond}$ & 191 $\pm$ 7 & 281 $\pm$ 8 & 279 $\pm$ 9 & 4.04 $\pm$ 0.52 & 3.94 $\pm$ 0.38 & 171 $\pm$ 4 & 247 $\pm$ 6 & 270 $\pm$ 7 \\
        \vspace{.05cm}
        \textbf{H-ATLAS J125632.4+233627}  & 194.13524 & 23.60758 & $3.565^{\diamond}$ & 209 $\pm$ 7 & 288 $\pm$ 8 & 264 $\pm$ 9 & 3.41 $\pm$ 8 & 3.39 $\pm$ 0.35 & 201 $\pm$ 5 & 261 $\pm$ 7 & 256 $\pm$ 6 \\
        \vspace{.05cm}
        \textbf{H-ATLAS J132427.0+284449}  & 201.11257 & 28.74716 & $1.676^{\diamond}$ & 342 $\pm$ 7 & 371 $\pm$ 8 & 251 $\pm$ 9 & 2.23 $\pm$ 0.22 & 2.27 $\pm$ 0.29 & 329 $\pm$ 7 & 342 $\pm$ 7 & 244 $\pm$ 5 \\
        \vspace{.05cm}
        H-ATLAS J125135.3+261457  & 192.89721 & 26.24942 & $3.675^{\diamond}$ & 158 $\pm$ 8 & 202 $\pm$ 8 & 207 $\pm$ 9 & 4.47 $\pm$ 0.44 & 4.30 $\pm$ 0.39 & 112 $\pm$ 4 & 177 $\pm$ 6 & 206 $\pm$ 6 \\
        \vspace{.05cm}
        \textbf{H-ATLAS J133008.5+245900}  & 202.53579 & 24.98334 & $3.111^{\diamond}$ & 271 $\pm$ 7 & 278 $\pm$ 8 & 203 $\pm$ 8 & 2.29 $\pm$ 0.31 & 2.33 $\pm$ 0.29 & 255 $\pm$ 6 & 269 $\pm$ 7 & 196 $\pm$ 5 \\
        \vspace{.05cm}
        \textbf{H-ATLAS J133649.9+291800}  & 204.20819 & 29.3     & $2.202^{\diamond}$ & 294 $\pm$ 7 & 286 $\pm$ 8 & 194 $\pm$ 8 & 1.79 $\pm$ 0.17 & 1.83 $\pm$ 0.27 & 291 $\pm$ 7 & 265 $\pm$ 6 & 162 $\pm$ 4 \\
        \vspace{.05cm}
        \textbf{H-ATLAS J133542.9+300401}  & 203.92905 & 30.06709 & $2.685^{\diamond}$ & 137 $\pm$ 7 & 146 $\pm$ 8 & 125 $\pm$ 8 & 2.21 $\pm$ 0.60 & 2.25 $\pm$ 0.29 & 130 $\pm$ 6 & 134 $\pm$ 7 & 95 $\pm$ 5 \\
        \vspace{.05cm}
        \textbf{H-ATLAS J132859.2+292327}  & 202.2468  & 29.39074 & $2.778^{\diamond}$ & 268 $\pm$ 7 & 296 $\pm$ 7 & 249 $\pm$ 8 & 2.83 $\pm$ 0.44 & 2.85 $\pm$ 0.32 & 233 $\pm$ 5 & 275 $\pm$ 6 & 234 $\pm$ 5 \\
        \vspace{.05cm}
        H-ATLAS J125652.5+275900  & 194.21857 & 27.98339 & $2.79^{\circ}$ & 134 $\pm$ 8 & 164 $\pm$ 8 & 132 $\pm$ 9 & 2.76 $\pm$ 0.66 & 2.78 $\pm$ 0.32 & 129 $\pm$ 5 & 151 $\pm$ 6 & 126 $\pm$ 5 \\
        \vspace{.05cm}
        H-ATLAS J130118.0+253708  & 195.32517 & 25.61902 & $4.04^{\circ}$ & 60 $\pm$ 7 & 101 $\pm$ 8 & 102 $\pm$ 8 & 3.87 $\pm$ 2.23 & 3.79 $\pm$ 0.37 & 57 $\pm$ 4 & 80 $\pm$ 5 & 85 $\pm$ 6 \\
        \vspace{.05cm}
        H-ATLAS J132301.7+341649  & 200.75701 & 34.28036 & $2.19^{\circ}$ & 124 $\pm$ 7 & 145 $\pm$ 8 & 137 $\pm$ 9 & 3.35 $\pm$ 9 & 3.33 $\pm$ 0.35 & 102 $\pm$ 5 & 131 $\pm$ 6 & 127 $\pm$ 6 \\
        \vspace{.05cm}
        \textbf{H-ATLAS J132504.4+311534}  & 201.26821 & 31.25951 & $1.84^{\circ}$ & 241 $\pm$ 7 & 227 $\pm$ 8 & 165 $\pm$ 9 & 1.79 $\pm$ 0.21 & 1.83 $\pm$ 0.27 & 233 $\pm$ 8 & 212 $\pm$ 7 & 130 $\pm$ 4 \\
        \vspace{.05cm}
        \textbf{H-ATLAS J133846.5+255055}  & 204.69389 & 25.84855 & $2.34^{\circ}$ & 159 $\pm$ 8 & 183 $\pm$ 8 & 138 $\pm$ 9 & 2.42 $\pm$ 0.61 & 2.46 $\pm$ 0.30 & 151 $\pm$ 7 & 165 $\pm$ 7 & 125 $\pm$ 5 \\
		\hline
        $\mu_{1/2}$ & & & & & & & -0.017 & -0.003 & & & \\
        $\mu$ & & & & & & & 0.004 & 0.009 & & & \\
        $\sigma$ & & & & & & & 0.143 & 0.138 & & & \\
        \hline
    \end{tabular}}
    \begin{flushleft}
    \tiny ${\star}$ from \cite{Pearson2013} \\
    \tiny ${\diamond}$ from \cite{Negrello2017} \\
    \tiny ${\circ}$ from \cite{Bakx2017}
    \end{flushleft}
\end{sidewaystable*}

\begin{sidewaystable*}
  \centering      
  \caption{As in Table~\ref{tab:spectroscopic} but for fluxes and redshifts estimated using the backsub maps and convolving the images of the sources with the SPIRE MFs ($z_{MF}^{b}$ and $S_{MF}^{b}$), or using the fbacksub maps and convolving the images of the sources with the SPIRE MFs ($z_{MF}^{fb}$ and $S_{MF}^{fb}$), or  using the fbacksub maps and convolving the images of the sources with the SPIRE PSFs ($z_{PSF}^{fb}$ and $S_{PSF}^{fb}$).}
  \label{tab:spectroscopic1}
  \tabcolsep=0.35cm
        \resizebox{\textwidth}{!}{\begin{tabular}{cccccccccccccr}
        \hline
		H-ATLAS name & $z_{spec}$ & $z_{MF}^{b}$ & & $S_{MF}^{b}$ (mJy) & & $z_{MF}^{fb}$ & & $S_{MF}^{fb}$ (mJy) & & $z_{PSF}^{fb}$ & & $S_{PSF}^{fb}$ (mJy) & \\
        \cline{4-6} \cline{8-10} \cline{12-14}
        & & & 250 $\mu$m & 350 $\mu$m & 500 $\mu$m & & 250 $\mu$m & 350 $\mu$m & 500 $\mu$m & & 250 $\mu$m & 350 $\mu$m & 500 $\mu$m \\
		\hline
        \textbf{GAMA-9} &  & \\
        \textbf{H-ATLAS J083051.0+013225} & $3.634^{\diamond}$ & 2.50 $\pm$ 0.54 & 186 $\pm$ 5 & 206 $\pm$ 5 & 160 $\pm$ 4 & 2.13 $\pm$ 0.28 & 117 $\pm$ 4 & 119 $\pm$ 4 & 82 $\pm$ 3 & 3.19 $\pm$ 1.45 & 181 $\pm$ 5 & 228 $\pm$ 6 & 212 $\pm$ 5 \\
        \vspace{.05cm}
        H-ATLAS J085358.9+015537 & $2.091^{\star}$ & 1.47 $\pm$ 0.09 & 288 $\pm$ 6 & 233 $\pm$ 5 & 127 $\pm$ 3 & 1.14 $\pm$ 0.05 & 175 $\pm$ 4 & 123 $\pm$ 3 & 59 $\pm$ 2 & 1.89 $\pm$ 0.22 & 288 $\pm$ 5 & 272 $\pm$ 5 & 173 $\pm$ 3 \\
        \vspace{.05cm}
        \textbf{H-ATLAS J091043.1-000321} & $1.784^{\star}$ & 1.20 $\pm$ 0.05 & 290 $\pm$ 6 & 209 $\pm$ 4 & 103 $\pm$ 2 & 0.95 $\pm$ 0.05 & 113 $\pm$ 3 & 72 $\pm$ 2 & 32 $\pm$ 1 & 1.58 $\pm$ 0.15 & 216 $\pm$ 5 & 182 $\pm$ 4 & 104 $\pm$ 2 \\
        \vspace{.05cm}
        \textbf{H-ATLAS J090302.9-014127} & $2.308^{\star}$ & 1.64 $\pm$ 0.12 & 236 $\pm$ 6 & 204 $\pm$ 5 & 118 $\pm$ 3 & 1.23 $\pm$ 0.09 & 116 $\pm$ 4 & 85 $\pm$ 3 & 42 $\pm$ 2 & 1.97 $\pm$ 0.27 & 215 $\pm$ 5 & 208 $\pm$ 5 & 136 $\pm$ 3 \\
        \vspace{.05cm}
        \textbf{H-ATLAS J090740.0-004200} & $1.577^{\star}$ & 0.62 $\pm$ 0.02 & 347 $\pm$ 7 & 189 $\pm$ 4 & 75 $\pm$ 2 & 0 $\pm$ 0.01 & 202 $\pm$ 4 & 80 $\pm$ 2 & 25 $\pm$ 1 & 0.87 $\pm$ 0.03 & 349 $\pm$ 6 & 215 $\pm$ 4 & 93 $\pm$ 2 \\
        \vspace{.05cm}
        \textbf{H-ATLAS J090311.6+003906} & $3.037^{\star}$ & 2.96 $\pm$ 26 & 103 $\pm$ 4 & 125 $\pm$ 5 & 110 $\pm$ 4 & 3.47 $\pm$ 0.57 & 48 $\pm$ 2 & 64 $\pm$ 3 & 63 $\pm$ 3 & 3.57 $\pm$ 0.60 & 96 $\pm$ 4 & 129 $\pm$ 5 & 130 $\pm$ 5 \\
        \vspace{.05cm}
        \textbf{H-ATLAS J091305.0-005343} & $2.626^{\star}$ & 2.02 $\pm$ 0.33 & 91 $\pm$ 4 & 89 $\pm$ 4 & 59 $\pm$ 3 & 3.24 $\pm$ 1.01 & 30 $\pm$ 2 & 38 $\pm$ 2 & 35 $\pm$ 2 & 3.38 $\pm$ 1.24 & 62 $\pm$ 3 & 80 $\pm$ 4 & 78 $\pm$ 4 \\
        \vspace{.05cm}
        H-ATLAS J084933.4+021443 & $2.410^{\star}$ & 1.98 $\pm$ 0.31 & 138 $\pm$ 5 & 134 $\pm$ 5 & 88 $\pm$ 3 & 1.91 $\pm$ 0.42 & 71 $\pm$ 3 & 67 $\pm$ 3 & 43 $\pm$ 2 & 2.49 $\pm$ 1.38 & 141 $\pm$ 5 & 156 $\pm$ 5 & 120 $\pm$ 4 \\
        \vspace{.05cm}
        \textbf{H-ATLAS J091840.8+023047} & $2.581^{\star}$ & 2.80 $\pm$ 5 & 76 $\pm$ 5 & 90 $\pm$ 6 & 76 $\pm$ 5 & 3.14 $\pm$ 1.32 & 35 $\pm$ 3 & 44 $\pm$ 3 & 40 $\pm$ 3 & 3.31 $\pm$ 1.35 & 73 $\pm$ 4 & 93 $\pm$ 6 & 89 $\pm$ 5\\ 
        
        \textbf{GAMA-12} &  & \\
        \textbf{H-ATLAS J114637.9-001132} & $3.259^{\star}$ & 1.97 $\pm$ 0.17 & 201 $\pm$ 5 & 194 $\pm$ 4 & 127 $\pm$ 3 & 2.24 $\pm$ 0.81 & 81 $\pm$ 2 & 84 $\pm$ 2 & 60 $\pm$ & 2.82 $\pm$ 1.11 & 155 $\pm$ 4 & 184 $\pm$ 4 & 156 $\pm$ 4 \\
        \vspace{.05cm}
        H-ATLAS J114023.0-001043 & $0.844^{\star}$ & 0.27 $\pm$ 0.08 & 51 $\pm$ 7 & 23 $\pm$ 3 & 8 $\pm$ 1 & 0.58 $\pm$ 0.19 & 22 $\pm$ 4 & 12 $\pm$ 2 & 5 $\pm$ 1 & 0.69 $\pm$ 0.18 & 43 $\pm$ 6 & 24 $\pm$ 4 & 10 $\pm$ 2 \\
        \vspace{.05cm}
        H-ATLAS J115820.2-013753 & $2.191^{\star}$ & 2.27 $\pm$ 0.91 & 61 $\pm$ 5 & 64 $\pm$ 5 & 46 $\pm$ 4 & 2.31 $\pm$ 2 & 33 $\pm$ 3 & 35 $\pm$ 3 & 26 $\pm$ 2 & 2.61 $\pm$ 4.5 & 66 $\pm$ 4 & 75 $\pm$ 5 & 60 $\pm$ 4\\
        \vspace{.05cm}
		H-ATLAS J113243.0-005108 & $2.578^{\star}$ & 3.06 $\pm$ 14 & 42 $\pm$ 4 & 52 $\pm$ 4 & 47 $\pm$ 4 & 2.77 $\pm$ 2.18 & 30 $\pm$ 3 & 35 $\pm$ 4 & 29 $\pm$ 3 & 3.85 $\pm$ 3 & 51 $\pm$ 4 & 72 $\pm$ 5 & 76 $\pm$ 5\\
        \vspace{.05cm}
		\textbf{H-ATLAS J113526.3-014605} & $3.128^{\star}$ & 1.79 $\pm$ 0.19 & 198 $\pm$ 5 & 181 $\pm$ 5 & 111 $\pm$ 3 & 1.78 $\pm$ 0.24 & 106 $\pm$ 3 & 97 $\pm$ 3 & 59 $\pm$ 2 & 2.07 $\pm$ 0.56 & 190 $\pm$ 5 & 189 $\pm$ 5 & 128 $\pm$ 3\\
        
        \textbf{GAMA-15} &  & \\
        H-ATLAS J142935.3-002836 & $1.026^{\star}$ & 0 $\pm$ 0.01 & 578 $\pm$ 6 & 230 $\pm$ 2 & 72 $\pm$ 1 & 0 $\pm$ 0.01 & 247 $\pm$ 3 & 98 $\pm$ 1 & 31 $\pm$ 1 & 0.18 $\pm$ 0.01 & 469 $\pm$ 5 & 204 $\pm$ 2 & 68 $\pm$ 1\\
        \vspace{.05cm}
        \textbf{H-ATLAS J142413.9+022303} & $4.243^{\star}$ & 3.89 $\pm$ 0.57 & 86 $\pm$ 4 & 122 $\pm$ 5 & 130 $\pm$ 5 & 4.72 $\pm$ 0.38 & 34 $\pm$ 2 & 56 $\pm$ 3 & 67 $\pm$ 4 & 4.83 $\pm$ 0.36 & 69 $\pm$ 3 & 117 $\pm$ 5 & 143 $\pm$ 6\\
        \vspace{.05cm}
        \textbf{H-ATLAS J141351.9-000026} & $2.478^{\star}$ & 2.36 $\pm$ 0.53 & 119 $\pm$ 5 & 128 $\pm$ 5 & 95 $\pm$ 4 & 2.12 $\pm$ 0.76 & 60 $\pm$ 3 & 62 $\pm$ 3 & 42 $\pm$ 2 & 2.89 $\pm$ 5.6 & 112 $\pm$ 4 & 134 $\pm$ 5 & 115 $\pm$ 4\\
        \vspace{.05cm}
        H-ATLAS J141148.9-011439 & $0.857^{\star}$ & 1.30 $\pm$ 0.59 & 35 $\pm$ 8 & 27 $\pm$ 6 & 14 $\pm$ 3 & 1.12 $\pm$ 0.50 & 20 $\pm$ 4 & 14 $\pm$ 3 & 7 $\pm$ 1 & 1.42 $\pm$ 0.75 & 37 $\pm$ 8 & 29 $\pm$ 6 & 16 $\pm$ 3\\
        
        \textbf{NGP} &  & \\
        \textbf{H-ATLAS J134429.4+303034} & $2.301^{\diamond}$ & 1.72 $\pm$ 0.09 & 341 $\pm$ 6 & 304 $\pm$ 5 & 182 $\pm$ 3 & 2.03 $\pm$ 0.27 & 171 $\pm$ 3 & 169 $\pm$ 3 & 112 $\pm$ 2 & 2.40 $\pm$ 0.79 & 312 $\pm$ 5 & 338 $\pm$ 5 & 254 $\pm$ 4\\
        \vspace{.05cm}
        \textbf{H-ATLAS J132630.1+334407} & $2.951^{\diamond}$ & 3.86 $\pm$ 0.45 & 118 $\pm$ 3 & 165 $\pm$ 5 & 176 $\pm$ 5 & 5.73 $\pm$ 0.25 & 35 $\pm$ 1 & 75 $\pm$ 3 & 103 $\pm$ 4 & 5.40 $\pm$ 0.23 & 84 $\pm$ 2 & 165 $\pm$ 5 & 216 $\pm$ 6\\
        \vspace{.05cm}
        \textbf{H-ATLAS J125632.4+233627} & $3.565^{\diamond}$ & 3.27 $\pm$ 1.46 & 144 $\pm$ 4 & 184 $\pm$ 5 & 174 $\pm$ 5 & 4.97 $\pm$ 0.26 & 47 $\pm$ 2 & 82 $\pm$ 3 & 102 $\pm$ 4 & 4.93 $\pm$ 0.26 & 97 $\pm$ 3 & 169 $\pm$ 5 & 209 $\pm$ 6\\
        \vspace{.05cm}
        \textbf{H-ATLAS J132427.0+284449} & $1.676^{\diamond}$ & 1.87 $\pm$ 0.16 & 241 $\pm$ 6 & 226 $\pm$ 5 & 142 $\pm$ 3 & 2.00 $\pm$ 0.38 & 103 $\pm$ 3 & 100 $\pm$ 3 & 66 $\pm$ 2 & 2.31 $\pm$ 0.53 & 205 $\pm$ 5 & 217 $\pm$ 5 & 159 $\pm$ 4\\
        \vspace{.05cm}
        H-ATLAS J125135.3+261457 & $3.675^{\diamond}$ & 4.16 $\pm$ 0.43 & 84 $\pm$ 3 & 124 $\pm$ 5 & 139 $\pm$ 5 & 4.27 $\pm$ 0.37 & 44 $\pm$ 2 & 67 $\pm$ 3 & 76 $\pm$ 3 & 4.67 $\pm$ 0.33 & 84 $\pm$ 3 & 137 $\pm$ 5 & 164 $\pm$ 6\\
        \vspace{.05cm}
        \textbf{H-ATLAS J133008.5+245900} & $3.111^{\diamond}$ & 1.84 $\pm$ 0.18 & 195 $\pm$ 5 & 181 $\pm$ 5 & 113 $\pm$ 3 & 2.33 $\pm$ 1.52 & 69 $\pm$ 3 & 74 $\pm$ 3 & 54 $\pm$ 2 & 2.46 $\pm$ 0.60 & 131 $\pm$ 4 & 144 $\pm$ 5 & 111 $\pm$ 4\\
        \vspace{.05cm}
        \textbf{H-ATLAS J133649.9+291800} & $2.202^{\diamond}$ & 1.44 $\pm$ 0.10 & 212 $\pm$ 6 & 169 $\pm$ 5 & 91 $\pm$ 3 & 1.84 $\pm$ 0.32 & 75 $\pm$ 3 & 70 $\pm$ 3 & 44 $\pm$ 2 & 2.03 $\pm$ 0.41 & 160 $\pm$ 5 & 157 $\pm$ 5 & 105 $\pm$ 3\\
        \vspace{.05cm}
        \textbf{H-ATLAS J133542.9+300401} & $2.685^{\diamond}$ & 2.21 $\pm$ 0.64 & 89 $\pm$ 5 & 92 $\pm$ 5 & 65 $\pm$ 4 & 3.13 $\pm$ 1.53 & 32 $\pm$ 3 & 40 $\pm$ 3 & 36 $\pm$ 3 & 2.68 $\pm$ 2.46 & 72 $\pm$ 4 & 83 $\pm$ 5 & 68 $\pm$ 4\\
        \vspace{.05cm}
        \textbf{H-ATLAS J132859.2+292327}  & $2.778^{\diamond}$ & 2.21 $\pm$ 0.23 & 177 $\pm$ 4 & 183 $\pm$ 4 & 129 $\pm$ 3 & 2.19 $\pm$ 0.62 & 78 $\pm$ 2 & 81 $\pm$ 2 & 57 $\pm$ 2 & 2.96 $\pm$ 1.15 & 146 $\pm$ 4 & 177 $\pm$ 4 & 156 $\pm$ 4 \\
        \vspace{.05cm}
        H-ATLAS J125652.5+275900  & $2.79^{\circ}$ & 2.45 $\pm$ 0.66 & 98 $\pm$ 5 & 108 $\pm$ 5 & 82 $\pm$ 4 & 3.34 $\pm$ 0.89 & 46 $\pm$ 3 & 60 $\pm$ 3 & 57 $\pm$ 3 & 3.26 $\pm$ 1.75 & 91 $\pm$ 4 & 116 $\pm$ 5 & 110 $\pm$ 5 \\
        \vspace{.05cm}
        H-ATLAS J130118.0+253708  & $4.04^{\circ}$ & 6.24 $\pm$ 0.60 & 17 $\pm$ 1 & 43 $\pm$ 3 & 62 $\pm$ 5 & 5.62 $\pm$ 0.54 & 15 $\pm$ 1 & 30 $\pm$ 2 & 41 $\pm$ 3 & 5.23 $\pm$ 0.60 & 34 $\pm$ 2 & 63 $\pm$ 4 & 81 $\pm$ 5 \\
        \vspace{.05cm}
        H-ATLAS J132301.7+341649  & $2.19^{\circ}$ & 2.90 $\pm$ 5 & 81 $\pm$ 4 & 97 $\pm$ 5 & 84 $\pm$ 4 & 3.28 $\pm$ 1.23 & 32 $\pm$ 2 & 41 $\pm$ 3 & 39 $\pm$ 3 & 4.29 $\pm$ 0.74 & 55 $\pm$ 3 & 83 $\pm$ 5 & 94 $\pm$ 6 \\
        \vspace{.05cm}
        \textbf{H-ATLAS J132504.4+311534}  & $1.84^{\circ}$ & 1.46 $\pm$ 0.14 & 170 $\pm$ 6 & 137 $\pm$ 5 & 75 $\pm$ 3 & 2.33 $\pm$ 2.12 & 62 $\pm$ 3 & 66 $\pm$ 3 & 49 $\pm$ 3 & 2.42 $\pm$ 0.66 & 118 $\pm$ 5 & 128 $\pm$ 5 & 97 $\pm$ 4 \\
        \vspace{.05cm}
        \textbf{H-ATLAS J133846.5+255055}  & $2.34^{\circ}$ & 2.07 $\pm$ 0.42 & 115 $\pm$ 5 & 115 $\pm$ 5 & 78 $\pm$ 4 & 3.13 $\pm$ 1.24 & 46 $\pm$ 3 & 57 $\pm$ 4 & 52 $\pm$ 3 & 3.18 $\pm$ 1.88 & 94 $\pm$ 4 & 119 $\pm$ 6 & 110 $\pm$ 5 \\
        \hline
        $\mu_{1/2}$ & & -0.128 & & & & 0.045 & & & & 0.115 & & &\\
        $\mu$ & & -0.092 & & & & -0.016 & & & & 0.088 & & &\\
        $\sigma$ & & 0.201 & & & & 0.272 & & & & 0.230 & & &\\
        \hline
    \end{tabular}}
    \begin{flushleft}
    \tiny ${\star}$ from \cite{Pearson2013} \\
    \tiny ${\diamond}$ from \cite{Negrello2017} \\
    \tiny ${\circ}$ from \cite{Bakx2017}
    \end{flushleft}
\end{sidewaystable*}

\begin{sidewaystable*}
  \centering      
    \caption{As in Table~\ref{tab:spectroscopic} but for redshifts estimated by applying a Pearson's $\chi^2$ test statistic considering flux measurements from backsub ($S_{\chi^2}^{b}$) and fbacksub ($S_{\chi^2}^{fb}$) maps, respectively, as the observed data and the fluxes obtained directly with the \protect{\cite{Pearson2013}} SED (Eq. \ref{eq:temp_model}) as the theoretical data ($z_{\chi^2}^{b}$ and $z_{\chi^2}^{fb}$); and by applying a $\chi^2$ statistic considering the tabulated fluxes from the H-ATLAS catalog as the observed data and the fluxes obtained directly with the \protect{\cite{Pearson2013}} SED as the theoretical data ($z_{\chi^2}^{t}$).}
  \label{tab:spectroscopic2}
        \tabcolsep=0.65cm
        \resizebox{\textwidth}{!}{\begin{tabular}{ccccccccccr}
        \hline
		H-ATLAS name & $z_{spec}$ & $z_{\chi^2}^{b}$ & & $S_{\chi^2}^{b}$ (mJy) & & $z_{\chi^2}^{fb}$ & & $S_{\chi^2}^{fb}$ (mJy) & & $z_{\chi^2}^{t}$ \\
        \cline{4-6} \cline{8-10}
        & & & 250 $\mu$m & 350 $\mu$m & 500 $\mu$m & & 250 $\mu$m & 350 $\mu$m & 500 $\mu$m & \\
		\hline
        \textbf{GAMA-9} &  & \\
        \textbf{H-ATLAS J083051.0+013225} & $3.634^{\diamond}$ & 3.22 $\pm$ 1 & 219 $\pm$ 12 & 283 $\pm$ 11 & 258 $\pm$ 12 & 3.02 $\pm$ 1 & 247 $\pm$ 9 & 303 $\pm$ 9 & 269 $\pm$ 9 & 3.01 $\pm$ 1 \\
        \vspace{.05cm}
        H-ATLAS J085358.9+015537 & $2.091^{\star}$ & 2.02 $\pm$ 1 & 348 $\pm$ 12 & 347 $\pm$ 11 & 223 $\pm$ 12 & 1.82 $\pm$ 1 & 394 $\pm$ 9 & 356 $\pm$ 8 & 227 $\pm$ 9 & 1.84 $\pm$ 1 \\
        \vspace{.05cm}
        \textbf{H-ATLAS J091043.1-000321} & $1.784^{\star}$ & 1.66 $\pm$ 1 & 406 $\pm$ 10 & 339 $\pm$ 9 & 214 $\pm$ 10 & 1.70 $\pm$ 1 & 416 $\pm$ 8 & 361 $\pm$ 8 & 220 $\pm$ 8 & 1.70 $\pm$ 1\\
        \vspace{.05cm}
        \textbf{H-ATLAS J090302.9-014127} & $2.308^{\star}$ & 1.99 $\pm$ 1 & 316 $\pm$ 11 & 296 $\pm$ 10 & 207 $\pm$ 11 & 2.00 $\pm$ 1 & 337 $\pm$ 8 & 327 $\pm$ 8 & 218 $\pm$ 8 & 1.94 $\pm$ 1\\
        \vspace{.05cm}
        \textbf{H-ATLAS J090740.0-004200} & $1.577^{\star}$ & 1.35 $\pm$ 1 & 424 $\pm$ 12 & 321 $\pm$ 11 & 172 $\pm$ 11 & 1.18 $\pm$ 1 & 467 $\pm$ 9 & 323 $\pm$ 9 & 169 $\pm$ 9 & 1.16 $\pm$ 1\\
        \vspace{.05cm}
        \textbf{H-ATLAS J090311.6+003906} & $3.037^{\star}$ & 3.81 $\pm$ 1 & 114 $\pm$ 11 & 171 $\pm$ 10 & 161 $\pm$ 11 & 3.48 $\pm$ 1 & 131 $\pm$ 8 & 184 $\pm$ 8 & 166 $\pm$ 8 & 3.43 $\pm$ 1\\
        \vspace{.05cm}
        \textbf{H-ATLAS J091305.0-005343} & $2.626^{\star}$ & 2.47 $\pm$ 1 & 119 $\pm$ 9  & 132 $\pm$ 8  & 100 $\pm$ 9 & 2.59 $\pm$ 1 & 117 $\pm$ 7 & 135 $\pm$ 7 & 105 $\pm$ 7 & 2.58 $\pm$ 1\\
        \vspace{.05cm}
        H-ATLAS J084933.4+021443 & $2.410^{\star}$ & 2.95 $\pm$ 1 & 183 $\pm$ 11 & 223 $\pm$ 10 & 193 $\pm$ 11 & 2.71 $\pm$ 1 & 214 $\pm$ 8 & 246 $\pm$ 8 & 206 $\pm$ 8 & 2.72 $\pm$ 1\\
        \vspace{.05cm}
        \textbf{H-ATLAS J091840.8+023047} & $2.581^{\star}$ & 3.19 $\pm$ 1 & 118 $\pm$ 12 & 150 $\pm$ 12 & 138 $\pm$ 12 & 2.99 $\pm$ 1 & 120 $\pm$ 10 & 151 $\pm$ 10 & 128 $\pm$ 10 & 2.87 $\pm$ 1\\ 
        
        \textbf{GAMA-12} &  & \\
        \textbf{H-ATLAS J114637.9-001132} & $3.259^{\star}$ & 2.64 $\pm$ 1 & 293 $\pm$ 9 & 333 $\pm$ 9 & 271 $\pm$ 10 & 2.63 $\pm$ 1 & 313 $\pm$ 7 & 346 $\pm$ 8 & 291 $\pm$ 9 & 2.64 $\pm$ 1\\
        \vspace{.05cm}
        H-ATLAS J114023.0-001043 & $0.844^{\star}$ & 1.24 $\pm$ 1 & 71 $\pm$ 11 & 46 $\pm$ 10 & 29 $\pm$ 11 & 0.95 $\pm$ 1 & 66 $\pm$ 8 & 44 $\pm$ 8 & 20 $\pm$ 8 & 0.91 $\pm$ 1\\
        \vspace{.05cm}
        H-ATLAS J115820.2-013753 & $2.191^{\star}$ & 2.79 $\pm$ 1 & 101 $\pm$ 11 & 115 $\pm$ 10 & 101 $\pm$ 11 & 2.52 $\pm$ 1 & 115 $\pm$ 10 & 124 $\pm$ 9 & 101 $\pm$ 10 & 2.42 $\pm$ 1\\
        \vspace{.05cm}
		H-ATLAS J113243.0-005108 & $2.578^{\star}$ & 3.34 $\pm$ 1 & 77 $\pm$ 12 & 100 $\pm$ 9 & 96 $\pm$ 10 & 4.03 $\pm$ 1 & 67 $\pm$ 9 & 106 $\pm$ 8 & 101 $\pm$ 8 & 3.97 $\pm$ 1\\
        \vspace{.05cm}
		\textbf{H-ATLAS J113526.3-014605} & $3.128^{\star}$ & 2.40 $\pm$ 1 & 240 $\pm$ 11 & 263 $\pm$ 10 & 195 $\pm$ 12 & 2.31 $\pm$ 1 & 265 $\pm$ 7 & 287 $\pm$ 8 & 204 $\pm$ 9 & 2.19 $\pm$ 1\\
        
        \textbf{GAMA-15} &  & \\
        H-ATLAS J142935.3-002836 & $1.026^{\star}$ & 0.79 $\pm$ 1 & 733 $\pm$ 9 & 415 $\pm$ 9 & 195 $\pm$ 10 & 0.77 $\pm$ 1 & 782 $\pm$ 8 & 428 $\pm$ 8 & 206 $\pm$ 8 & 0.72 $\pm$ 1\\
        \vspace{.05cm}
        \textbf{H-ATLAS J142413.9+022303} & $4.243^{\star}$ & 4.18 $\pm$ 1 & 111 $\pm$ 11 & 162 $\pm$ 11 & 185 $\pm$ 12 & 4.35 $\pm$ 1 & 111 $\pm$ 8 & 177 $\pm$ 9 & 192 $\pm$ 9 & 4.37 $\pm$ 1\\
        \vspace{.05cm}
        \textbf{H-ATLAS J141351.9-000026} & $2.478^{\star}$ & 2.84 $\pm$ 1 & 164 $\pm$ 11 & 196 $\pm$ 10 & 165 $\pm$ 11 & 2.78 $\pm$ 1 & 179 $\pm$ 8 & 212 $\pm$ 8 & 175 $\pm$ 9 & 2.67 $\pm$ 1\\
        \vspace{.05cm}
        H-ATLAS J141148.9-011439 & $0.857^{\star}$ & 1.81 $\pm$ 1 & 57 $\pm$ 20 & 48 $\pm$ 19 & 35 $\pm$ 18 & 1.11 $\pm$ 1 & 59 $\pm$ 19 & 44 $\pm$ 18 & 18 $\pm$ 17 & 1.06 $\pm$ 1\\
        
        \textbf{NGP} &  & \\
        \textbf{H-ATLAS J134429.4+303034} & $2.301^{\diamond}$ & 2.06 $\pm$ 1 & 459 $\pm$ 12 & 439 $\pm$ 11 & 313 $\pm$ 11 & 2.22 $\pm$ 1 & 457 $\pm$ 8 & 463 $\pm$ 8 & 340 $\pm$ 8 & 2.21 $\pm$ 1\\
        \vspace{.05cm}
        \textbf{H-ATLAS J132630.1+334407} & $2.951^{\diamond}$ & 3.59 $\pm$ 1 & 187 $\pm$ 12 & 246 $\pm$ 11 & 256 $\pm$ 11 & 3.95 $\pm$ 1 & 186 $\pm$ 9 & 279 $\pm$ 8 & 278 $\pm$ 8 & 3.87 $\pm$ 1\\
        \vspace{.05cm}
        \textbf{H-ATLAS J125632.4+233627} & $3.565^{\diamond}$ & 3.15 $\pm$ 1 & 216 $\pm$ 12 & 267 $\pm$ 11 & 251 $\pm$ 11 & 3.54 $\pm$ 1 & 202 $\pm$ 9 & 287 $\pm$ 9 & 261 $\pm$ 9 & 3.45 $\pm$ 1\\
        \vspace{.05cm}
        \textbf{H-ATLAS J132427.0+284449} & $1.676^{\diamond}$ & 2.13 $\pm$ 1 & 338 $\pm$ 12 & 344 $\pm$ 11 & 235 $\pm$ 12 & 2.23 $\pm$ 1 & 339 $\pm$ 10 & 357 $\pm$ 9 & 250 $\pm$ 9 & 2.25 $\pm$ 1\\
        \vspace{.05cm}
        H-ATLAS J125135.3+261457 & $3.675^{\diamond}$ & 3.39 $\pm$ 1 & 148 $\pm$ 12 & 177 $\pm$ 11 & 193 $\pm$ 11 & 3.61 $\pm$ 1 & 150 $\pm$ 9 & 201 $\pm$ 8 & 206 $\pm$ 9 & 3.46 $\pm$ 1\\
        \vspace{.05cm}
        \textbf{H-ATLAS J133008.5+245900} & $3.111^{\diamond}$ & 2.21 $\pm$ 1 & 259 $\pm$ 12 & 257 $\pm$ 11 & 194 $\pm$ 11 & 2.29 $\pm$ 1 & 264 $\pm$ 8 & 274 $\pm$ 8 & 204 $\pm$ 8 & 2.24 $\pm$ 1\\
        \vspace{.05cm}
        \textbf{H-ATLAS J133649.9+291800} & $2.202^{\diamond}$ & 1.97 $\pm$ 1 & 277 $\pm$ 10 & 267 $\pm$ 10 & 174 $\pm$ 11 & 2.02 $\pm$ 1 & 290 $\pm$ 8 & 281 $\pm$ 9 & 192 $\pm$ 9 & 2.03 $\pm$ 1\\
        \vspace{.05cm}
        \textbf{H-ATLAS J133542.9+300401} & $2.685^{\diamond}$ & 3.04 $\pm$ 1 & 117 $\pm$ 11 & 149 $\pm$ 10 & 127 $\pm$ 11 & 2.60 $\pm$ 1 & 135 $\pm$ 8 & 143 $\pm$ 8 & 125 $\pm$ 8 & 2.57 $\pm$ 1\\
        \vspace{.05cm}
        \textbf{H-ATLAS J132859.2+292327}  & $2.778^{\diamond}$ & 2.80 $\pm$ 1 & 239 $\pm$ 9 & 281 $\pm$ 9 & 237 $\pm$ 9 & 2.68 $\pm$ 1 & 258 $\pm$ 8 & 286 $\pm$ 8 & 247 $\pm$ 7 & 2.61 $\pm$ 1 \\
        \vspace{.05cm}
        H-ATLAS J125652.5+275900  & $2.79^{\circ}$ & 3.04 $\pm$ 1 & 122 $\pm$ 11 & 143 $\pm$ 11 & 138 $\pm$ 11 & 2.98 $\pm$ 1 & 128 $\pm$ 8 & 162 $\pm$ 8 & 136 $\pm$ 8 & 2.82 $\pm$ 1 \\
        \vspace{.05cm}
        H-ATLAS J130118.0+253708  & $4.04^{\circ}$ & 4.34 $\pm$ 1 & 56 $\pm$ 10 & 99 $\pm$ 9 & 89 $\pm$ 10 & 4.39 $\pm$ 1 & 59 $\pm$ 8 & 100 $\pm$ 8 & 101 $\pm$ 8 & 4.37 $\pm$ 1 \\
        \vspace{.05cm}
        H-ATLAS J132301.7+341649  & $2.19^{\circ}$ & 2.77 $\pm$ 1 & 126 $\pm$ 12 & 131 $\pm$ 11 & 130 $\pm$ 11 & 2.99 $\pm$ 1 & 123 $\pm$ 9 & 140 $\pm$ 9 & 136 $\pm$ 9 & 2.99 $\pm$ 1 \\
        \vspace{.05cm}
        \textbf{H-ATLAS J132504.4+311534}  & $1.84^{\circ}$ & 2.21 $\pm$ 1 & 206 $\pm$ 12 & 207 $\pm$ 11 & 152 $\pm$ 12 & 2.05 $\pm$ 1 & 237 $\pm$ 8 & 218 $\pm$ 9 & 164 $\pm$ 9 & 2.05 $\pm$ 1 \\
        \vspace{.05cm}
        \textbf{H-ATLAS J133846.5+255055}  & $2.34^{\circ}$ & 2.51 $\pm$ 1 & 146 $\pm$ 12 & 162 $\pm$ 11 & 127 $\pm$ 12 & 2.59 $\pm$ 1 & 156 $\pm$ 9 & 184 $\pm$ 9 & 136 $\pm$ 9 & 2.54 $\pm$ 1 \\
		\hline
        $\mu_{1/2}$ & & 0.028 & & & & 0.008 & & & & -0.002 \\
        $\mu$ & & 0.041 & & & & 0.024 & & & & 0.011 \\
        $\sigma$ & & 0.151 & & & & 0.137 & & & & 0.135 \\  
        \hline
    \end{tabular}}
    \begin{flushleft}
    \tiny ${\star}$ from \cite{Pearson2013} \\
    \tiny ${\diamond}$ from \cite{Negrello2017} \\
    \tiny ${\circ}$ from \cite{Bakx2017}
    \end{flushleft}
\end{sidewaystable*}

\begin{figure*}
\minipage{0.33\textwidth}
  \includegraphics[width=\linewidth]{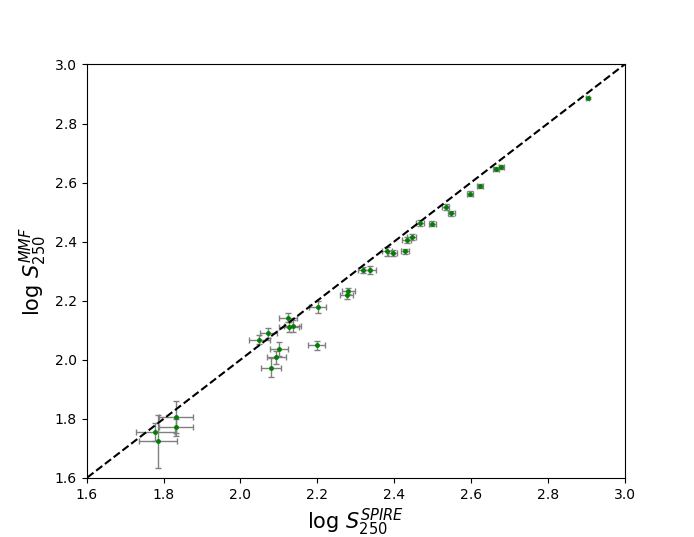}
\endminipage\hfill
\minipage{0.33\textwidth}
  \includegraphics[width=\linewidth]{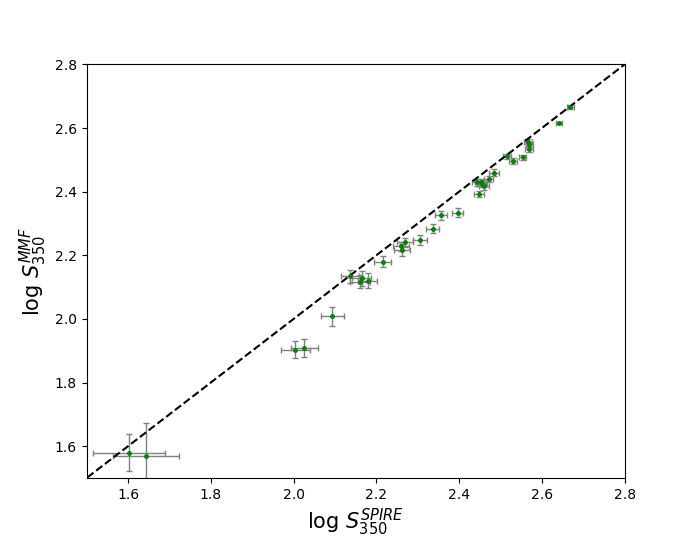}
\endminipage\hfill
\minipage{0.33\textwidth}
  \includegraphics[width=\linewidth]{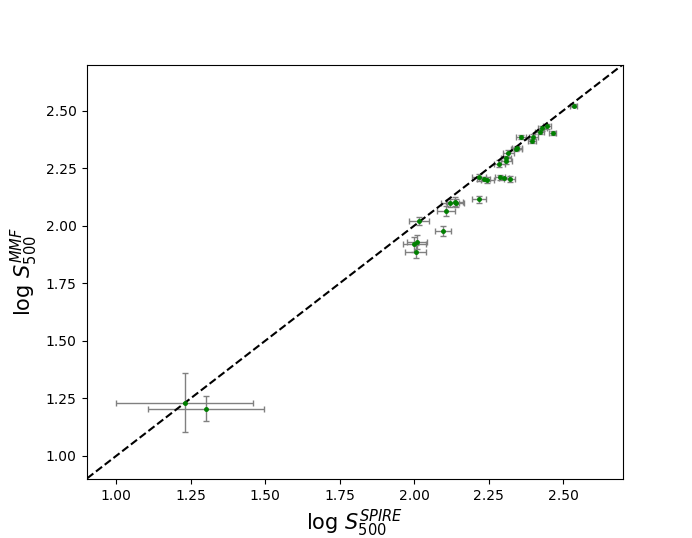}
\endminipage
\caption{Flux measurements in the 250 (left), 350 (middle) and 500 $\mu$m (right) SPIRE channels derived with the MMF versus the corresponding flux measurements from the H-ATLAS catalog in logarithmic scale for the 32 known-spectroscopic sources from Table~\ref{tab:spectroscopic}. The dashed line marks perfect correlation. All fluxes are in units of mJy/beam according to the beam profile of the respective channel.} \label{fig:log_spec_fluxes}
\end{figure*}

Focusing now our attention on flux densities, practically all recovered fluxes with the MMF ($S_{MMF}$) are consistent with the corresponding tabulated fluxes from H-ATLAS catalog ($S_{SPIRE}$), in the sense that the IR peak is recovered at the right corresponding wavelength for 29 out of the 32 sources considered from Table~\ref{tab:spectroscopic}. On the other hand, and as expected, all IR peaks are recovered in the right band for the flux densities $S_{\chi^2}^{fb}$ taken from the fbacksub maps (see Table~\ref{tab:spectroscopic2}). The comparison between our MMF estimates of the flux densities and those from the H-ATLAS catalog is shown in Fig.~\ref{fig:log_spec_fluxes}. It can be seen how our flux estimations seem to be systematically below the values from the H-ATLAS catalog. This slight underestimate is expected since the noise reduction carried out by the MMF must lead to flux densities lower than the H-ATLAS ones. The average flux underestimates between the flux densities estimated from the MMF method and the H-ATLAS fluxes are 17 $\pm$ 13 mJy at 250 $\mu$m, 18 $\pm$ 9 mJy at 350 $\mu$m and 14 $\pm$ 14 mJy at 500 $\mu$m.

\section{High-$z$ candidates in H-ATLAS} \label{sec:results}
%%%%%%%%%%%%%%%%%%%%%%%%%%%%%%%%%%%%%%%%%%%%%%%%%%%%%%%%%%%%
In this section we describe our strategy to find high-$z$ candidates in the H-ATLAS data. In the two subsections below, we explore two different strategies. A first subsample (section \ref{sample_selection}) is defined, where the candidates have to be visible in all three bands in SPIRE (this will define our bright subsample of high-$z$ candidates) and for which a more reliable estimate of the redshift can be obtained. 
In the second subsample (section \ref{sec:risers}), we focus on the 500-$\mu$m risers where the highest flux is found in the band with the longest wavelength. Although this does not guarantee that the source is at high redshift, all of the most distant objects in H-ATLAS will be 500-$\mu$m risers, as the peak of the IR emission will be at wavelengths longer than 500-$\mu$m. 

\subsection{The bright subsample: High-$z$ candidates with photometric redshift estimates} \label{sample_selection}
%---------------------------------------------------------
In order to define a sample of reliable high-$z$ candidates in the H-ATLAS, and motivated by the work exposed in \cite{Negrello2017}, we follow a strategy based on applying different cuts to the official photometric catalog and the results of running the MMF. This strategy is based on flux cuts, color cuts, and agreement of fit between the photometric measurements and an assumed SED. By applying a series of cuts to the full sample, we reduce the number of candidates until we arrive to a small subsample of objects which meet all our criteria. Since we focus on high-$z$ candidates, some of the cuts are designed to remove low-$z$ sources. We describe these cuts in detail below:\\*

\begin{figure*}
    \begin{subfigure}{.5\textwidth}
    \centering
    \includegraphics[width=.8\linewidth]{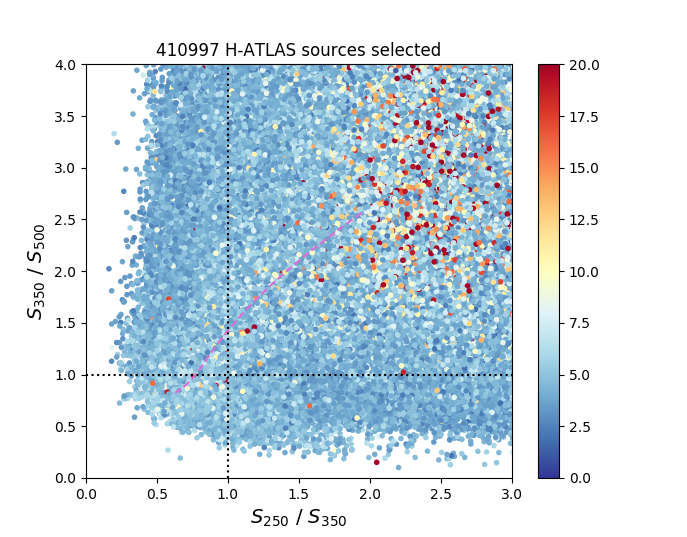}
    \caption{Removal of stars and extended sources.}
    \label{fig:cut_1}
  \end{subfigure}\hfil
  \begin{subfigure}{.5\textwidth}
    \centering
    \includegraphics[width=.8\linewidth]{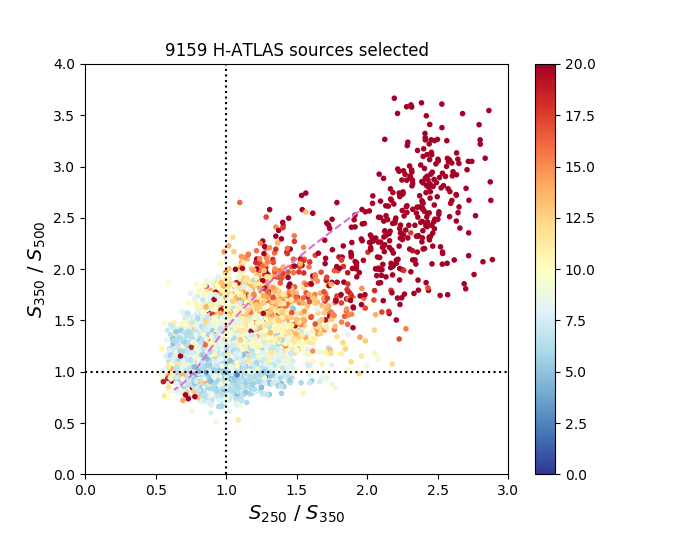}
    \caption{S/N $\geq$ 5$\sigma$ in the three SPIRE channels.}
    \label{fig:cut_2}
  \end{subfigure}\hfil
  \begin{subfigure}{.5\textwidth}
    \centering
    \includegraphics[width=.8\linewidth]{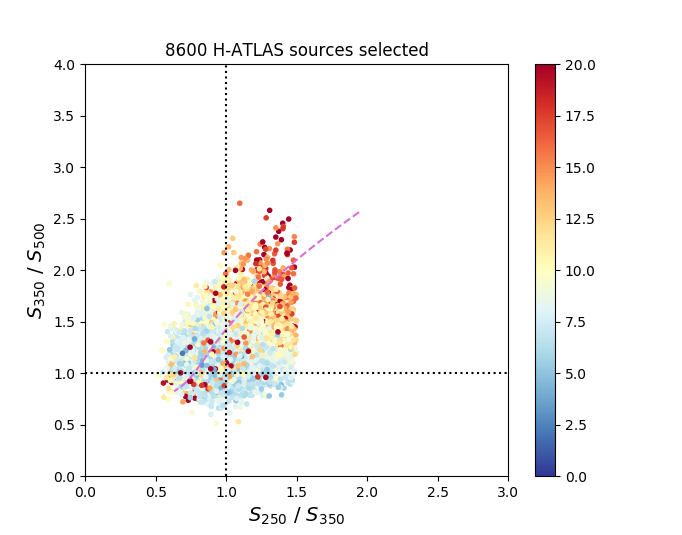}
    \caption{Exclusion of local galaxies at low redshifts.}
    \label{fig:cut_3}
  \end{subfigure}\hfil
  \begin{subfigure}{.5\textwidth}
    \centering
    \includegraphics[width=.8\linewidth]{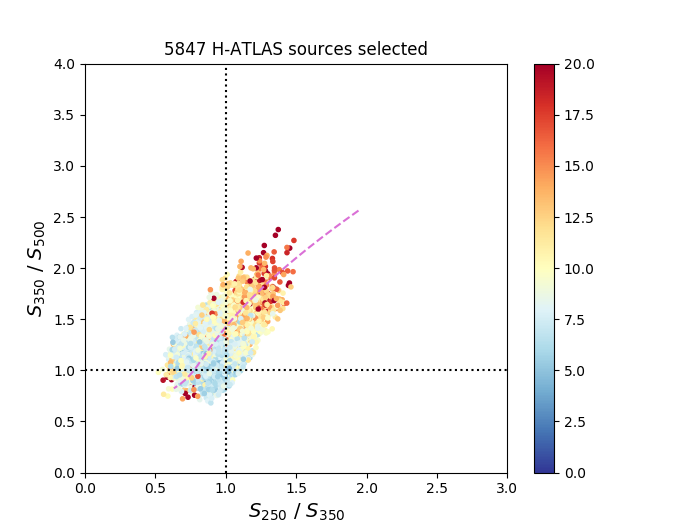}
    \caption{Photometric behavior close to the SED model.}
    \label{fig:cut_4}
  \end{subfigure}\hfil
  \begin{subfigure}{.5\textwidth}
    \centering
    \includegraphics[width=.8\linewidth]{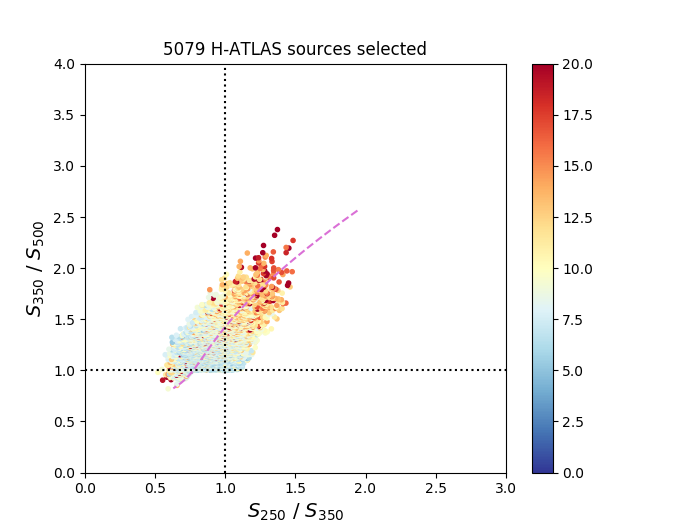}
    \caption{Exclusion of possible blazars.}
    \label{fig:cut_5}
  \end{subfigure}\hfil
  \begin{subfigure}{.5\textwidth}
    \centering
    \includegraphics[width=.8\linewidth]{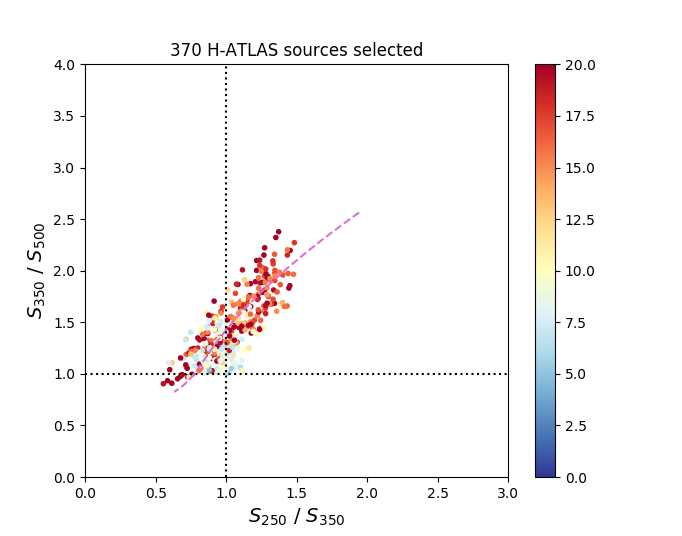}
    \caption{S/N $\geq$ 15$\sigma$ except if source has an association.}
    \label{fig:cut_6}
  \end{subfigure}\hfil
\caption{Evolution of the color-color diagram of the H-ATLAS sources studied as cuts are applied in order to get a sample of robust high-$z$ candidates. The dashed pink line is the track of the \protect\cite{Pearson2013} SED for redshifts in the range [0.5,4.5] (in increasing order from the top-right to the bottom-left corner). The vertical lateral coloured bar present in all plots is a scale of the S/N of the sources exhibited, achieved with the MMF.}
\label{fig:fig}
\end{figure*}

i) Since our method only works for the case of strictly point sources and high-$z$ galaxies will appear as unresolved, first we discard all sources we know for sure are non-pointlike, that is, sources which have aperture fluxes at the three SPIRE wavelength different from the point source fluxes. We retain only those sources for which their aperture radius has `-99'-value in the H-ATLAS catalog, which means that the aperture flux and the point source flux are the same. We also remove sources identified as stars and those with null or negative fluxes in any of the channels. This results in a sample of 410997 objects from H-ATLAS (see Fig.~\ref{fig:cut_1}), on all of which our MMF method is applied.

\begin{figure}
	\includegraphics[width=\columnwidth]{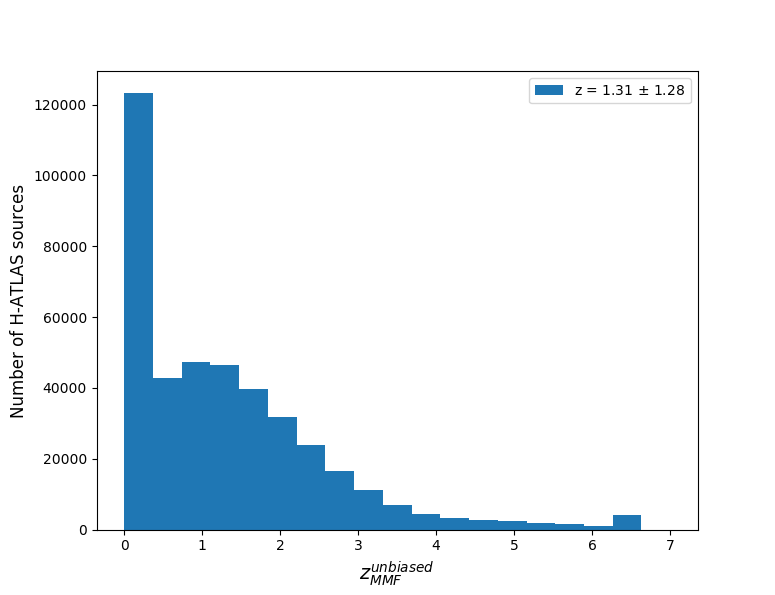}
    \caption{Redshift distribution, according to the redshift estimates obtained with the MMF method, of the 410997 sources from H-ATLAS selected after removing the non-point objects and the ones identified as stars.}
    \label{fig:histogram_410997}
\end{figure}

The redshift distribution found for these H-ATLAS objects is shown in Fig.~\ref{fig:histogram_410997}. An important peak close to zero-redshift can be seen. This figure shows clearly that there are a lot of sources that could be either, a) low redshift sources or b) sources for which their frequency dependence does not resemble the Pearson model considered in Eq. (\ref{eq:temp_model}), and hence are not adequate to be used with our method (resulting in erroneous redshifts and fluxes). \\

\begin{figure}
	\includegraphics[width=\columnwidth]{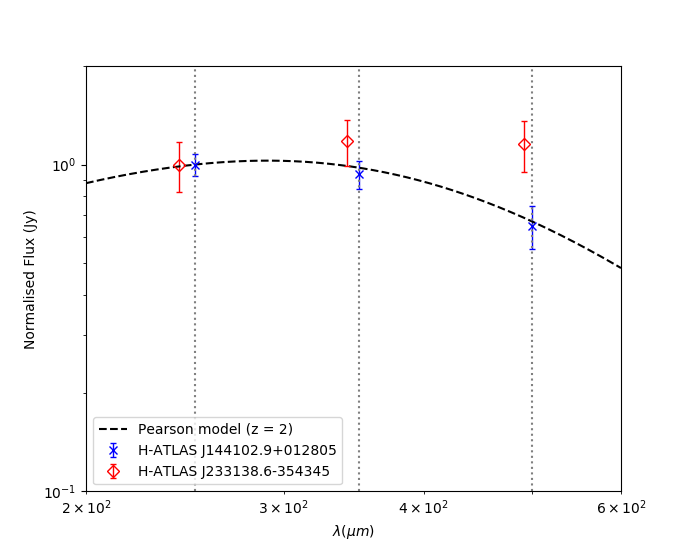}
    \caption{Normalized SED from \protect{\cite{Pearson2013}}, as defined in Eq. (\ref{eq:temp_model}), at $z=2$ in contrast with the normalized tabulated fluxes at 250, 350 and 500 $\mu$m (vertical dotted) of two H-ATLAS sources at $z_{MMF}=2$: one (J144102.9+012805) that fits well to the model and other one (J233138.6-354345), whose points have been slightly displaced in x-axis to get better clarity, that does not fit properly to the model according to our criteria.}
    \label{fig:two_sources}
\end{figure}

ii) After removing stars and non-point sources, we proceed to make a preselection using the photometric information of the H-ATLAS catalog. We want our sources to be at high redshift and bright enough so our redshift estimations are robust. Thus we select those which have a S/N greater than 5$\sigma$ in all three SPIRE channels, which leaves us with just 9159 sources, 2.2\% of the total (see Fig.~\ref{fig:cut_2}). This is the most stringent cut. \\

iii) Good candidates are required to have a flux ratio between 250 and 350 $\mu$m bands less than or equal to 1.5 (i.e $S_{250}$/$S_{350}$ $\leq 1.5$). This cut has the effect of excluding local galaxies at low redshifts. (see Fig.~\ref{fig:cut_3}). \\

iv) One important requirement for our preselection is to ensure that the chosen sources have a photometric behavior close to the response offered by the Pearson model (Eq.~\ref{eq:temp_model}) used to estimate their redshifts, since, as we discussed earlier in Section~\ref{sec:method}, the method does not work equally well for all H-ATLAS sources (see Fig.~\ref{fig:two_sources}). This can clearly be seen reflected in the large number of sources far away from the \cite{Pearson2013} SED model in Fig.~\ref{fig:cut_1}. We discard sources that are at a distance larger than 0.3 from the Pearson model in the color-color diagram (see Fig.~\ref{fig:cut_4}), according to:

\begin{equation}
  \sqrt{\left(\frac{S_{250}^{SPIRE}}{S_{350}^{SPIRE}} - \frac{S_{250}^{MMF}}{S_{350}^{MMF}}\right)^2 + \left(\frac{S_{350}^{SPIRE}}{S_{500}^{SPIRE}} - \frac{S_{350}^{MMF}}{S_{500}^{MMF}}\right)^2} < 0.3
   \label{eq:fit_to_Pearson}
\end{equation}

The number 0.3 is a compromise between a more stringent requirement that would result in a smaller number of candidates and a more relaxed requirement that would increase the number of candidates but at the expense of increasing the number of sources with unreliable redshift estimations. \\

v) The last requirement in our preselection is to exclude the presence of possible blazars. As showed in \cite{Negrello2017}, the leaking of blazars into a catalog of high-$z$ candidates can be reduced by demanding our sources to have $S_{350}$/$S_{500}$ $> 1$, unless this ratio is already above the Pearson model in the color-color diagram. (see Fig.~\ref{fig:cut_5}). \\

At this point, after the cuts i) through v), 5079 sources remain in the sample. Since this sample will be used later, we denote it the ``full high-$z$'' sample. These cuts are not perfect at removing low-$z$ objects but the sample should be dominated by high-$z$ candidates. \\

vi) Finally, to reduce this level of contamination, and select high-S/N sources for which photometry is expected to be robust (and consequently the photometric redshift as well), we impose the condition that the S/N defined in Eq. (\ref{eq:snr}) must be greater than or equal to 15$\sigma$ in the filtered image after our MMF has been applied, except if the H-ATLAS SPIRE position has an association with a galaxy of known spectroscopic redshift within a separation of 5 arcsec. This cut leaves 370 objects (see Fig.~\ref{fig:cut_6}). This selection constitutes our ``robust high-$z$'' sample of high-$z$ candidates, with a redshift distribution of $\overline{z}$ = 2.13 and $\sigma$ = 0.65, and will be used later to identify possible lensed systems. \\

This sample is partially shown in Appendix \ref{sec:appendixA} including the estimated redshifts and flux densities. The entire catalog is included on-line. We have performed an additional study for the objects within this sample by comparing the Pearson's $\chi^2$ value obtained considering only the SPIRE flux densities with the one obtained taking into account also the PACS flux densities. For this we have used the flux densities from the H-ATLAS catalog as the observed data and the frequency dependence provided by the Pearson SED (Eq.~\ref{eq:temp_model}), at the photometric redshift estimated by our MMF, as the theoretical data. The result of this study is shown within the on-line catalog through the flag `Reliability'. Those sources for which the $\chi^2$ improves or remains the same when using PACS fluxes are flagged with a `0', whether the $\chi^2$ worse slightly but it is still acceptable they are flagged with a `1', if the $\chi^2$ is much worse they are flagged with a `2' and if the source does not have PACS flux densities we can use, it is flagged with a `-1'.

\begin{figure}
	\includegraphics[width=\columnwidth]{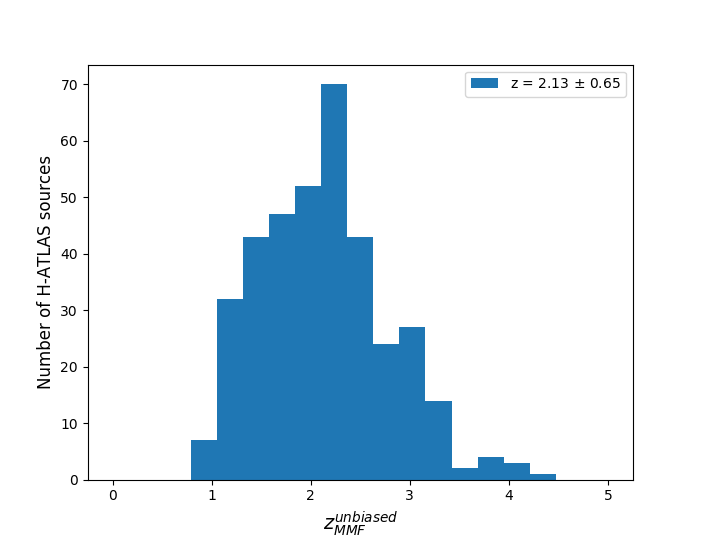}
    \caption{Redshift distribution, according to the redshift estimates obtained with the MMF method, of the 370 high-$z$ H-ATLAS sources from the robust high-$z$ sample selected after imposing all cuts enumerated in Section~\ref{sample_selection}.}
    \label{fig:histogram_370}
\end{figure}

Within this sample, 201 candidates are in the GAMA fields (60 in the GAMA9, 58 in the GAMA12 and 83 in the GAMA15), 82 in the NGP and 87 in the SGP. The number density of sources in the GAMA fields is higher than in NGP and SGP after this last cut because the number of associations with objects having spectroscopic redshifts is higher in the GAMA fields. 21 out of the 32 spectroscopic redshift sources shown in \cref{tab:spectroscopic,tab:spectroscopic1,tab:spectroscopic2} (highlighted in bold) are included in this sample. Among all the objects of this sample we find 35 QSOs.  Fig.~\ref{fig:histogram_370} shows the redshift distribution of the robust high-$z$ sample in order to compare it with the redshift distribution of the initial sample shown in Fig.\ref{fig:histogram_410997}. \\

\begin{figure*}
\minipage{0.33\textwidth}
  \includegraphics[width=\linewidth]{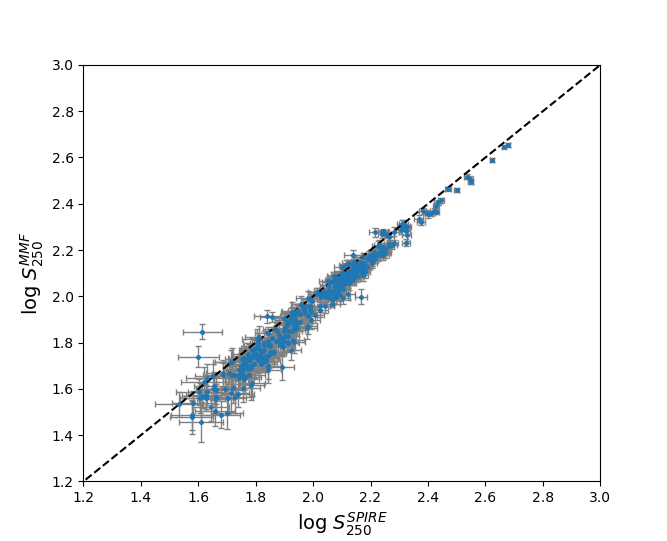}
\endminipage\hfill
\minipage{0.33\textwidth}
  \includegraphics[width=\linewidth]{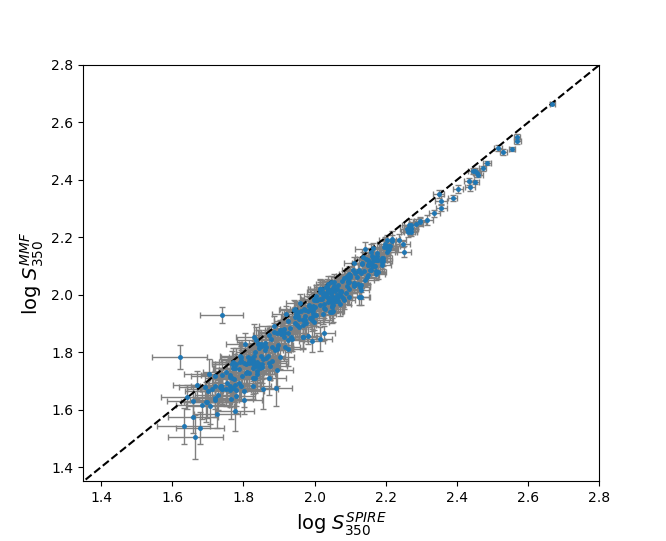}
\endminipage\hfill
\minipage{0.33\textwidth}
  \includegraphics[width=\linewidth]{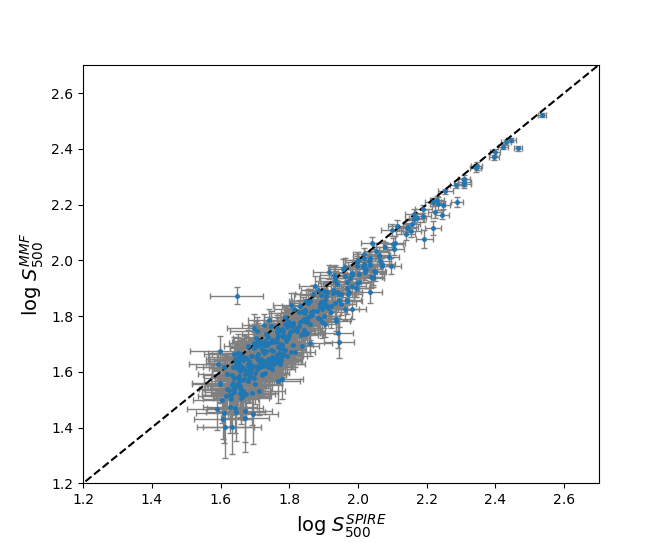}
\endminipage
\caption{Flux measurements in the 250 (left), 350 (middle) and 500 $\mu$m (right) SPIRE channels after our MMF has been applied versus the corresponding tabulated flux measurements from the H-ATLAS catalog in logarithmic scale for the 370 high-$z$ H-ATLAS from the robust high-$z$ sample (see Appendix \ref{sec:appendixA}). The linear behavior with zero-intercept is drawn with a black dashed line. All fluxes are in units of mJy/beam according to the beam profile of the respective channel.}
\label{fig:log_370_fluxes}
\end{figure*}

A direct comparison between our estimates of the flux densities in all channels and the tabulated fluxes from the H-ATLAS catalog is shown in Fig.~\ref{fig:log_370_fluxes}. A clear linear trend is observed and the accordance is pretty good. As happens with the spectroscopic redshift sources from Fig.~\ref{fig:log_spec_fluxes}, an overall underestimation of our MMF flux densities, greater for fainter sources, can be observed. As the number of sources is greater, here the effect is most remarkable. The average flux underestimates between the flux densities estimated from the MMF method and the H-ATLAS fluxes are 10 $\pm$ 9 mJy at 250 $\mu$m, 12 $\pm$ 9 mJy at 350 $\mu$m and 9 $\pm$ 8 mJy at 500 $\mu$m. Since the MMF combines information from all three wavelengths, which allows to reduce the background and boost the signal, instrumental, foreground and confusion noises are better removed so flux density estimates are less affected by Eddington bias than H-ATLAS flux densities. This underestimation with respect to H-ATLAS flux densities is stronger toward low flux densities, which supports the Eddington bias hypothesis, but is also observed to a lesser extent for high flux densities, suggesting a possible degradation of the MMF photometry that could be related to the way we re-pixelize the 350 and 500 $\mu$m images and combine them during the multifiltering step (see Section~\ref{sec:data}). \\

\begin{figure*}
\minipage{0.33\textwidth}
  \includegraphics[width=\linewidth]{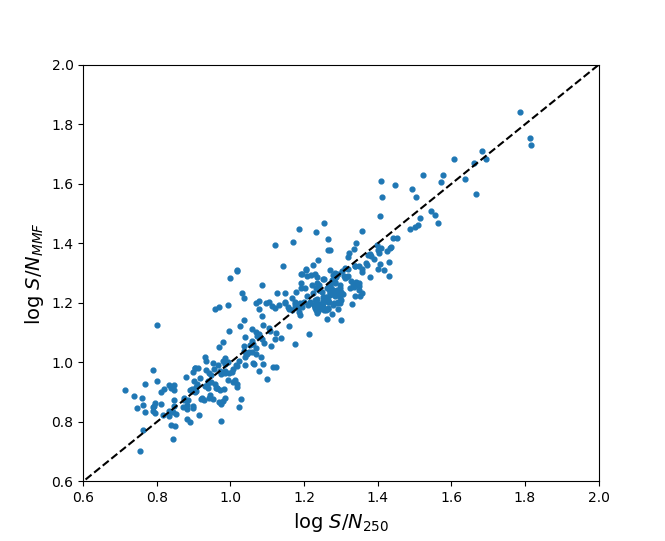}
\endminipage\hfill
\minipage{0.33\textwidth}
  \includegraphics[width=\linewidth]{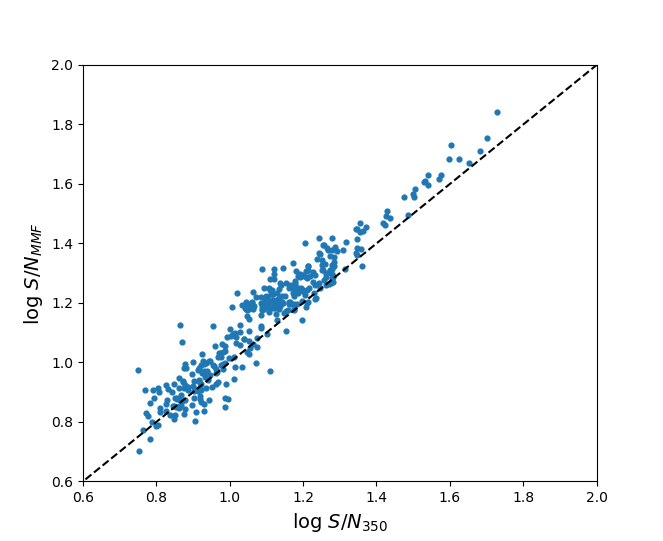}
\endminipage\hfill
\minipage{0.33\textwidth}
  \includegraphics[width=\linewidth]{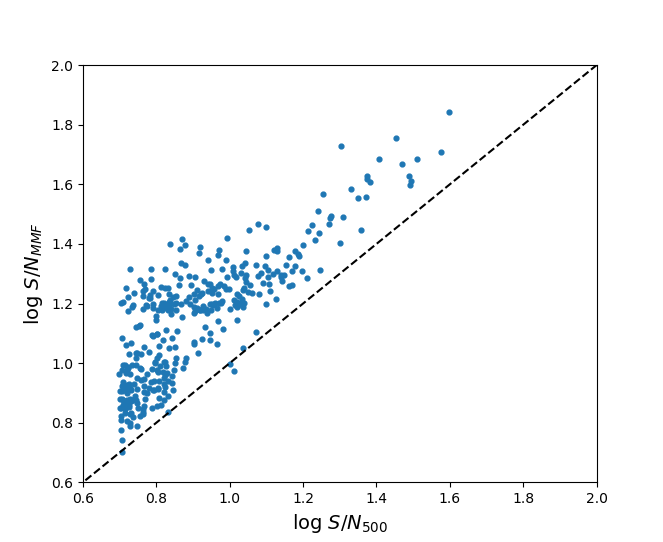}
\endminipage
\caption{S/N in the filtered image after our MMF has been applied versus the S/Ns in the 250 (left), 350 (middle) and 500 $\mu$m (right) SPIRE channels in logarithmic scale for the 370 high-$z$ H-ATLAS sources from the robust high-$z$ sample. The linear behavior with zero-intercept is drawn with a black dashed line.}
\label{fig:log_370_snr}
\end{figure*}

Fig.~\ref{fig:log_370_snr} shows the improvement in S/N achieved with our MMF method for the robust high-$z$ sample in contrast with the S/N of the three $\mu$m SPIRE bands. An average improvement of 76\% in the S/N has been achieved for this sample with our MMF technique compared to the 500 $\mu$m band. Besides, an average improvement of 16\% and a slight improve of 0.2\% have been obtained for the 350 $\mu$m and 250 $\mu$m, respectively. \\

In the end, we have a selected sample that includes several hundreds of interesting objects from H-ATLAS which both agree with the \cite{Pearson2013} SED used to estimate the redshift and have high redshifts and signal-to-noise ratios.

\subsection{Faint subsample: ``500 $\mu$m-risers''} \label{sec:risers}
%-----------------------------------------------------------------------------
Apart from the robust high-$z$ sample explained above in Section~\ref{sample_selection}, we also looked for faint sources at 250 and 350 $\mu$m but bright at 500 $\mu$m in the H-ATLAS data, the so-called ``500 $\mu$-risers''. Our selection criterion looks for sources whose detection is at least barely significant at 500 $\mu$m and that are not clearly detected at 250 and 350 $\mu$m in the H-ATLAS catalog. We select objects with $\mathrm{S/N}_{500} \geq 4\sigma$, $\mathrm{S/N}_{250}$ $\leq$ 4$\sigma$ and $\mathrm{S/N}_{350}$ $\leq$ 4$\sigma$ in the H-ATLAS catalog and apply our multifrequency MMF filter to them, in order to enhance the statistical signification of those detection candidates. Those sources with S/N $\geq$ 5$\sigma$ after the MMF filtering and that satisfy the condition $S_{250}$ $\leq$ $S_{350}$ $\leq$ $S_{500}$ are considered to be statistically significant enough to be firm candidates to be 500 $\mu$m-risers. This way, we get a sample of 695 reddened SPIRE objects. We must not forget the limitations of the \cite{Pearson2013} SED (Eq.~\ref{eq:temp_model}) used to estimate the redshifts so by selecting again the sources which fit better to the model in the color-color diagram (Eq.~\ref{eq:fit_to_Pearson}) we are left out with 237 objects. This selection constitutes our 500 $\mu$m-riser sample of robust high-$z$ candidates, with a redshift distribution of $\overline{z}$ = 4.62 and $\sigma$ = 0.71. This sample is partially shown in Appendix \ref{sec:appendixA} including redshift and flux density estimates. The entire catalog is included on-line. The same additional $\chi^2$ study, considering PACS flux densities, performed for the ``robust high-$z$'' sample has been applied to this sample, and the result is shown within the on-line catalog through the same flags explained in Section \ref{sample_selection}. Within this sample, 97 objects are from the GAMA fields (27 in the GAMA9, 37 in the GAMA12 and 33 in the GAMA15), 68 from the NGP and 135 from the SGP.

\begin{figure*}
\minipage{0.33\textwidth}
  \includegraphics[width=\linewidth]{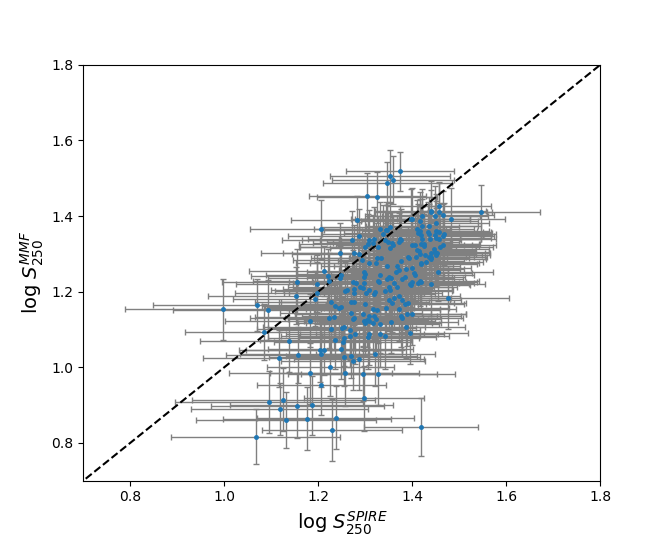}
\endminipage\hfill
\minipage{0.33\textwidth}
  \includegraphics[width=\linewidth]{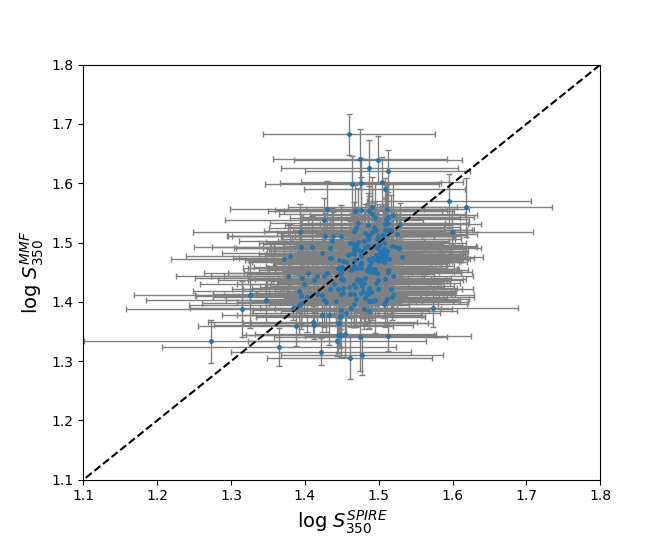}
\endminipage\hfill
\minipage{0.33\textwidth}
  \includegraphics[width=\linewidth]{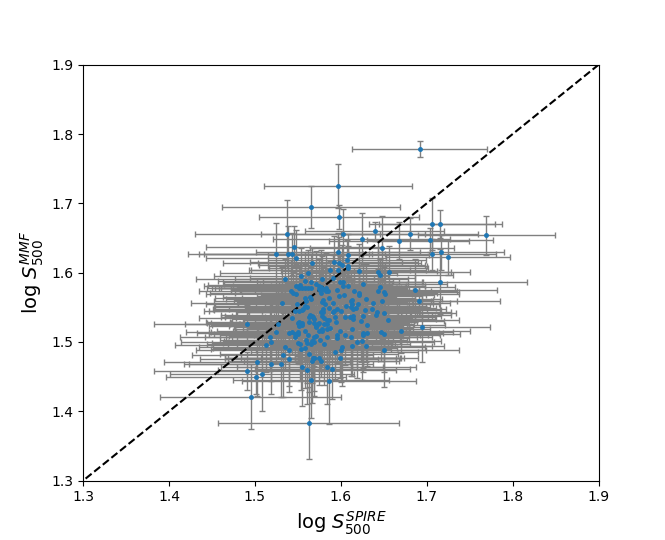}
\endminipage
\caption{Flux measurements in the 250 (left), 350 (middle) and 500 $\mu$m (right) SPIRE channels after our MMF has been applied versus the corresponding tabulated flux measurements from the H-ATLAS catalog in logarithmic scale for the 237 high-$z$ H-ATLAS from the 500 $\mu$m-riser sample (see Appendix \ref{sec:appendixA}). The linear behavior with zero-intercept is drawn with a black dashed line. All fluxes are in units of mJy/beam according to the beam profile of the respective channel.}
\label{fig:log_237_fluxes}
\end{figure*}

The comparison between our estimates of the flux densities in all channels and the tabulated fluxes from the H-ATLAS catalog for the 500 $\mu$m-riser sample is shown in Fig.~\ref{fig:log_237_fluxes}. A much larger scattering than the one seen in Fig.~\ref{fig:log_370_fluxes} for the robust high-$z$ sample can be observed. But this behavior was expected as we are aiming to sources which have a barely significant detection at 500 $\mu$m and are not detected at 250 and 350 $\mu$m. The average flux underestimates between the flux densities estimated from the MMF method and the H-ATLAS fluxes are 4 $\pm$ 4 mJy at 250 $\mu$m, 0.4 $\pm$ 4 mJy at 350 $\mu$m and 3 $\pm$ 5 mJy at 500 $\mu$m. \\

\begin{figure*}
\minipage{0.33\textwidth}
  \includegraphics[width=\linewidth]{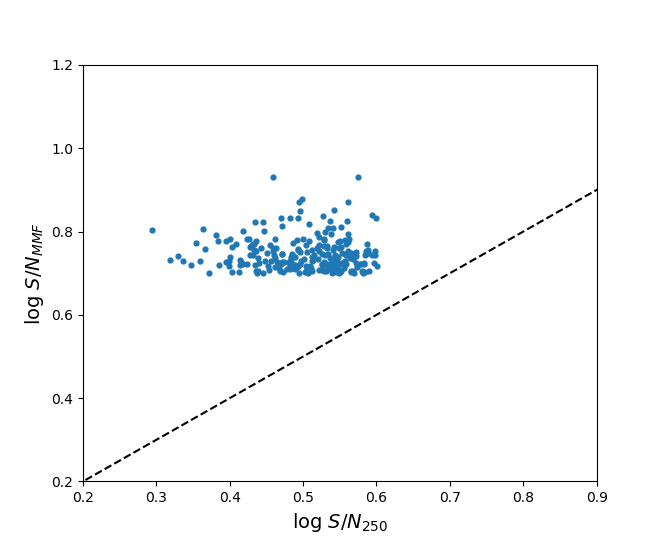}
\endminipage\hfill
\minipage{0.33\textwidth}
  \includegraphics[width=\linewidth]{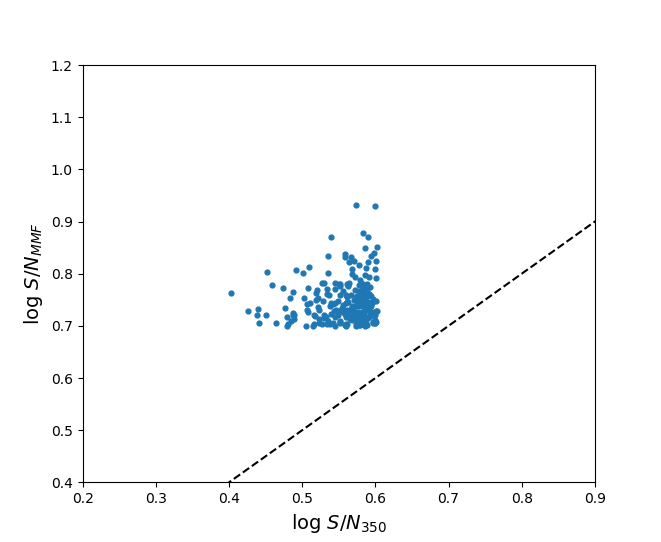}
\endminipage\hfill
\minipage{0.33\textwidth}
  \includegraphics[width=\linewidth]{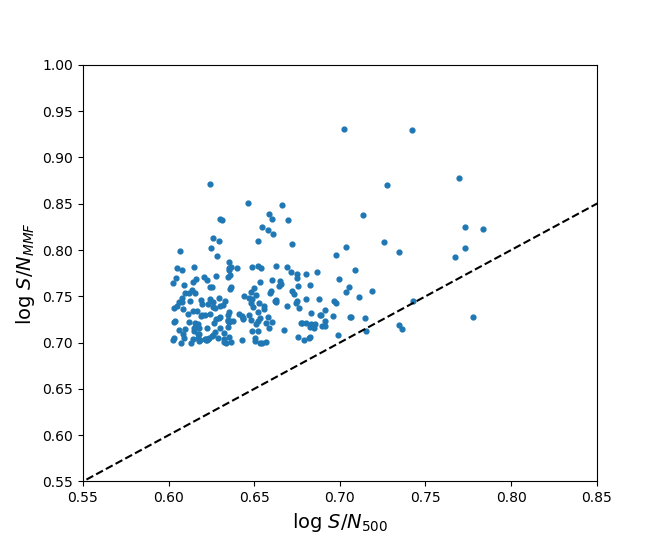}
\endminipage
\caption{S/N in the filtered image after our MMF has been applied versus the S/Ns in the 250 (left), 350 (middle) and 500 $\mu$m (right) SPIRE channels in logarithmic scale for the 237 high-$z$ H-ATLAS from the 500 $\mu$m-riser sample. The linear behavior with zero-intercept is drawn with a black dashed line.}
\label{fig:log_237_snr}
\end{figure*}

Fig.~\ref{fig:log_237_snr} shows the comparison between the S/N reached with our MMF method and the S/N in all three SPIRE channels. It seems logical that the improvement achieved with our method in S/N for these 500 $\mu$m-riser objects (Fig.~\ref{fig:log_237_snr}) should be better than for the objects from the robust high-$z$ sample (Fig.~\ref{fig:log_370_snr}), as they are near the H-ATLAS detection limit. This is confirmed since we have achieved average improvements of 25\%, 55\% and 76\% in the S/N for the 500, 350 and 250 $\mu$m, respectively. This clearly reflects that it is in this kind of faint objects where our MMF method accomplishes bigger impact in terms of signal significance. \\

Unlike the previous robust high-$z$ sample that sought to select bright objects in all bands, among which the probability of finding lensed systems is relatively high, now we pursue faint high-$z$ objects. Most of them are not expected to be lensed by foreground sources but to be intrinsically luminous. These robust high-$z$ and 500 $\mu$m-riser samples are, in fact, built from starting requirements mutually excluding. Sources from the robust high-$z$ sample are initially required to have $\mathrm{S/N}_{250}$, $\mathrm{S/N}_{350}$ and $\mathrm{S/N}_{500}$ greater than 4$\sigma$ while sources from the 500 $\mu$m-riser sample are demanded to have $\mathrm{S/N}_{500}$ $\geq$ 4$\sigma$, $\mathrm{S/N}_{250}$ $\leq$ 4$\sigma$, and $\mathrm{S/N}_{350}$ $\leq$ 4$\sigma$. But this does not mean, for example, that there are not candidates to lensed sources among H-ATLAS 500 $\mu$m-riser galaxies. It should be pointed out, for instance, the case of J090045.4+004125 ($\alpha$ = 135.191, $\delta$ = 0.6897), a dusty star-forming galaxy at $z$ = 6 revealed by strong gravitational lensing and detected in the GAMA field being part of a subsample of the H-ATLAS 500 $\mu$m-riser galaxies \citep{Zavala2018}. This object does not appear in the robust high-$z$ sample but is part of our 500 $\mu$m-riser sample under the identifying name J090045.5+004131 with a redshift estimation of $z$ = 6.35 via our MMF technique. \\

An important effort was made recently by \cite{Ivison16} in order to take advantage of the 250, 350 and 500 $\mu$m \textit{Herschel}-ATLAS imaging survey and select extremely red objects. That work focused on studying the space density of luminous dusty star-forming galaxies (DSFGs) at $z$ > 4 by selecting galaxies from the H-ATLAS survey with extremely red far-infrared colors and faint 350 and 500 $\mu$m flux densities, called ultra-red galaxies. It is important to bear in mind that they used a modified version of the MADX algorithm to identify their sources, so some of their sources are not in the official H-ATLAS catalog. This fact explains why we have been able to locate in the H-ATLAS catalog only 78 of the 109 sources shown in their sample. \\

None of these 78 red sources is, of course, in our robust high-$z$ sample since they were detected at a S/N $\geq$ 3.5$\sigma$ at 500 $\mu$m, being mostly brighter in this band than in the others, and our robust high-$z$ sample was built with relatively bright sources in all bands. Instead, and as expected, there is an overlap between our 500 $\mu$m-riser sample and these red sources from \cite{Ivison16}. These 78 objects were required to be above 3.5$\sigma$ in any of the three SPIRE bands and our 500 $\mu$m-riser sample was built demanding $\mathrm{S/N}_{500} \geq 4\sigma$, $\mathrm{S/N}_{250} \leq 4\sigma$ and $\mathrm{S/N}_{350} \leq 4\sigma$. To begin with, 54 out of these 78 objects are not included in our 500 $\mu$m-riser sample because all they have $\mathrm{S/N}_{350}$ $>$ 4 and thus are excluded by our criterion, which leaves us with 24 possible objects. Only nine of these remaining 24 objects from \cite{Ivison16} (J090045.5+004131, J090304.5-004616, J114038.8-022804, J114350.3-005210, J114353.5+001250, J114412.1+001812, J115614.0+013900, J142710.6+013806 and J004615.0-321825) are included in our 500 $\mu$m-riser sample. If sources were not demanded to behave like the Pearson SED, there would be 16 objects. \\

In the next section we will study the correlations between the high-redshift sources and their possible lenses. We will focus on the robust high-$z$ sample because we find no significant correlation for the case of the 500 $\mu$m-riser sample. This suggests us that this last sample is mostly not lensed as expected due to the lower flux criterion used to select its sources.

\section{Possible lensed galaxies} \label{candidate_lensed_sample}
%%%%%%%%%%%%%%%%%%%%%%%%%%%%%%%%%%%%%%%%%%%%%%%%%%%%%%%%%%%%%%%%%%%%%

\subsection{Preliminary comparison with previous works}

In the previous section we presented our samples of high-redshift candidates (full, robust high-$z$, and 500 $\mu$m-riser samples). Since the optical depth of strong lensing grows with the redshift of the background source, these samples of high-$z$ candidates may contain some lensed galaxies. In fact, this is known to be particularly true for \textit{Herschel} sources, where the brightest high-$z$ sources correspond to SLGs \citep{Negrello2007,Negrello2010,Negrello2014,Negrello2017,Cox2011,Conley2011,Lapi11,Gonzalez-Nuevo2012,Messias2014,Dye2015}. \\

Here we compare the robust high-$z$ sample (370 candidates) with similar catalogs found in the literature. Our robust high-$z$ sample contains 62 of the 80 candidate SLGs with flux density above 100 mJy at 500 $\mu$m presented in \cite{Negrello2017}. 17 of the candidates in the robust high-$z$ sample are part of the sample of 20 confirmed SLGs \citep{Negrello2017}. The only three confirmed strong lens systems that are not included in our sample are J085358.9+015537 (flagged as a star), J142935.3-002836 (which is a major merger system at $z$ = 1.027 \citep{Messias2014} and is excluded by our cut iii) and J125135.3+261457 (excluded by our cut v). Among the confirmed lensed galaxies, J114637.9-001132 at $z$ = 3.26 is interesting since it is associated to a candidate high-$z$ proto-cluster\footnote{We have recently submitted an ALMA proposal to confirm the redshifts of the objects associated to this possible proto-cluster.}\citep{Fu12,herranz13,Clements16,Greenslade18}. 

In addition, six of the eight objects labeled in \cite{Negrello2017} as likely to be lensed and 39 of the 51 objects defined as unclear are included in our robust high-$z$ sample of 370 candidates. The two missed objects labeled as likely to be lensed were excluded in our cut iv. The only one object from \cite{Negrello2017}
confirmed to not be a strongly lensed galaxy (J084933.4+021442) is nor part of our sample because it is flagged as a star in the H-ATLAS catalog. It is indeed a binary system of Hyper Luminous Infrared Galaxies (HyLIRGs) at $z$ = 2.410 \citep{Ivison2013}. Our sample also contains five sources from the SGP field (J004736.0-272951, J011424.0-333614, J235623.1-354119, J001010.5-360237 and J014849.3-331820) which meet the flux criterion demanded by \cite{Negrello2017} but are not in their proposal. \\

In \cite{Gonzalez-Nuevo2012}, the authors applied to the H-ATLAS Science Demonstration Phase field ($\simeq$ 14.4 deg$^{2}$), which covers part of the GAMA9 field, a method for efficiently selecting faint candidate SLGs. This method was called HALOS (\textit{Herschel}-ATLAS Lensed Objects Selection). They found 31 candidate SLGs, whose respective candidate lenses were identified in the VIKING near-infrared catalog and proposed that the application of HALOS over the full H-ATLAS surveyed area would increase the size of the sample up to $\sim$ 1000 SLGs. Eight of these sources are included in our robust high-$z$ sample of 370 sources: J090302.9-014127, J090311.6+003906, J090740.0-004200, J091043.0-000321, J091304.9-005343 (all of them confirmed as strongly lensed in \cite{Negrello2017}), J085855.3+013728, J090957.6-003619 and J091331.4-003644. \\

The H-ATLAS catalog can be used to find potential lens systems (lens plus lensed galaxy) using the already available optical associations with SDSS \citep{SLOANIV} for each SPIRE source \citep{Bourne2016}. These associations are sought via a Likelihood-Ratio analysis of optical candidates within 10 arcsec of all SPIRE sources with S/N $\geq$ 4 at 250 $\mu$m. \cite{Bourne2014} studied the fact that redder and brighter submm sources have optical associations with greater positional offsets than would expected if they were due to random positional errors. They concluded that lensing is the most plausible cause for increased offsets of red submm sources and that the problem of misidentifying a galaxy in a lensing structure as the counterpart to a higher redshift submillimeter galaxy may be more common than previously thought.
Most of these optical associations do not have spectroscopic information (i.e secure redshift), however, there are 180 objects in our robust high-$z$ sample for which this condition is fulfilled (mostly because of the cut vi). Spectroscopic redshifts are obtained from many different surveys, like SDSS DR7, SDSS DR10, 6dFGS, 2SLAQ or GAMA. 138 sources out of these 180 have a reliable spectroscopic redshift (Z_QUAL $\geq$ 3) in the range 0.1 $\leq$ $z$ $\leq$ 1.1, that is significantly smaller than the photometric redshift estimated by our MMF method. Hence, these associations may correspond to possible lens systems since the redshifts of the alleged lens and the high-$z$ candidate are so different. In those cases where the optical association is not the same object as the SPIRE source, it will be an object at a smaller redshift and close (in angular separation) to the SPIRE source. The conditions would be given for the lens effect to take place and these cases should be studied in detail to verify it. However, since these associations are already given in the catalogs themselves and their spectroscopic redshifts come from many different sources, we are going to proceed to look for our own associations. \\

The above discussion shows how our robust high-$z$ sample has the potential to host many unknown lensed galaxies. Most of the previously known \textit{Herschel} lensed galaxies were unveiled by the 500 $\mu$m flux density criterion ($S_{500}>100$ mJy), which has proven to be a simple (but powerful) method of selecting strongly lensed candidates. Here we rely on a cross-correlation study based on matching distant IR sources with foreground potential lenses located at distances that make them consistent with being a lens system.

\subsection{Statistical lensing analysis. Correlation analysis with SDSS} \label{sec:Correlation}
%%%%%%%%%%%%%%%%%%%%%%%%%%%%%%%%%%%%%%%%%%%%%%%%%%%%%%%%%%%%%%%%%%%%%%%%%
Additional evidence for significant lensing in our two samples (full and robust high-$z$) can be obtained through a simple correlation analysis with a catalog of foreground galaxies. If the \textit{Herschel} sources are tracing the magnification pattern produced by a population of lenses at $z<1$, one would expect an excess of IR sources detected around regions of magnification larger than one. Alternatively, the alleged high-$z$ source could be instead a lower redshift associated with the lens. In this case, the excess found in the correlation would be produced by contamination of our sample (i.e low-$z$ sources being misinterpreted as high-$z$ sources). 

For the catalog of potential foreground lenses ($z<1$), we use lenses extracted from the SDSS. By lenses, we mean here either individual galaxies or groups of galaxies (see below). Since SDSS does not cover the SGP field, we consider only the IR sources which come from the GAMA and NGP fields. After removing IR high-$z$ candidates from the SGP field, the full high-$z$ sample is reduced to 2828 candidates while the robust high-$z$ sample is left with 283 candidates. For a simple estimation of the correlation, we compare the number of matches found within an aperture and for different aperture radii, $N_{m}(R)$, with the expected number from a random distribution ($N_{r}(R)$, see Eq.~\ref{Eq_Random}). This random number is obtained by the following equation:

\begin{equation}
  N_{r}\left(R\right) = N_H\frac{A_c\left(R\right)}{A_H}
  \label{Eq_Random}
\end{equation}
\noindent 
where $N_H$ is the number of H-ATLAS high-$z$ candidates, $A_c(R)$ is the total area covered (within the footprint of H-ATLAS) by the disks of radius $R$ around the SDSS sources, and $A_H = 341.65$ square degrees is the total area of H-ATLAS survey excluding the SGP field. By construction, $N_{r}(R) \leq N_H$. \\

On the other hand, the number of matches ($N_{m}(R)$) between the H-ATLAS sources and the SDSS lenses is obtained by computing the number of associations between both catalogs as a function of radius by centering disks of radius $R$ on the SDSS lenses and counting the number of H-ATLAS sources which fall within the disk. Any significant excess over the expected value in the random case is evidence for either lensing or contamination. The uncertainty, or significance, with respect to the random case is given by the Poissonian error (i.e., the uncertainty is given by $\sigma_r(R)=\sqrt{N_{r}(R)}$). If the excess is due to contamination, this hypothesis can be tested, since one would expect the separation between the positions of the \textit{Herschel} sources and the SDSS lenses to be comparable to the positional error in \textit{Herschel} (which is significantly larger than the corresponding error in SDSS), that is 2--3 arcsec. In these cases, the \textit{Herschel} source may actually be the SDSS lens. If, on the contrary, a high-$z$ \textit{Herschel} candidate is found at more than 3 arcsec from the SDSS source, lensing is possibly responsible for that association. Some of the associations should be due to pure random alignments but this number can be estimated by the Poissonian expectation number discussed above. \\

For the SDSS lenses, we use two catalogs of potential lenses derived from SDSS. The first catalog focuses on rare but massive potential lenses at $z$ $\leq$ 0.6 while the second catalog focuses on less massive, but more abundant, potential lenses with $z$ $\leq$ 1.1. We set a lower limit to the redshift of the potential lenses since below certain redshift, strong lensing becomes inefficient due to the increase in the critical surface mass density ($z_{min}\sim 0.1$). \\

For the association with massive lenses, we use the SDSS DR8 redMaPPer cluster catalog with 26111 objects \citep{Rykoff2014}. This catalog is the result of applying the Red-sequence Matched-filter Probabilistic Percolation (redMaPPer) cluster finding algorithm to the 10400 $\mathrm{deg^{2}}$ of photometric data from the Eighth Data Release  \citep[DR8,][]{Aihara2011} of the SDSS. The redMaPPer algorithm has been designed to handle an arbitrary photometric galaxy catalog, with an arbitrary number of photometric bands ($\geq$ 3), and performs well provided the photometric bands span the 4000 {\AA} break over the redshift range of interest. It adapts therefore well to a survey such as the Sloan Digital Sky Survey. Because the number of objects in this catalog is not very large, we use all of them in the cross-correlation which cover a range of 0.08 $\leq$ $z$ $\leq$ 0.6 in redshift and 19.85 $\leq$ $\lambda$ $\leq$ 299.46 in cluster richness. $N_H$ = 881 of the 26111 halos fall in the footprint of H-ATLAS. We find no significant excess when cross-correlating redMaPPer with our catalog of high-$z$ H-ATLAS sources. Given the fact that 17 of the 20 strongly lensed candidates from \cite{Negrello2017} (confirmed as such) are in our selected sample, and none of them has a match with redMaPPer, this confirms that the lenses in \cite{Negrello2017} are not massive halos, but rather relatively small halos (like elliptical galaxies for instance). \\

\begin{figure}
	\includegraphics[width=\columnwidth]{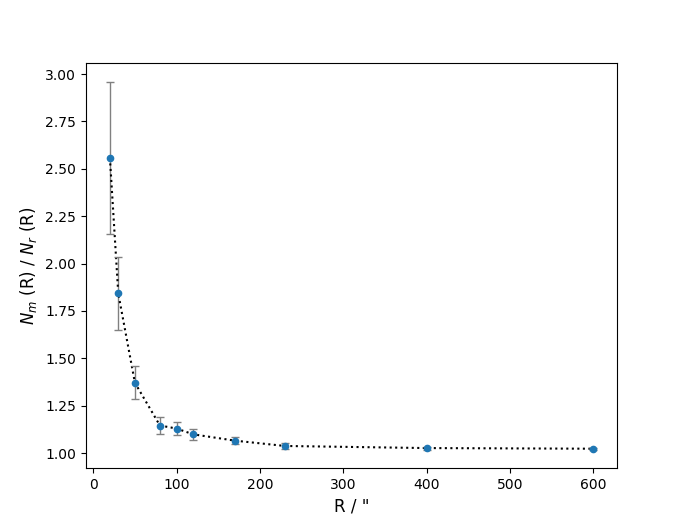}
    \caption{Ratio between the number of matches found ($N_{m}(R)$), after cross-matching the full high-$z$ subsample of 2828 H-ATLAS sources with a sample of 1776242 known-redshift galaxies from SDSS DR14, and the number of matches expected ($N_{r}(R)$) from a random distribution (Eq.~\ref{Eq_Random}) for different aperture radii $R$ around the SDSS sources.}
    \label{fig:excess_plot_middle}
\end{figure}

\begin{figure}
	\includegraphics[width=\columnwidth]{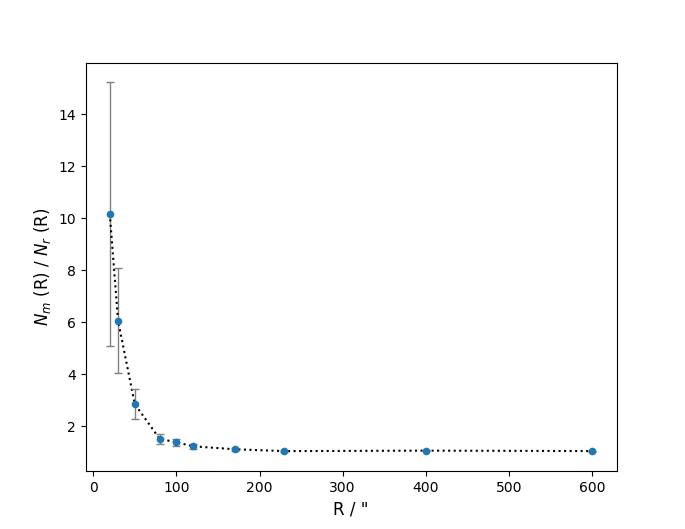}
    \caption{Ratio between the number of matches found ($N_{m}(R)$), after cross-matching the robust high-$z$ subsample of 283 candidates with a sample of 1776242 known-redshift galaxies from SDSS DR14, and the number of matches expected ($N_{r}(R)$) from a random distribution (Eq.~\ref{Eq_Random}) for different aperture radii $R$ around the SDSS sources.}
    \label{fig:excess_plot_final}
\end{figure}

Our second search for potential lensed galaxies focuses on the low-mass regime of the lenses. From SDSS DR14 \citep{SDSS14} we select a larger catalog of galaxies with known spectroscopic redshifts. We focus on galaxies with known redshift in order to minimize possible contamination from galaxies that are misinterpreted as having  $z>0.15$ and also to reduce the computation time. The sample contains 1776242 galaxies from the Sloan Digital Sky Survey Data Release 14 with 0.15 $\leq$ $z$ $\leq$ 1.1. As mentioned above, we limit the minimum redshift to 0.15 since below this redshift most galaxies are expected to be subcritical (and not produce strong lensing effects). Among all of them, $N_H$ = 50175 are the galaxies that fall in the footprint of H-ATLAS. We cross-correlate our full high-$z$ subsample of 2828 H-ATLAS sources with the SDSS catalog of galaxies and compare it with the expected number in the case of no correlation (i.e, the random case described above). The ratio of the observed ($N_{m}(R)$) and random matches ($N_{r}(R)$) between this catalog and our full high-$z$ selection sample is shown in Fig.~\ref{fig:excess_plot_middle} for different radii. In Fig.~\ref{fig:excess_plot_final} we exhibit the same but for our robust high-$z$ subsample of 283 candidates. Both Fig.~\ref{fig:excess_plot_middle} and Fig.~\ref{fig:excess_plot_final} show a non-one signal for aperture radii of several arcminutes which is unexpected and an example of the lensing-induced cross-correlations between high-$z$ submillimeter galaxies and low-$z$ galaxy population \citep{Wang2011,gnuevo14,gnuevo17,Bourne2014}. \\

There is a clear increase in significance when considering the robust high-$z$ subsample of 283 candidates. A sharp increase in the excess of matches is found at distances below 60 arcsec. The smaller amplitude of the excess in the full high-$z$ sample with 2828 sources suggests that this sample may be more contaminated by low-$z$ candidates. \\

\begin{table*}
	\centering
	\caption{List of 40 matches found at a separation radius less than or equal to 20 arcsec after cross-matching the robust high-$z$ subsample of 283 high-$z$ candidates with a sample of 1776242 known-low-redshift galaxies from SDSS DR14. The object J145420.6-005203 is identified as a QSO in the H-ATLAS catalog. For each H-ATLAS source the following information is provided: our redshift estimate after revising the bias ($z_{MMF}^{unbiased}$), spectroscopic redshift of the optical association assigned in the H-ATLAS catalog ($z_{spec}^{H-ATLAS}$), identifying name (SDSS ID) of the association found by us, its spectroscopic redshift ($z_{spec}^{SDDS}$), its separation radius in arcsec ($R$) and a rough estimation of Einstein radius that the lensed system would have according to the calculations and assumptions described in Section \ref{sec:Correlation}.}
	\label{tab:40_matches}
	\resizebox{\textwidth}{!}{\begin{tabular}{cccccccccccc} % four columns, alignment for each
		\hline
		\addstackgap[9pt]{No.} & H-ATLAS name & $\alpha$ ($\mathrm{deg}$) & $\delta$ ($\mathrm{deg}$) & $z_{MMF}^{unbiased}$ & $z_{spec}^{H-ATLAS}$ & SDSS ID & $\alpha$ ($\mathrm{deg}$) & $\delta$ ($\mathrm{deg}$) & $z_{spec}^{SDSS}$ & $R$ (arcsec) & $\theta_{E}$ (arcsec) \\
		\hline \\
1 & J083726.1+015641 & 129.358797 & 1.944871 & 1.86 $\pm$ 0.28 & 0.396 & 1237653622837216254 & 129.359078 & 1.944271 & 0.396 & 2.4 & 3.9 \\
2 & J085229.0+010217 & 133.121072 & 1.038117 & 2.46 $\pm$ 0.30 & 0.492 & 1237650797288031084 & 133.120142 & 1.038521 & 0.492 & 3.7 & 6.9 \\
3 & J085250.9-010000 & 133.21219 & -1.000204 & 1.13 $\pm$ 0.26 & 0.581 & 1237673709866321335 & 133.213142 & -1.000520 & 0.583 & 3.6 & 2.3 \\
4 & J085855.4+013729 & 134.731174 & 1.624899 & 2.36 $\pm$ 0.30 & 0.665 & 1237651753990489111 & 134.729564 & 1.626498 & 0.658 & 8.1 & 26.6 \\
5 & J090038.0-003522 & 135.158643 & -0.589715 & 1.54 $\pm$ 0.27 & 0.407 & 1237674460412707797 & 135.154630 & -0.586598 & 0.570 & 18.3 & 4.5 \\
6 & J090311.6+003907 & 135.798682 & 0.652064 & 3.17 $\pm$ 0.34 & 0.300 & 1237648722285101574 & 135.798225 & 0.651815 & 0.300 & 1.9 & 1.4 \\
7 & J091331.4-003644 & 138.380891 & -0.612341 & 1.49 $\pm$ 0.27 & 0.296 & 1237648720675668446 & 138.383247 & -0.610143 & 0.499 & 11.6 & 3.4 \\
8 & J092348.4+011907 & 140.952056 & 1.318877 & 1.44 $\pm$ 0.27 & 0.590 & 1237653664181126111 & 140.952085 & 1.319323 & 0.590 & 1.6 & 5.4 \\
9 & J114512.9-002446 & 176.303787 & -0.412797 & 3.13 $\pm$ 0.34 & 0.444 & 1237671140943987623 & 176.308407 & -0.412486 & 0.629 & 16.7 & 5.6 \\
10 & J115101.7-020024 & 177.757213 & -2.006799 & 1.89 $\pm$ 0.28 & 0.264 & 1237650762389586324 & 177.760981 & -2.006413 & 0.454 & 13.6 & 6.3 \\
11 & J115819.5-000127 & 179.581642 & -0.024356 & 2.32 $\pm$ 0.29 & 0.597 & 1237648721230561969 & 179.579745 & -0.026204 & 0.597 & 9.5 & 6.8 \\
12 & J120656.4+012641 & 181.735018 & 1.444872 & 2.11 $\pm$ 0.28 & 0.637 & 1237651752400454138 & 181.733129 & 1.443833 & 0.637 & 7.8 & 5.6 \\
13 & J120932.7+002517 & 182.386595 & 0.421524 & 1.51 $\pm$ 0.27 & 0.470 & 1237674651003454270 & 182.385894 & 0.420998 & 0.471 & 3.2 & 5.4 \\
14 & J125233.5+331031 & 193.139528 & 33.175366 & 2.05 $\pm$ 0.28 & 0.490 & 1237665331465945161 & 193.139024 & 33.175429 & 0.490 & 1.5 & 4.8 \\
15 & J130138.7+302808 & 195.411281 & 30.468764 & 2.03 $\pm$ 0.28 & 0.166 & 1237665443126575176 & 195.410385 & 30.468297 & 0.166 & 3.3 & 2.3 \\
16 & J130152.8+245012 & 195.469976 & 24.836605 & 2.28 $\pm$ 0.29 & 0.519 & 1237667911671153185 & 195.469452 & 24.836544 & 0.519 & 1.7 & 3.7 \\
17 & J131407.0+271208 & 198.529016 & 27.202121 & 0.96 $\pm$ 0.27 & 0.189 & 1237667443513229416 & 198.529224 & 27.203110 & 0.189 & 3.6 & 1.3 \\
18 & J131533.9+233254 & 198.891415 & 23.54824 & 1.35 $\pm$ 0.26 & - & 1237667910598787245 & 198.892569 & 23.550384 & 0.551 & 8.6 & 4.2 \\
19 & J131635.1+332406 & 199.14629 & 33.401613 & 3.33 $\pm$ 0.35 & 0.463 & 1237665127475380722 & 199.147629 & 33.401386 & 0.463 & 4.1 & 5.6 \\
20 & J131715.6+322643 & 199.314904 & 32.44538 & 1.25 $\pm$ 0.26 & 0.185 & 1237665227847565637 & 199.313765 & 32.446109 & 0.185 & 4.3 & 11.7 \\
21 & J132111.2+265009 & 200.296539 & 26.835849 & 2.08 $\pm$ 0.28 & - & 1237667323262992805 & 200.293255 & 26.836073 & 0.542 & 10.6 & 0.6 \\
22 & J132355.2+282319 & 200.979945 & 28.388498 & 2.33 $\pm$ 0.29 & 0.406 & 1237665440981254703 & 200.978406 & 28.389180 & 0.406 & 5.5 & 3.1 \\
23 & J132453.6+244909 & 201.223168 & 24.819245 & 1.82 $\pm$ 0.27 & 0.569 & 1237667448346903078 & 201.221548 & 24.818992 & 0.569 & 5.4 & 3.2 \\
24 & J133020.9+240249 & 202.587121 & 24.046861 & 3.00 $\pm$ 0.33 & 0.610 & 1237667911673971962 & 202.586077 & 24.046954 & 0.610 & 3.5 & 4.2 \\
25 & J133231.6+350843 & 203.131806 & 35.145221 & 1.82 $\pm$ 0.27 & 0.189 & 1237664852029079667 & 203.131680 & 35.145198 & 0.189 & 0.4 & 2.0 \\
26 & J133806.8+351530 & 204.528535 & 35.258382 & 1.90 $\pm$ 0.28 & 0.339 & 1237664852566409566 & 204.528530 & 35.257533 & 0.339 & 3.1 & 0.6 \\ 
27 & J134324.5+240202 & 205.852117 & 24.034017 & 2.15 $\pm$ 0.29 & 0.500 & 1237667912212153027 & 205.854422 & 24.035555 & 0.500 & 9.4 & 2.6 \\
28 & J134429.5+303034 & 206.12285 & 30.509484 & 2.29 $\pm$ 0.29 & - & 1237665329860248195 & 206.122041 & 30.504826 & 0.672 & 17.0 & 0.9 \\
29 & J134853.0+270011 & 207.220753 & 27.002946 & 2.49 $\pm$ 0.30 & 0.175 & 1237665533319643218 & 207.223273 & 27.003177 & 0.175 & 8.1 & 3.0 \\
30 & J141351.9-000026 & 213.466627 & -0.007454 & 2.72 $\pm$ 0.31 & - & 1237674603753243273 & 213.466488 & -0.009563 & 0.548 & 7.6 & 1.2 \\
31 & J141550.4+012750 & 213.960027 & 1.464006 & 2.47 $\pm$ 0.30 & 0.296 & 1237651752951415050 & 213.959966 & 1.465131 & 0.296 & 4.1 & 1.5 \\
32 & J141605.5+011828 & 214.023078 & 1.308023 & 3.15 $\pm$ 0.34 & 0.588 & 1237651735235003221 & 214.024281 & 1.307622 & 0.588 & 4.6 & 3.3 \\
33 & J141827.4-001703 & 214.614453 & -0.284336 & 1.85 $\pm$ 0.28 & 0.250 & 1237648704053903588 & 214.614920 & -0.284529 & 0.250 & 1.8 & 3.6 \\
34 & J141832.9+010212 & 214.637488 & 1.036847 & 4.14 $\pm$ 0.39 & 0.668 & 1237651752414871842 & 214.636719 & 1.036710 & 0.668 & 2.8 & 3.2 \\
35 & J142008.9-001434 & 215.03741 & -0.242847 & 2.27 $\pm$ 0.29 & 0.615 & 1237648704054035361 & 215.037730 & -0.242879 & 0.615 & 1.2 & 5.9 \\
36 & J142233.9+023413 & 215.641415 & 2.570521 & 2.20 $\pm$ 0.29 & 0.747 & 1237651736846270820 & 215.641393 & 2.568986 & 0.747 & 5.5 & 0.2 \\
37 & J143358.4-012718 & 218.493585 & -1.455029 & 1.89 $\pm$ 0.28 & 0.239 & 1237655693012369781 & 218.49700 & 1.456681 & 0.238 & 13.7 & 2.1 \\
38 & J143845.8+013503 & 219.691249 & 1.584351 & 1.62 $\pm$ 0.27 & 0.498 & 1237651735774364011 & 219.691389 & 1.583848 & 0.498 & 1.9 & 5.3 \\
39 & J145420.6-005203 & 223.586163 & -0.867644 & 2.58 $\pm$ 0.31 & 0.765 & 1237648720176087748 & 223.586516 & -0.872245 & 0.551 & 16.6 & 8.4 \\
40 & J145653.4-000720 & 224.222595 & -0.122431 & 1.58 $\pm$ 0.27 & 0.647 & 1237648721250092410 & 224.222216 & 0.1229648 & 0.647 & 2.4 & 1.5 \\
	\\ \hline
	\end{tabular}}
\end{table*}

Focusing on the smaller radii, we find 40 associations at a separation lower than 20 arcsec, between the 50175 known-spectroscopic-redshift galaxies from SDSS DR14 that fall in H-ATLAS footprint and our robust high-$z$ subsample of 283 H-ATLAS sources. One of them (J145420.6-005203) is identified as a QSO in the H-ATLAS catalog. We chose to consider this separation since most of the lenses would have an Einstein radius less than 20 arcsec, which is the radius around which the strongest magnifications are expected. From among these matches, 28 have a separation greater than the positional error in \textit{Herschel} ($>$3 arcsec) so lensing is possibly responsible for that association. $\sim$4 associations should be caused due to pure random alignments so it is expected that a considerable number of these associations are lensed. These 40 matches are shown in Table~\ref{tab:40_matches}. And snapshots of them, centered on the SDSS DR14 sources, are shown in \cref{fig:postage_1,fig:postage_2,fig:postage_3} from Appendix \ref{sec:appendixB}. \\

We have used the SDSS DR14 asinh magnitudes in the $r$-band of these 40 low redshift optical sources shown in Table~\ref{tab:40_matches} to get a rough estimation of the Einstein radius of each possible lensed system. Firstly, we have estimated from \cite{van_Uitert2015} the corresponding corrections for the redshift of their spectra (i.e., the k-correction) and for the intrinsic evolution of their luminosity (i.e., the e-correction) in order to correct the $r$-band magnitudes. Once the magnitudes of the optical sources are corrected, we have calculated their fluxes and then their luminosities (through their luminosity distances $D_{L}$). At this point, we used the luminosity-to-halo mass relation $M_{eff}$ $=$ $M_{0,L}(L/L_{0})^{\beta_{L}}$ parametrized in \cite{van_Uitert2015} to estimate the mass of each SDSS galaxy for the corresponding luminosity previously obtained. The pivot luminosity $L_{0}$ is the same for every object while the $M_{0,L}$ and $\beta_{L}$ parameters depend on the spectroscopic redshift of the galaxy. Finally, we supposed the galaxy behaves as a singular isothermal sphere to estimate the Einstein radius (see \cite{lectures_lensing_2008}). We assumed the virial radius of the galaxy to be $r$ $=$ $1.3(M/10^{15}M_{\odot})^{1/3}$ Mpc in order to estimate its velocity dispersion $\sigma_{v}$ $=$ $\sqrt{GM/2r}$. The Einstein radius can be then estimated through $\theta_{E}$ $=$ ($4\pi\sigma_{v}^{2}D_{ds}$)/($c^{2}D_{s}$), where $D_{ds}$ and $D_{s}$ are the angular diameter distances between the lens and the source, and observer and source, respectively. These distances are calculated with the spectroscopic redshift of the SDSS galaxy acting as lens ($z_{spec}^{SDSS}$) and the photometric redshift of the source estimated with our MMF ($z_{MMF}^{unbiased}$). These Einstein radius estimates are included in Table~\ref{tab:40_matches} and endorse the choice of 20 arcsec as upper limit for the separation between H-ATLAS sources and their partners from SDSS DR14. 15 of the 40 matches have a separation radius smaller than the corresponding estimation of the Einstein radius. For these 15 we will expect to find counter-images outside the Einstein radius. Possible reasons why we do not find those could be: i) the foreground galaxy is not a lens, ii) there is substructure around the lens that can create fluctuations in the flux or iii) the Einstein radius is overestimated. Einstein rings corresponding to this radii have been drawn in the snapshots from Appendix \ref{sec:appendixB}. \\

All H-ATLAS objects from Table~\ref{tab:40_matches}, except four of them, had a previous association with a galaxy with spectroscopic redshift at less than 10 arcsec. In our search for possible lens systems, we extend our association radius to 20 arcsec. We find that 2 of the previously unmatched H-ATLAS sources (J132111.2+265009 and J134429.5+303034) can now be matched with a galaxy having spectroscopic redhsift within 20 arcsec. For the other two remaining unmatched sources (J131533.9+233254 and J141351.9-000026), we find associations within 10 arcsec with SDSS galaxies having spectroscopic redshift. These last two associations may have not been included in the H-ATLAS catalog because SDSS DR7 and SDSS DR10 were used and we cross-match our robust high-$z$ sample with known-low-redshift galaxies from SDSS DR14. 

There are five sources for which there is a mismatch between the redshift published in the H-ATLAS catalog and the redshift of the match found in this work.
These sources are J090038.0-003522, J091331.4-003644 (SLG candidate from \cite{Gonzalez-Nuevo2012}), J114512.9-002446, J115101.7-020024 (labeled as unclear in \cite{Negrello2017}) and J145420.6-005203. In all cases, our matches are beyond the 10 arcsec radius used in the H-ATLAS association. All these five sources lie in the GAMA fields for which more redshift information is available other than the SDSS redshifts (and that was not used in our association). In fact, we can confirm that spectroscopic redshifts for the first four sources come from the GAMA project \citep{Driver2009,Driver2016} but the last one comes from the WiggleZ Dark Energy Survey. \\

After visual inspection, among the 40 matches we identify interesting cases that are likely to be lensed. Three of them (J090311.6+003907, J134429.5+303034 and J141351.9-000026) are already confirmed as strongly lensed systems in \cite{Negrello2017}. Some other SDSS stamps reveal overdensities of galaxies at $z<1$ that could indicate the presence of a group of galaxies acting as a lens for the corresponding candidate. These are mainly J083726.1+015641, J085229.0+010217, J085250.9-010000, J115819.5-000127, J130138.7+302808, J131407.0+271208, J131533.9+233254, J133231.6+350843, J133806.8+351530, J141832.9+010212 and J143845.8+013503. \\

The properties and complex nature of the strong gavitationally lensed system J090311.6+003907 (proposed as candidate SLGs in \cite{Gonzalez-Nuevo2012}) have been studied in detail in \cite{Dye2015} by modeling ALMA long baseline imaging. This source has an optical association in the H-ATLAS catalog at 1.87 arcsec with $z$=0.3 which is compatible with both the assocciation we found after the cross-matching and the association identified in \cite{Negrello2017}. \\

The source J134429.5+303034 has no association in the H-ATLAS catalog (within 10 arcsec) with an optical counterpart with known-redshift. Although we find an SDSS galaxy (with unknown redshift) at 0.43 arcsec distance. At larger radii, we find an association with a SDSS galaxy having $z_{spec}$ = 0.67 at 16.96 arcsec. In \cite{Negrello2017} they find a potential lens with $z_{spec}$ = 0.6721 but much closer to the H-ATLAS source (0.43 arcsec). A closer look at the SDSS images reveals how the potential lens can be barely resolved in SDSS images. We propose that the small lens indentified in \cite{Negrello2017} is part of a larger halo (or lens) at $z=0.67$. \\

The source J141351.9-000026 resembles the situation of the previous source. It has no optical association with any galaxy having known spectroscopic redshift in the H-ATLAS catalog but we find a galaxy (with unknown redshift) in the SDSS images at 1.12 arcsec distance. For this source we find an association with a SDSS galaxy at 7.61 arcsec at $z=0.547$. This association differs from the one in \cite{Negrello2017} which finds a much fainter galaxy at 1.12 arcsec (also at $z=0.547$). A closer inspection of the optical high resolution image in \cite{Negrello2017} reveals a giant arc at the position of the IR source. The morphology of the arc (with an arc radius of order 10 arcsec and aligned in a tangential direction with respect to our lens candidate) suggests that our association is the correct one. However, the galaxy found in \cite{Negrello2017} could still be contributing to the magnification since it seems to lie at, or near, the critical curve (for the redshifts of the lens and IR source). \\

The object J115101.7-020024 is also present in \cite{Negrello2017} proposal. It is labeled as an unclear case because no optical association was found. But we have found an association for this source with $z_{spec}$ = 0.4543 at 13.63 arcsec which differs from the optical association assigned in H-ATLAS catalog. \\

J131533.9+233254 is an interesting example. The H-ATLAS catalog contains no optical association. We find an association in SDSS at 8.61 arcsec and with $z=0.551$. The snapshot of this association in Fig.~\ref{fig:postage_2} reveals a number of galaxies in the vicinity of the SDSS galaxy, suggesting that the lens may be a group or a cluster at z=0.55. The stamp for J141351.9-000026 (discussed above) resembles this situation although with fewer SDSS galaxies. Other interesting candidates are J083726.1+015641, J085229.0+010227 and J115819.5-000127 which show a concentration of SDSS galaxies near the IR source. The already mentioned object J091331.4-003644 and J085855.4+013729 are the only ones candidates from \cite{Gonzalez-Nuevo2012} presented in the robust high-$z$ sample for which we found associations with spectroscopic redshifts. \\

The selection shown in Appendix \ref{sec:appendixB} contains only a small sample of potential candidates to be lensing systems. Our search for this type of systems is limited by the fact that we restrict ourselves to SDSS galaxies with spectroscopic redshifts. Future surveys with many more galaxies having spectroscopic redshifts (like EUCLID), or even photometric redshifts with accurate photo-$z$ (like J-PAS) can increase significantly the number of lensed systems.

\section{Conclusions} \label{sec:conclusions} 
%%%%%%%%%%%%%%%%%%%%%%%%%%%%%%%%%%%%%%%%%%%%%%%
In this paper we have extended previous work on the multifrequency matched multifilter (MMF) technique to submm astronomy, in particular the \textit{Herschel}-ATLAS fields. Our multifrequency technique allows us to, on one hand, boost the statistical significance and S/N of the detections and, on the other hand, provide an estimation of the photometric redshift. This approach relies on semi-analytic modeling of the thermal peak of the SED of dusty galaxies \citep[see e.g.,][]{Pearson2013}. We remark that the use of SED templates is not strictly necessary for the MMF technique\footnote{As discussed in \cite{Lanz2010,Lanz2013}, the MMF can work with arbitrary SED points taken as free random variables, but some physical modeling is often both justified and useful.}, but it becomes very useful for the purpose of redshift estimation. We have tested our technique with both realistic simulations and also with H-ATLAS sources for which their redshifts are known, concluding that the MMF indeed leads to an improved S/N with respect to single-frequency detection and that the photometric redshifts are relatively accurate, specially for those H-ATLAS sources with a photometric behavior close the SED used in the range 0.8 < $z$ < 4.3. Outside of this range, we have studied the bias in the MMF photo-$z$ estimation and provided correction factors up to $z \sim 7.0$. \\

410997 point-like sources from the DR1 and DR2 H-ATLAS fields \citep{Valiante2016,Maddox2018} have been re-analyzed with this multifrequency filter. Two different sets of selection criteria based on flux, color, S/N, and agreement between observed photometry and the SED model have been carried out in order to select the H-ATLAS sources for which the MMF method performs the best and illustrate its potential usefulness. A total of 607 H-ATLAS sources, for which new photometry and photometric redshifts are provided, are found. They are split in two samples of interest:

\begin{itemize}
\item A sample of 370 bright high-$z$ objects with $\bar{z}$ = 2.13 and $\sigma_z$ = 0.65. This sample contains 62 of the 80 strongly lensed candidates identified in \cite{Negrello2017}, and including 17 out of the 20 confirmed strongly lensed galaxies. An average improvement of 76\% in the S/N has been achieved for this sample with our MMF method compared to the 500-$\mu$m band. Besides, an average improvement of 16\% and a slight improve of 0.2\% have been obtained for the 350-$\mu$m and 250-$\mu$m, respectively. By cross-correlating this high-$z$ sample with a catalog of low-known-redshift galaxies from SDSS DR14, we find 40 potential candidates to being gravitational lens systems which have been shown and discussed. Some SDSS stamps around these candidates reveal overdensities of galaxies at $z<1$ that could indicate the presence of a group of galaxies acting as a lens. \\

\item A second sample of 237 faint high significance candidates to 500-$\mu$m risers (with $\bar{z}$ = 4.62 and $\sigma_z$ = 0.71) which were previously near the H-ATLAS detection limit but are now confirmed with the MMF as high significance detections. We have achieved average improvements of 25\%, 55\% and 76\% in the S/N for the 500-$\mu$m, 350-$\mu$m and 250-$\mu$m, respectively. This clearly reflects that is in this kind of faint objects where our MMF method reaches its full potential in terms of S/N improvement.

\end{itemize}

\newpage

The MMF has proven to be able to enhance the sensitivity of fainter objects and improve its photometry. According to the comparative analysis of the photometric redshift estimates with known-redshift H-ATLAS sources, it may also return more robust photometric redshifts than the ones that would be obtained using the fluxes of the H-ATLAS catalog, as proven by the mean $\mu$ and median $\mu_{1/2}$ values shown in \cref{tab:spectroscopic,tab:spectroscopic2}. \\

Running a non-blind search, using the already-known source positions in the public H-ATLAS catalog, allows us to directly compare our method with the detection pipeline of the public H-ATLAS catalog. However, given that, in the end, H-ATLAS detection image simply reduces to the 250 $\mu$m channel \citep{Valiante2016}, in the future we plan to apply this method in a blind search mode over the H-ATLAS maps. Thereby we will be able to exploit the information from the other wavelengths and be able to detect faint sources that we could be missing now. \\

The work done here could also be extended in the future focusing only in the sources already confirmed. One of the possible deeper analysis or improvements would be to repeat the MMF execution using other SEDs different from the \cite{Pearson2013} model. For instance, SEDs derived from ultraluminous dusty galaxies Arp200; SMM J2135-0102, ``The Cosmic Eyelash" at $z=2.3$ \citep{Ivison2010} or H-ATLAS J142413.9+022304 alias G15.141 at $z=4.23$ \citep{Cox2011}, which are characteristic of the local submillimeter galaxies. This procedure would show us how much the redshifts and flux densities change depending on the SED used and would help us to select sources in a more robust way. Nevertheless, we think that the \cite{Pearson2013} model is the best model choice since it is not a certain SED from a single source but a empirical model based on 40 H-ATLAS sources and it has already been used in several previous works \citep{Eales15, Ivison16,Bianchini2016,Bianchini2018,Negrello2017,Fudamoto2017,Bakx2017,Donevski18}. This further analysis would also allow to check whether the methodology bias observed in simulations is because of the particular shape of the function employed.

\section*{Acknowledgements}

We acknowledge partial financial support from the Spanish `Ministerio de Econom\'ia y Competitividad' (MINECO) project AYA2015-64508-P. DH also acknowledges support from the RADIOFOREGROUNDS project, funded by the European Comission's H2020 Research Infrastructures under the Grant Agreement 687312. 
JGN and LB also acknowledge financial support from the I+D 2015 project AYA2015-65887-P (MINECO/FEDER). JGN also acknowledges financial support from the Spanish MINECO for a `Ramon y Cajal' fellowship (RYC-2013-13256).
We acknowledge Santander Supercomputacion support group at the University of Cantabria who provided access to the supercomputer Altamira Supercomputer at the Institute of Physics of Cantabria (IFCA-CSIC), member of the Spanish Supercomputing Network, for performing simulations and analyses.\\

The \textit{Herschel}-ATLAS is a project with \textit{Herschel}, which is an ESA space observatory with science instruments provided by European-led Principal Investigator consortia and with important participation from NASA. The H-ATLAS website is \url{http://www.h-atlas.org/}.\\ 

GAMA is a joint European-Australasian project based around a spectroscopic campaign using the Anglo-Australian Telescope. The GAMA input catalog is based on data taken from the Sloan Digital Sky Survey and the UKIRT Infrared Deep Sky Survey. Complementary imaging of the GAMA regions is being obtained by a number of independent survey programs including GALEX MIS, VST KiDS, VISTA VIKING, WISE, Herschel-ATLAS, GMRT and ASKAP providing UV to radio coverage. GAMA is funded by the STFC (UK), the ARC (Australia), the AAO, and the participating institutions. The GAMA website is \url{http://www.gama-survey.org/}.\\

Funding for the SDSS and SDSS-II has been provided by the Alfred P. Sloan Foundation, the Participating Institutions, the National Science Foundation, the U.S. Department of Energy, the National Aeronautics and Space Administration, the Japanese Monbukagakusho, the Max Planck Society, and the Higher Education Funding Council for England. The SDSS Web Site is \url{http://www.sdss.org/}. The SDSS is managed by the Astrophysical Research Consortium for the Participating Institutions. The Participating Institutions are the American Museum of Natural History, Astrophysical Institute Potsdam, University of Basel, University of Cambridge, Case Western Reserve University, University of Chicago, Drexel University, Fermilab, the Institute for Advanced Study, the Japan Participation Group, Johns Hopkins University, the Joint Institute for Nuclear Astrophysics, the Kavli Institute for Particle Astrophysics and Cosmology, the Korean Scientist Group, the Chinese Academy of Sciences (LAMOST), Los Alamos National Laboratory, the Max-Planck-Institute for Astronomy (MPIA), the Max-Planck-Institute for Astrophysics (MPA), New Mexico State University, Ohio State University, University of Pittsburgh, University of Portsmouth, Princeton University, the United States Naval Observatory, and the University of Washington.

%%%%%%%%%%%%%%%%%%%%%%%%%%%%%%%%%%%%%%%%%%%%%%%%%%
\bibliographystyle{aa}
\bibliography{manjon}
%%%%%%%%%%%%%%%%%%%%%%%%%%%%%%%%%%%%%%%%%%%%%%%%%%

%%%%%%%%%%%%%%%%% APPENDICES %%%%%%%%%%%%%%%%%%%%%

\begin{appendix}

\section{Appendix A: Selected high redshift sources} \label{sec:appendixA}

We present here two tables: the first one with the 20 first objects out of the 370 high-$z$ H-ATLAS sources from the robust high-$z$ sample and the second one containing the 20 first objects out of the 237 high-$z$ H-ATLAS sources from the 500 $\mu$m-riser sample.

\begin{table*}[h!]
	\centering
	\renewcommand\thetable{A.1}
	\begin{tabular}{lcccccccccccr} % four columns, alignment for each
		\hline \\
		No. & H-ATLAS name & $\alpha$ ($\mathrm{deg}$) & $\delta$ ($\mathrm{deg}$) & $z_{MMF}^{unbiased}$ & & $S_{MMF}$ (mJy/beam) & \\
        \cline{6-8}
        & & & & & 250 $\mu$m & 350 $\mu$m & 500 $\mu$m \\
        \hline \\
1 & HATLAS J083051.0+013225 & 127.712684 & 1.540284 & 3.15 $\pm$ 0.34 & 230 $\pm$ 5 & 287 $\pm$ 7 & 265 $\pm$ 6 \\
2 & HATLAS J083218.0+002527 & 128.075027 & 0.424236 & 1.08 $\pm$ 0.26 & 119 $\pm$ 7 & 80 $\pm$ 5 & 37 $\pm$ 2 \\
3 & HATLAS J083344.9+000109 & 128.437396 & 0.019282 & 3.16 $\pm$ 0.34 & 58 $\pm$ 5 & 72 $\pm$ 6 & 67 $\pm$ 6 \\
4 & HATLAS J083726.1+015641 & 129.358797 & 1.944871 & 1.71 $\pm$ 0.27 & 118 $\pm$ 8 & 103 $\pm$ 7 & 61 $\pm$ 4 \\
5 & HATLAS J083904.5+022848 & 129.768948 & 2.480177 & 2.54 $\pm$ 0.30 & 48 $\pm$ 7 & 53 $\pm$ 7 & 41 $\pm$ 6 \\
6 & HATLAS J083929.4+023536 & 129.872679 & 2.593495 & 1.66 $\pm$ 0.27 & 122 $\pm$ 8 & 104 $\pm$ 7 & 60 $\pm$ 4 \\
7 & HATLAS J084010.0+014336 & 130.04181 & 1.726677 & 2.71 $\pm$ 0.31 & 40 $\pm$ 6 & 46 $\pm$ 7 & 38 $\pm$ 6 \\
8 & HATLAS J084055.2+001819 & 130.230288 & 0.305532 & 1.37 $\pm$ 0.26 & 113 $\pm$ 7 & 86 $\pm$ 6 & 44 $\pm$ 3 \\
9 & HATLAS J084306.5+012342 & 130.777236 & 1.395141 & 3.15 $\pm$ 0.34 & 45 $\pm$ 5 & 56 $\pm$ 7 & 51 $\pm$ 6 \\
10 & HATLAS J084615.9+012004 & 131.566339 & 1.334708 & 1.49 $\pm$ 0.27 & 114 $\pm$ 6 & 91 $\pm$ 5 & 49 $\pm$ 3 \\
11 & HATLAS J084641.1+002738 & 131.671655 & 0.460763 & 2.42 $\pm$ 0.30 & 54 $\pm$ 5 & 59 $\pm$ 6 & 44 $\pm$ 4 \\
12 & HATLAS J084658.7+021112 & 131.744841 & 2.186738 & 0.99 $\pm$ 0.27 & 205 $\pm$ 8 & 132 $\pm$ 5 & 59 $\pm$ 2 \\
13 & HATLAS J084723.8+015430 & 131.849524 & 1.908484 & 2.34 $\pm$ 0.30 & 58 $\pm$ 7 & 61 $\pm$ 7 & 45 $\pm$ 5 \\
14 & HATLAS J084740.8+002421 & 131.920354 & 0.405969 & 1.85 $\pm$ 0.28 & 56 $\pm$ 6 & 51 $\pm$ 6 & 32 $\pm$ 3 \\
15 & HATLAS J084859.1-015438 & 132.246479 & -1.910578 & 2.91 $\pm$ 0.32 & 64 $\pm$ 4 & 77 $\pm$ 4 & 66 $\pm$ 4 \\
16 & HATLAS J085015.2+005658 & 132.563354 & 0.949446 & 2.37 $\pm$ 0.30 & 83 $\pm$ 7 & 89 $\pm$ 7 & 65 $\pm$ 5 \\
17 & HATLAS J085034.0+003231 & 132.641952 & 0.541958 & 2.82 $\pm$ 0.32 & 46 $\pm$ 6 & 54 $\pm$ 7 & 45 $\pm$ 6 \\
18 & HATLAS J085111.7+004934 & 132.79885 & 0.826131 & 1.77 $\pm$ 0.27 & 115 $\pm$ 6 & 103 $\pm$ 6 & 61 $\pm$ 3 \\
19 & HATLAS J085133.1+014940 & 132.888113 & 1.828026 & 2.58 $\pm$ 0.31 & 43 $\pm$ 6 & 48 $\pm$ 7 & 38 $\pm$ 5 \\
20 & HATLAS J085144.1+014214 & 132.933947 & 1.703908 & 1.35 $\pm$ 0.26 & 129 $\pm$ 9 & 98 $\pm$ 7 & 50 $\pm$ 3 \\
	\\ \hline
	\end{tabular}
    \caption{20 first objects of the 370 high-$z$ H-ATLAS sources that form our robust high-$z$ sample.}
	\label{tab:sample_370_a}
\end{table*}

\begin{table*}[h!]
	\centering
	\renewcommand\thetable{A.2}
	\begin{tabular}{lcccccccccccr} % four columns, alignment for each
		\hline \\
		No. & H-ATLAS name & $\alpha$ ($\mathrm{deg}$) & $\delta$ ($\mathrm{deg}$) & $z_{MMF}^{unbiased}$ & & $S_{MMF}$ (mJy/beam) & \\
        \cline{6-8}
        & & & & & 250 $\mu$m & 350 $\mu$m & 500 $\mu$m \\
        \hline \\
1 & HATLAS J083359.4+005008 & 128.497828 & 0.835735 & 3.81 $\pm$ 0.37 & 22 $\pm$ 4 & 31 $\pm$ 6 & 33 $\pm$ 6 \\
2 & HATLAS J083436.7+012907 & 128.653175 & 1.485354 & 4.41 $\pm$ 0.39 & 17 $\pm$ 3 & 28 $\pm$ 5 & 33 $\pm$ 6 \\
3 & HATLAS J083700.1+013111 & 129.250571 & 1.519785 & 6.03 $\pm$ 0.34 & 14 $\pm$ 2 & 39 $\pm$ 5 & 60 $\pm$ 7 \\
4 & HATLAS J084116.1+002619 & 130.317356 & 0.438641 & 5.01 $\pm$ 0.40 & 15 $\pm$ 3 & 29 $\pm$ 5 & 37 $\pm$ 7 \\
5 & HATLAS J084213.4-000531 & 130.556008 & -0.091959 & 3.76 $\pm$ 0.37 & 26 $\pm$ 5 & 36 $\pm$ 6 & 38 $\pm$ 7 \\
6 & HATLAS J084644.6+002352 & 131.686002 & 0.397871 & 3.96 $\pm$ 0.38 & 18 $\pm$ 4 & 26 $\pm$ 5 & 28 $\pm$ 6 \\
7 & HATLAS J085340.9+024044 & 133.420602 & 2.679046 & 5.14 $\pm$ 0.39 & 13 $\pm$ 2 & 26 $\pm$ 5 & 35 $\pm$ 7 \\
8 & HATLAS J085534.2+005258 & 133.892785 & 0.882992 & 5.29 $\pm$ 0.39 & 12 $\pm$ 2 & 26 $\pm$ 5 & 35 $\pm$ 7 \\
9 & HATLAS J090029.8+001653 & 135.12445 & 0.281402 & 5.17 $\pm$ 0.39 & 11 $\pm$ 2 & 22 $\pm$ 4 & 29 $\pm$ 5 \\
10 & HATLAS J090045.5+004131 & 135.189985 & 0.69217 & 6.35 $\pm$ 0.30 & 8 $\pm$ 1 & 25 $\pm$ 4 & 40 $\pm$ 6 \\
11 & HATLAS J090304.5-004616 & 135.769163 & -0.771248 & 6.35 $\pm$ 0.30 & 7 $\pm$ 1 & 23 $\pm$ 4 & 37 $\pm$ 6 \\
12 & HATLAS J090329.6+000753 & 135.873578 & 0.131548 & 4.89 $\pm$ 0.40 & 18 $\pm$ 3 & 33 $\pm$ 5 & 43 $\pm$ 6 \\
13 & HATLAS J090540.0+020347 & 136.417044 & 2.063127 & 3.65 $\pm$ 0.36 & 32 $\pm$ 5 & 44 $\pm$ 7 & 45 $\pm$ 7 \\
14 & HATLAS J090908.6-010140 & 137.286131 & -1.027792 & 4.33 $\pm$ 0.39 & 20 $\pm$ 4 & 31 $\pm$ 6 & 36 $\pm$ 7 \\
15 & HATLAS J091019.4+011755 & 137.581142 & 1.298769 & 4.04 $\pm$ 0.38 & 16 $\pm$ 3 & 24 $\pm$ 4 & 26 $\pm$ 5 \\
16 & HATLAS J091025.9+002116 & 137.60821 & 0.354684 & 4.34 $\pm$ 0.39 & 16 $\pm$ 3 & 25 $\pm$ 5 & 29 $\pm$ 5 \\
17 & HATLAS J091617.1+010036 & 139.071573 & 1.010031 & 4.97 $\pm$ 0.40 & 14 $\pm$ 3 & 26 $\pm$ 5 & 34 $\pm$ 6 \\
18 & HATLAS J091733.4+005054 & 139.389273 & 0.848382 & 4.34 $\pm$ 0.39 & 21 $\pm$ 3 & 33 $\pm$ 5 & 39 $\pm$ 6 \\
19 & HATLAS J091902.6-003905 & 139.760959 & -0.651409 & 5.80 $\pm$ 0.36 & 8 $\pm$ 2 & 21 $\pm$ 4 & 31 $\pm$ 6 \\
20 & HATLAS J092215.4+002920 & 140.564326 & 0.489096 & 3.93 $\pm$ 0.38 & 22 $\pm$ 4 & 32 $\pm$ 6 & 35 $\pm$ 7 \\
	\\ \hline
	\end{tabular}
    \caption{20 first objects of the 237 high-$z$ H-ATLAS sources that form our 500 $\mu$m-riser sample.}
	\label{tab:sample_237_a}
\end{table*}

\section{Appendix B: Postage Stamps of Candidate Lensed Galaxies} \label{sec:appendixB}

We show here the postage stamps of the 40 candidate lensed galaxies selected after cross-matching the robust high-$z$ subsample of 283 high-$z$ candidates with a sample of known-low redshift galaxies from SDSS DR14.

\FloatBarrier

\begin{figure*}
        \renewcommand\thefigure{B.1}
	\includegraphics[width=15.5cm, height=24cm]{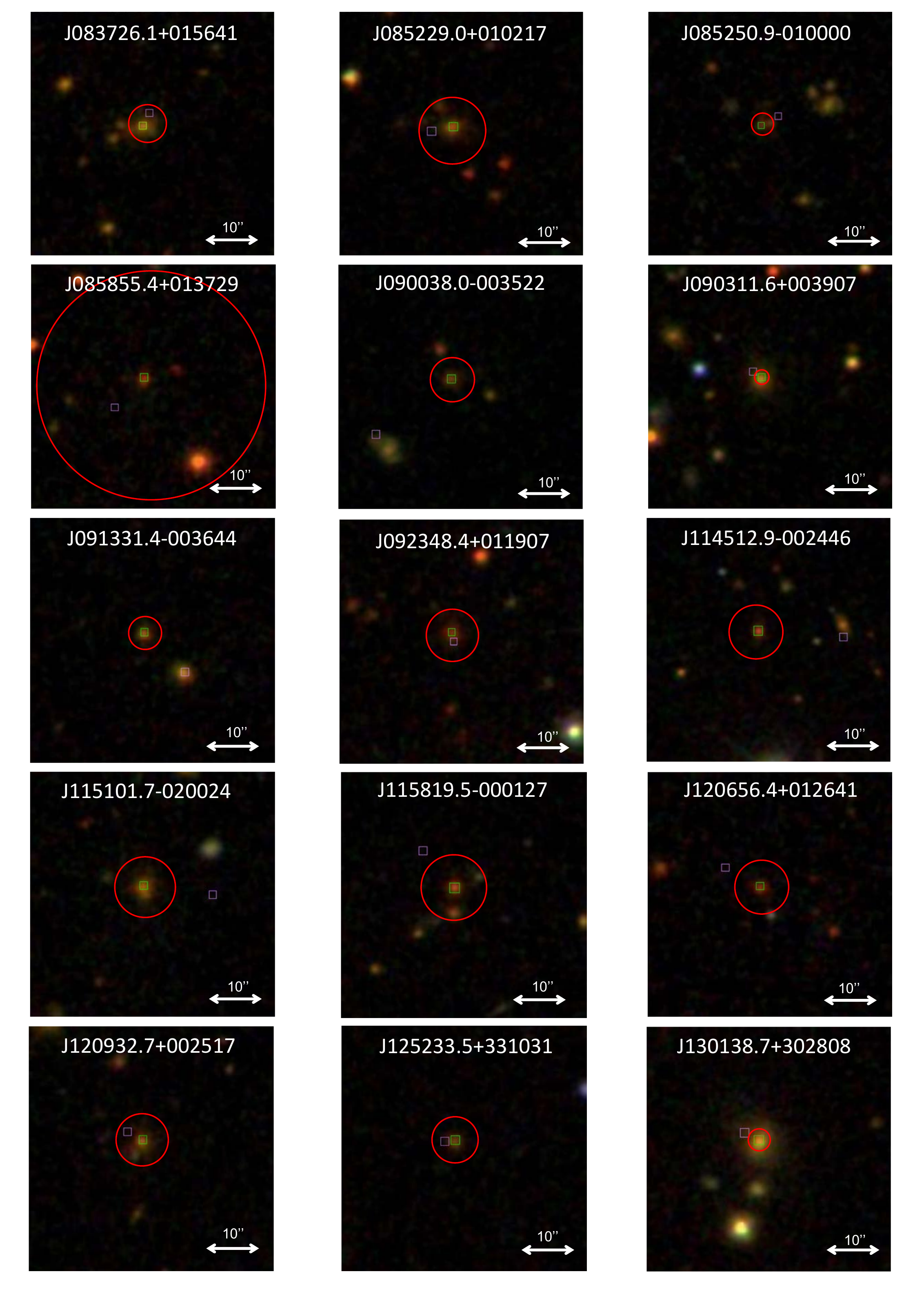}
    \caption{SDSS images (47 x 47 $\mathrm{arcsec^{2}}$) of part of the 40 associations found between the robust high-$z$ subsample of 283 high-$z$ candidates and sample of 1776242 known-low-redshift galaxies from SDSS DR14. H-ATLAS sources and SDSS associations are marked with a pink and a green square, respectively. Einstein rings according to the Einstein radii estimated in Table~\ref{tab:40_matches} are drawn as red circles.}
    \label{fig:postage_1}
\end{figure*}

\FloatBarrier

\begin{figure*}
        \renewcommand\thefigure{B.2}
	\includegraphics[width=15.5cm, height=24cm]{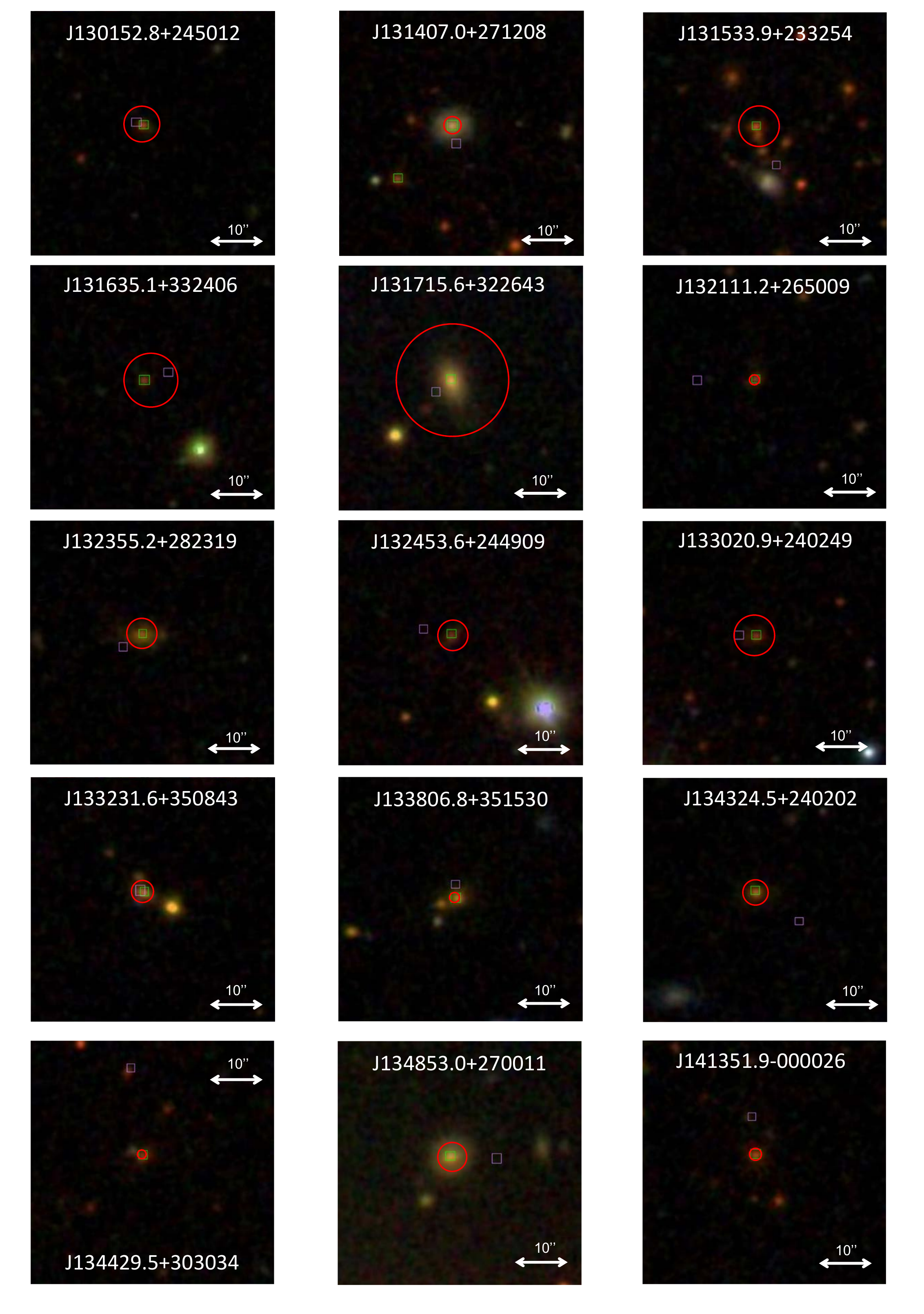}
    \caption{SDSS images (47 x 47 $\mathrm{arcsec^{2}}$) of part of the 40 associations found between the robust high-$z$ subsample of 283 high-$z$ candidates and sample of 1776242 known-low-redshift galaxies from SDSS DR14. H-ATLAS sources and SDSS associations are marked with a pink and a green square, respectively. Einstein rings according to the Einstein radii estimated in Table~\ref{tab:40_matches} are drawn as red circles.}
    \label{fig:postage_2}
\end{figure*}

\FloatBarrier

\begin{figure*}
        \renewcommand\thefigure{B.3}
	\includegraphics[width=16cm, height=19.6cm]{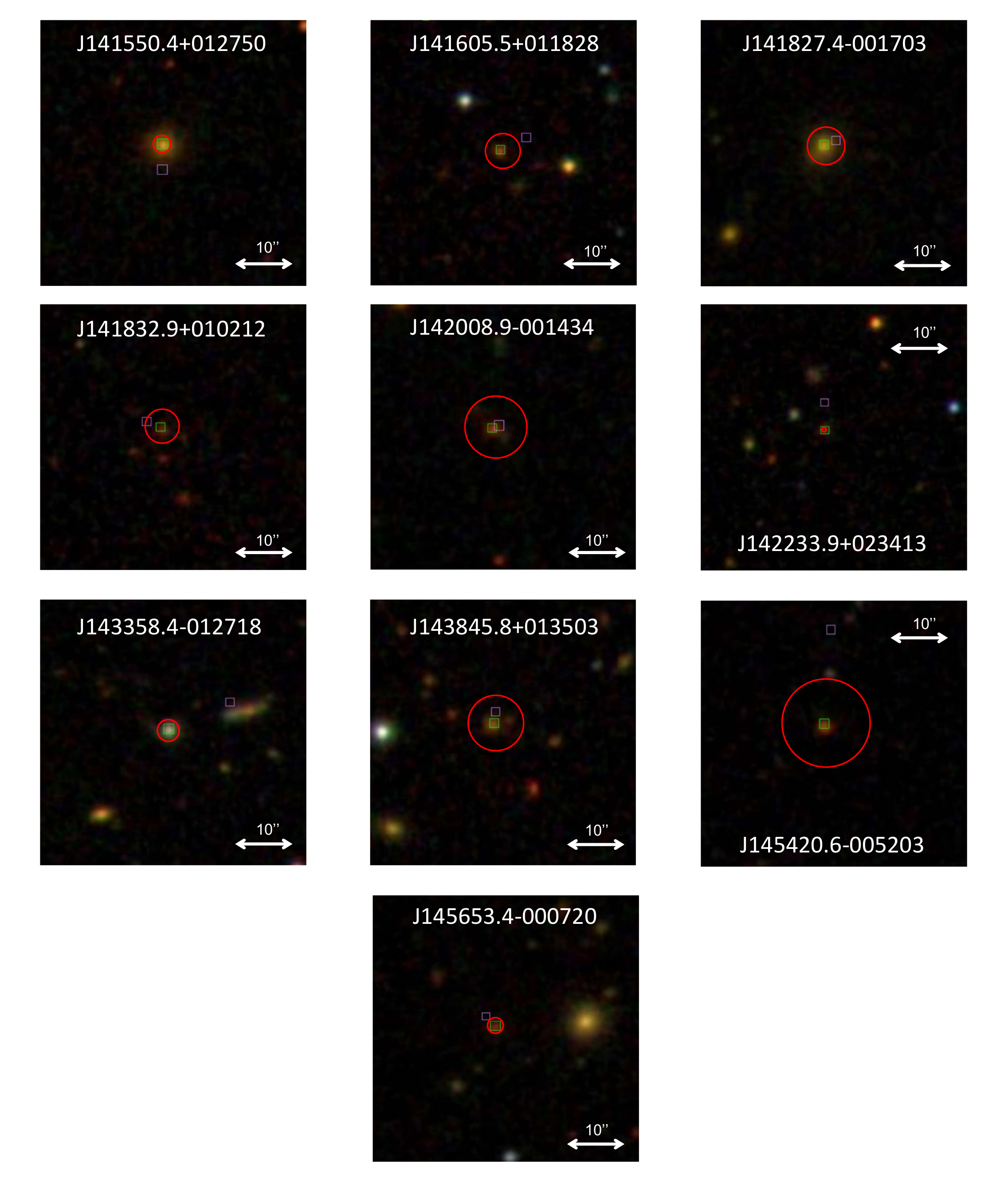}
    \caption{SDSS images (47 x 47 $\mathrm{arcsec^{2}}$) of part of the 40 associations found between the robust high-$z$ subsample of 283 high-$z$ candidates and sample of 1776242 known-low-redshift galaxies from SDSS DR14. H-ATLAS sources and SDSS associations are marked with a pink and a green square, respectively. The object J145420.6-005203 is identified as a QSO in the H-ATLAS catalog. Einstein rings according to the Einstein radii estimated in Table~\ref{tab:40_matches} are drawn as red circles.}
    \label{fig:postage_3}
\end{figure*}

\end{appendix}

%%%%%%%%%%%%%%%%%%%%%%%%%%%%%%%%%%%%%%%%%%%%%%%%%%

% Don't change these lines
%\bsp	% typesetting comment
\label{lastpage}
\end{document}